\documentclass[a4paper,fleqn,usenatbib]{mnras}
\usepackage[utf8]{inputenc}

\usepackage{acronym}
\usepackage{ae,aecompl}
\usepackage{booktabs}
\usepackage{dcolumn}
\usepackage{graphicx}
\usepackage{amsmath,amsfonts,amssymb}
\usepackage{mathrsfs}
\usepackage[normalem]{ulem}

\usepackage[T1]{fontenc}

\newcommand{\fornax}{F{\sc{ornax}}}

\newcolumntype{d}[1]{D{.}{.}{#1}}

\title{The Overarching Framework of Core-Collapse Supernova Explosions as Revealed by 3D \fornax\ Simulations}
\author[A.~Burrows, D.~Radice, D.~Vartanyan, et al.]{
Adam Burrows$^{1}$,
David Radice$^{1,2,3,4}$,
David Vartanyan$^{1}$,
Hiroki Nagakura$^{1}$,
\newauthor\phantom{} M. Aaron Skinner$^{5}$, and
Joshua C.~Dolence$^{6}$ \\
$^1$ Department of Astrophysical Sciences, Princeton University,
4 Ivy Lane, Princeton, NJ 08544, USA \\
$^2$ Institute for Advanced Study, 1 Einstein Drive,
Princeton, NJ 08540, USA \\
$^3$ Department of Physics, The Pennsylvania State University, University Park, PA 16802, USA\\
$^4$ Department of Astronomy \& Astrophysics, The Pennsylvania State University, University Park, PA 16802, USA\\
$^5$ Lawrence Livermore National Laboratory, 7000 East Ave.,
Livermore, CA 94550-9234 \\
$^6$ CCS-2, Los Alamos National Laboratory, P.O. Box 1663
Los Alamos, NM 87545 
}

\date{Accepted XXX. Received YYY; in original form ZZZ}

\pubyear{2019}

\begin{document}
\label{firstpage}
\pagerange{\pageref{firstpage}--\pageref{lastpage}}
\maketitle

\begin{abstract}
We have conducted nineteen state-of-the-art 3D core-collapse supernova
simulations spanning a broad range of progenitor masses.  This is the largest
collection of sophisticated 3D supernova simulations ever performed.
We have found that while the majority of these models explode, not all do, and that
even models in the middle of the available progenitor mass range may be less explodable.
This does not mean that those models for which we did not witness explosion would not explode
in Nature, but that they are less prone to explosion than others. One
consequence is that the ``compactness" measure is not a metric for explodability.
We find that lower-mass massive star progenitors likely experience lower-energy explosions,
while the higher-mass massive stars likely experience higher-energy explosions.
Moreover, most 3D explosions have a dominant dipole morphology, have a pinched,
wasp-waist structure, and experience simultaneous accretion and explosion.
We reproduce the general range of residual neutron-star masses inferred
for the galactic neutron-star population.  The most massive progenitor models, however,
in particular vis \`a vis explosion energy,  need to be continued for longer physical
times to asymptote to their final states.  We find that while the majority of the inner ejecta
have Y$_e = 0.5$, there is a substantial proton-rich tail.  This result has important
implications for the nucleosynthetic yields as a function of progenitor. Finally, we find
that the non-exploding models eventually evolve into compact inner configurations that experience
a quasi-periodic spiral SASI mode. We otherwise see little evidence of the SASI in the exploding
models.
\end{abstract}


%% -------------------------------------------------------
\section{Introduction}
%% -------------------------------------------------------

At the end of the quasi-static life of tens of millions to millions of years
of a star perhaps more massive than $\sim$8 M$_{\odot}$, its white-dwarf-like 
core is thought to experience the Chandrasekhar instability. This core 
would then dynamically implode to nuclear densities within less than a second,
giving birth in such a violent ``core collapse" to either a neutron 
star or ``stellar mass" black hole. It is thought that most of the time this scenario produces 
a gravitationally-powered supernova explosion, a core-collapse supernova (CCSN),
and that all Type IIp, IIb, IIn, Ib, and Ic supernovae, collectively the vast majority, 
originate in this context. The neutrino detections of SN 1987A 
\citep{1987PhRvL..58.1490H,1987PhRvL..58.1494B} support this general notion,
but the complexity of the theory and the heterogeneity of the observational database
mitigate against simple physical scenarios.

One ultimate goal of supernova theory is the credible mapping between progenitor
star and dynamical outcome.  Which massive stars end their lives in supernovae, with 
what properties, and why?  Inspired by this goal and using our new 
state-of-the-art radiation/hydrodynamic code \fornax (\S\ref{methods}), we have conducted
a suite of three-dimensional (3D) core-collapse and explosion simulations
of unprecedented breadth across most of the expected progenitor continuum to ascertain the 
differences in outcome as a function of initial core structure. This study
encompasses nineteen 3D simulations with competitive physical realism 
for progenitors with masses of 9-, 10-, 11-, 12-, 13-, 14-, 15-, 16-, 17-, 18-, 
19-, 20-, 25-, and 60-M$_{\odot}$.  These progenitors were all calculated
by  \citet{2016ApJ...821...38S}, except for the 25-M$_{\odot}$ progenitor 
which was taken from \citet{2018ApJ...860...93S}. This is by far the 
largest number of 3D simulations ever performed.  

Until recently, the complexity in 3D of factors and effects important to explosion had slowed progress in capturing
all the major processes and phenomena thought necessary to an ultimate resolution of the mechanism of core-collapse
supernova explosions.  These included avoiding the sloshing artifacts seen in two-dimensional (2D) axial simulations;
moving beyond the problematic ray-by-ray{+} transport simplification
(see \citet{2016ApJ...831...81S} and \S\ref{methods}); incorporating
all the important neutrino-matter interaction rates; capturing the post-shock turbulence
hydrodynamics to an acceptable degree; allowing simultaneous accretion and explosion (shown
to be important in maintaining neutrino driving), impossible in one dimension (1D)\footnote{but
also possible and seen in 2D}; naturally enabling (by calculating in multi-D all
the way to the center) the interior proto-neutron-star (PNS) convection that
can alter late-time neutrino luminosities \citep{2017ApJ...850...43R,2006ApJ...645..534D};
and including inelastic scattering and its associated matter heating effects
in the gain region \citep{1985ApJ...295...14B} behind the shock.
Now, with the advent of codes
%%% \footnote{such as \citet{2015ApJ...807L..31L}, \citet{melson:15b}, \citet{2016ApJ...829L..14K}, \citet{2016ApJ...831...98R},
%%% \citet{summa2018}, \citet{2018ApJ...865...81O}, and \citet{2019ApJ...873...45G}}
such as \fornax, albeit still evolving, 3D calculations that contain
the necessary realism are available to capture much of this complexity at a sufficient level of detail
and with respectable physical fidelity.

\begin{figure}
  \includegraphics[width=\columnwidth]{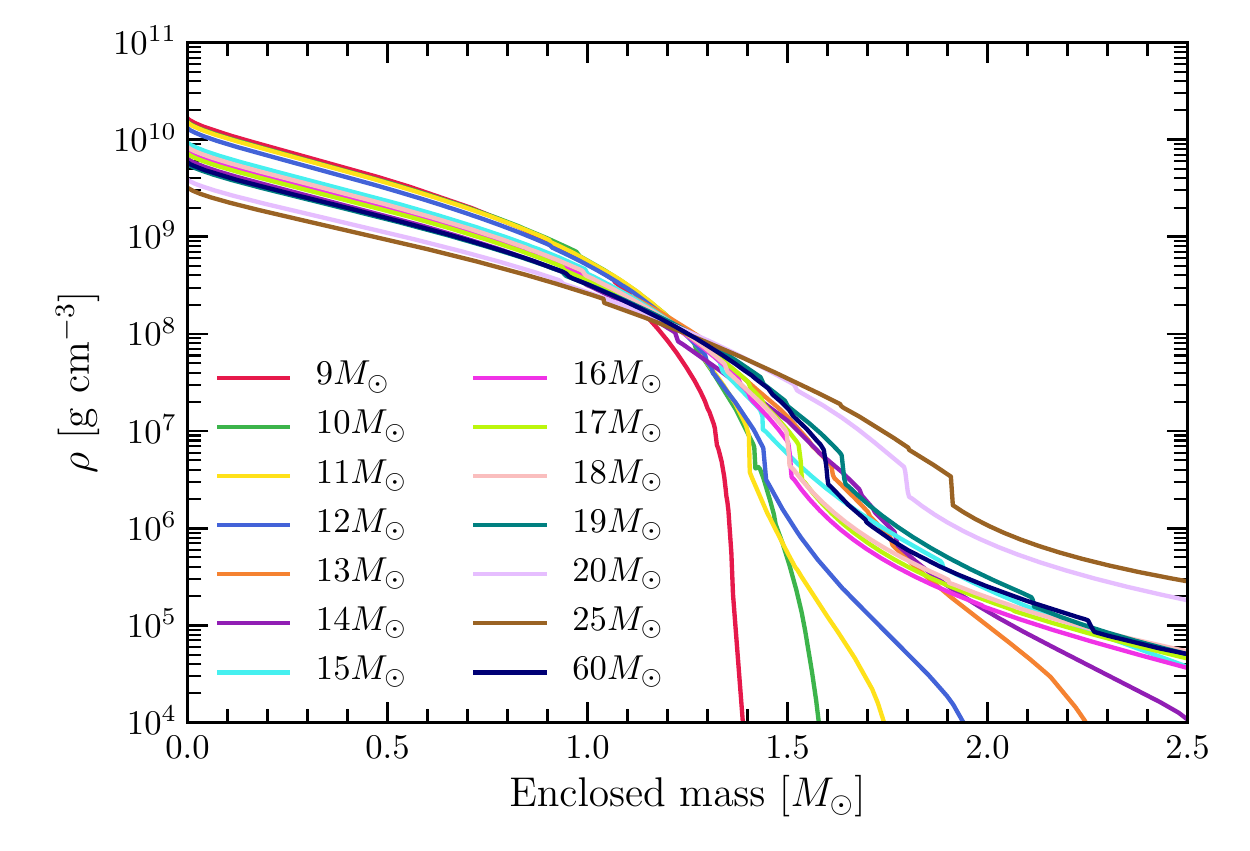}
  \caption{Density profiles for the progenitors considered in our study.
  We obtain successful explosions for both low-compactness stellar
  cores, such as those of the 9-M$_\odot$ and 10-M$_\odot$ progenitors,
  as well as for high-compactness stellar cores with sharp density drops
  at the Si/O interface, such as that of the 25-M$_\odot$ progenitor.}
  \label{density}
\end{figure}

\begin{table}
  \center{
  \begin{tabular}{lll}
    \hline
            Progenitor    & Envelope Binding Energy &  Compactness \\

            (M$_{\odot}$)         & ($10^{51}$ ergs) & (calculated at 1.75 M$_{\odot}$)  \\
    \hline
s9.0    &     0.002     &           $3.831\times{10^{-5}}$  \\
s10.0   &     0.012     &           $2.165\times{10^{-4}}$  \\
s11.0   &     0.025     &           $7.669\times{10^{-3}}$  \\
s12.0   &     0.050     &           $2.215\times{10^{-2}}$  \\
s13.0   &     0.072     &           $5.932\times{10^{-2}}$  \\
s14.0   &     0.110     &           0.1243  \\
s15.0   &     0.144     &           0.1674  \\
s16.0   &     0.212     &           0.1546  \\
s17.0   &     0.251     &           0.1644  \\
s18.0   &     0.309     &           0.1715  \\
s19.0   &     0.341     &           0.1783  \\
s20.0   &     0.413     &           0.2615  \\
s25.0   &     0.865     &           0.3010  \\
s60.0   &     0.513     &           0.1753  \\
    \hline
  \end{tabular}
  \caption{The binding energy (in units of $10^{51}$ ergs, one Bethe)
of the stellar envelope exterior to the 20,000 km outer boundary of the
computational domain and the so-called compactness of the progenitor,
calculated at 1.75 M$_{\odot}$. The former is the energy penalty an
explosion witnessed on the computational domain still has to pay to
eject that outer envelope material and reach ``infinity."  The compactness is
a crude, but ofttimes useful, metric of the shallowness of the mass density
profile of the inner progenitor.  It has been shown that while compactness is
not a measure of explodability it is roughly correlated with the envelope binding
energy \protect\citep{2018SSRv..214...33B}, a proposition this table supports.
  \label{sn_tab}
  }}
\end{table}

The mass density profiles of massive stars 
were once thought to be roughly monotonic with ZAMS mass.  This might have translated 
into a smooth dependence upon progenitor mass of the explosion characteristics, 
for a given metallicity.  However, recent 1D studies 
\citep{2016ApJ...821...38S,2018ApJ...860...93S,2019ApJ...878...49W} have called
this simple picture into question, with slight ``chaos" resulting in a non-monotonic dependence on 
the shallowness of the mass density profile in the crucial inner core.  
Figure \ref{density} depicts the mass density profiles 
for the initial models we employ for this study. In addition, Table \ref{sn_tab} 
provides for our model suite the ``compactness," a simple one-dimensional metric of
shallowness \citep{2013ApJ...762..126O}.  Compactness does affect
the evolution of the infall accretion rate, and, hence, the neutrino luminosities 
and neutrino energies.  Therefore, in the context of the neutrino-driven mechanism 
of explosion, it affects whether, when, and how a model explodes.  However, 
we have found in recent studies in 2D \citep{2018SSRv..214...33B} and 3D \citep{2019MNRAS.485.3153B} 
that this naive picture is not complete and that explodability is not correlated with compactness
in a simple way.  With this paper, we expand this notion and find that the 13-, 14-, and 15-M$_{\odot}$
models (not only the 13-M$_{\odot}$ model studied in \citealt{2019MNRAS.485.3153B}), 
fail to explode, when all the other models do\footnote{Importantly, unpublished 
higher angular-resolution (678$\times$256$\times$512, see \S\ref{methods}) simulations
we have recently performed of these same 13-M$_{\odot}$ and 15-M$_{\odot}$ progenitors, though their
mean shock radii do achieve slightly larger values, still do not explode.}.  This makes more firm the preliminary
conclusion in \citet{2019MNRAS.485.3153B} that there may be a mass gap in explodability
near the middle of the massive-star mass function.  In fact, we now find, and 
demonstrate in this paper, that both low and high compactness models explode, 
with the high compactness models likely exploding the most energetically, albeit later.  
It was once thought that low compactness and a steep initial density profile were 
prerequisites for explodability.  Our new results put this notion in doubt.   

To add further complexity, recent 3D stellar evolution studies reveal mixing
processes, gravity waves, and dynamics that the problematic mixing-length prescription
for convection can not capture and, therefore, that the progenitor landscape is still not fully understood
\citep{2015ApJ...808L..21C,2016A&A...593A..72J,2016ApJ...822...61C,muller2017,2019MNRAS.484.3307M,2019A&A...622A..74J,2019arXiv190307811Y}.
Moreover, the potential role of aspherical perturbations in the progenitor models in inaugurating
and maintaining turbulent convection behind the stalled shock wave
\citep{2015ApJ...799....5C,muller2017,2018SSRv..214...33B,2018MNRAS.477.3091V,2019MNRAS.484.3307M}, shown
to be important in igniting neutrino-driven explosions \citep{1995ApJ...450..830B}, highlights
the need to determine their magnitude and character.  One-dimensional stellar-evolution 
calculations are clearly not adequate. Therefore, the reader should keep in
mind the provisional character of current progenitor models employed by supernova theorists and the
ongoing need for further improvement. Nevertheless, such is the span of the mass density profiles of the model continuum we
incorporate in this study that though the derived mapping between ZAMS and outcome itself is probably provisional,
the range of behaviors in this wide progenitor range from 9 M$_{\odot}$, through 25 M$_{\odot}$, to 60 M$_{\odot}$
is probably, in the main, captured.

In this paper, in addition to determining whether and when these models explode, we present
the shock radius development, the integrated neutrino luminosities, the final
masses of exploding models, the neutrino heating rates, the spherical-harmonic
decompositions of the shock surface, the diagnostic explosion energy and its rate of climb,
the ejecta masses, the ejecta electron-fraction (Y$_e$) distributions, and approximate maps
of the putative ejecta $^{56}$Ni distributions. 
In \S\ref{methods}, we describe in detail the specifications of \fornax\ and the computational setup employed
for this paper. In \S\ref{results}, we provide our results, including explosion properties, the integrated 
neutrino luminosities, the spherical-harmonic decompositions of the shock surface, the 
diagnostic explosion energy and its rate of climb, the neutrino heating rates,
the final neutron-star masses of exploding models, the ejecta masses and the ejecta electron-fraction 
(Y$_e$) distributions, and approximate maps of the inferred ejecta $^{56}$Ni distributions.
In \S\ref{sensitivity}, we show how a model's hydrodynamic behavior might depend upon resolution, 
the Horowitz many-body correction, and employing a monopole term in place of a multipole expansion
to handle gravity.  We summarize our general results and conclusions in \S\ref{conclusions}.

%% -------------------------------------------------------
\section{Numerical Methods and Computational Setup}
\label{methods}
%% -------------------------------------------------------

The numerical and physical details incorporated into the code 
\fornax\ have been published in numerous papers in recent years
\citep{2016ApJ...831...81S,2018SSRv..214...33B,2018MNRAS.477.3091V,2019MNRAS.485.3153B,fornaxcode:19}.
In particular, \citet{fornaxcode:19} provided a challenging set
of hydrodynamic, radiation, and radiation-hydrodynamic tests and described the discritization,
reconstruction, solver, algorithms, and implementation specifics of \fornax.

Most of the code is written in C, with only a few Fortran 95 routines
for reading in microphysical data tables; we use an MPI/OpenMP
hybrid paralelism model. F{\sc{ornax}} employs spherical coordinates 
in one, two, and three spatial dimensions, solves the comoving-frame, 
multi-group, two-moment, velocity-dependent transport equations to O($v/c$),
and uses the M1 tensor closure for the second and third moments of the radiation
fields \citep{2011JQSRT.112.1323V}.  We do not use the dimensional reduction 
simplification known as ``ray-by-ray{+}" employed by most other groups, but follow
the vector flux densities of the first-moment equations in 3D.  The ray-by-ray+
approach, {though it addresses the lateral advective transport of lepton number,} 
has been shown to introduce artifacts in the results, particularly 
for 2D simulations \citep{2016ApJ...831...81S} and aspherical 3D simulations 
\citep{2019ApJ...873...45G}.  

Three species of neutrino
($\nu_e$, $\bar{\nu}_e$, and ``$\nu_{\mu}$" [$\nu_{\mu}$, $\bar{\nu}_{\mu}$,
$\nu_{\tau}$, and $\bar{\nu}_{\tau}$ lumped together]) are followed using
an explicit Godunov characteristic method applied to the radiation transport
operators, but an implicit solver for the radiation source terms.
In this way, the radiative transport and transfer
are handled locally, without the need for a global solution on the entire mesh.
This is also the recent approach taken by \citet{OConnor:2015rwy}, \citet{2019ApJ...873...45G}, 
and \citet{2018ApJ...865...81O}, though with some important differences. 
By addressing the transport operator with an explicit method, we significantly reduce 
the computational complexity and communication overhead of traditional multi-dimensional
radiative transfer solutions by bypassing the need for global iterative solvers that have proven
to be slow and/or problematic beyond $\sim$10,000 cores. Strong scaling of the transport solution in
three dimensions using F{\sc{ornax}} is excellent beyond 100,000 tasks on KNL 
and Cray architectures. The light-crossing time of a zone generally sets
the timestep, but since the speed of light and the speed of sound in the inner core are
not far apart in the core-collapse problem after bounce, this numerical stability
constraint on the timestep is similar to the CFL constraint of the explicit
hydrodynamics. Radiation quantities are reconstructed with linear profiles
and the calculated edge states are used to determine fluxes via an
HLLE solver.  In the non-hyperbolic regime, the HLLE fluxes are corrected to
reduce numerical diffusion \citep{2013ApJ...762..126O}. The momentum and energy transfer
between the radiation and the gas are operator-split and addressed implicitly.

The hydrodynamics in F{\sc{ornax}} is based on a directionally unsplit Godunov-type finite-volume
method.  Fluxes at cell faces are computed with the fast and accurate HLLC approximate
Riemann solver based on left and right states reconstructed from the underlying volume-averaged
states.  The reconstruction is accomplished via a novel algorithm we developed specifically for F{\sc{ornax}}
that uses moments of the coordinates within each cell and the volume-averaged states to
reconstruct TVD-limited parabolic profiles, while requiring one less ``ghost cell" than
the standard PPM approach.  The profiles always respect the cells' volume averages and,
in smooth parts of the solution away from extrema, yield third-order accurate states on
the faces.  To eliminate the carbuncle and related phenomenon \citep{carbuncle}, F{\sc{ornax}} specifically
detects strong, grid-aligned shocks and employs in neighboring cells HLLE, rather than HLLC,
fluxes that introduce a small amount of smoothing in the transverse direction. Currently,
we do not include the effects of nuclear burning.  Given the ejecta masses we obtain (\S\ref{ejecta}), 
we expect this usually to amount to no more than a $\sim$10\% effect 
on the explosion energies (however, see \S\ref{energy}), but this remains to be seen \citep{2013ApJ...771...27Y}. 

Without gravity, the coupled set of radiation/hydrodynamic equations conserves energy 
and momentum to machine accuracy.  Total lepton number is conserved by construction.  
With gravity, energy conservation is excellent before and after core bounce 
\citep{fornaxcode:19}. However, as with all other supernova codes, at bounce 
the total energy as defined in integral form glitches by $\ge 10^{49}$ ergs\footnote{Most
supernova codes jump in this quantity at this time by more than 10$^{50}$ ergs \citep{2010ApJS..189..104M}.}. This
is due to the fact that the gravitational terms are handled in the momentum and
energy equations as source terms and are not in conservative divergence form.

The code is written in a covariant/coordinate-independent fashion, with generalized connection
coefficients, and so can employ any coordinate mapping.  This facilitates the use of any
logically-Cartesian coordinate system and, if necessary, the artful distribution of zones.
In the interior, to circumvent Courant limits due to converging angular zones,
the code can deresolve in both angles ($\theta$ and $\phi$) independently
with decreasing radius, conserving hydrodynamic and radiative fluxes in a
manner similar to the method employed in AMR codes at refinement boundaries.
The use of such a ``dendritic grid," or ``static-mesh refinement," allows us to avoid 
angular Courant limits at the center, while maintaining accuracy and enabling us to
employ the useful spherical coordinate system natural for the supernova problem.

Importantly, the overheads for Christoffel symbol calculations are minimal, 
since the code uses {\it static} refinement, and, hence, the terms are
calculated only once (in the beginning).  Therefore, the overhead
associated with the covariant formulation is almost nonexistent. In the context
of a multi-species, multi-group, neutrino radiation hydrodynamics
calculation, the additional memory footprint is small (note that the
radiation requires hundreds of variables to be stored per zone).  In terms
of FLOPs, the additional costs are associated with occasionally
transforming between contravariant and covariant quantities and in the
evaluation of the geometric source terms.  Again, in the context of a
radiation/hydrodynamics calculation, the additional expense is extremely
small.

Gravity {can be} handled in 2D and 3D with a multipole solver 
\citep{1995CoPhC..89...45M}, where we {would} generally set 
the maximum spherical harmonic order necessary equal to twelve.
For these calculations {however, to gain a bit of speed,} we use the monopole only 
(see \S\ref{sensitivity}). {In all implementations of gravity,} the monopole gravitational term is 
altered to approximately accommodate general-relativistic gravity \citep{2006AA...445..273M}
and we employ the metric terms, $g_{rr}$ and $g_{tt}$, derived from this potential
in the neutrino transport equations to incorporate general relativistic redshift
effects (in the manner of \citet{2002A&A...396..361R}; see also \citet{2018SSRv..214...33B}).
We use for this extensive suite of simulations the SFHo EOS of \citet{2013ApJ...774...17S}.
This EOS is one of those still consistent with known laboratory nuclear physics 
constraints \citep{2017ApJ...848..105T}. {However, the study of the EOS dependence 
of core-collapse theory is one of the important topics for future research                           
\citep{2009ApJ...707.1495S,2012ApJ...748...70H,Couch:2012gh,Suwa:2012xd,2013ApJ...774...17S,2017arXiv170701527D,2018ApJ...854..136N,2019arXiv190602009S}.}

The neutrino-matter interaction cross sections and rates are taken from 
\citet{2006NuPhA.777..356B} and we use detailed balance to derive emissivities 
from absorption rates.  Many-body corrections to the axial-vector part of 
the neutrino-nucleon scattering rate are taken from \citet{2017PhRvC..95b5801H}. 
All our default simulations incorporated this correction.
We note that such corrections for both neutral-current scattering and charged-current 
absorption are still in play and have been shown to be potentially important 
\citep{2018SSRv..214...33B,1998PhRvC..58..554B}. Weak magnetism and recoil corrections to 
scattering and absorption rates off nucleons \`a la \citet{2002PhRvD..65d3001H} 
are employed.  Coherent Freedman scattering off nuclei is corrected using lepton 
screening and form-factor terms described in \citet{2006NuPhA.777..356B}.  Nucleon-nucleon bremmstrahlung
is handled using the rates found in \citet{2000PhRvC..62c5802T} and our approach to $e^+/e^-$ 
annihilation into neutrino pairs is found in both \citet{2000PhRvC..62c5802T} and \citet{2006NuPhA.777..356B}. 
Inelastic scattering of neutrinos and anti-neutrinos off electrons and free nucleons 
is addressed using the prescriptions described in \citet{2003ApJ...592..434T},
with more details of its implementation to be found in \citet{burrows_thompson2004}. 
Our approach to electron capture on heavy nuclei, most important on infall, is taken 
from \citet{2010NuPhA.848..454J}. 

To summarize, the advantages of \fornax\ are: 1) it does not employ the simplifying ``ray-by-ray{+}"
approximation used by many others \citep{2014ApJ...792...96T,2015ApJ...807L..31L,melson:15b,
bmuller_2015,2016MNRAS.461L.112T,muller2017,summa2018} that suppresses the 
important lateral transport \citep{2016ApJ...831...81S}; 2) it handles the 
spatial transport operator explicitly, thereby avoiding problematic global 
iterative solves; 3) the source terms are still handled implicitly, including inelastic energy 
redistribution, ensuring stability and speed; 4) we use static-mesh-refinement 
in the inner core and along the polar axis to thwart the significant Courant
timestep hit in the angular directions that would otherwise obtain there when using spherical 
coordinates; and 5) all important physical effects (except neutrino oscillations) 
are handled at some reasonable level of approximation. In short, all the necessary 
physical realism is included. The result is a code that is $\sim$5 times faster
than previous implementations, and this speedup is what enables the significant increase in simulation 
cadence represented by this paper.  Drawbacks of the current implementation of \fornax\
are that it incorporates approximate general-relativistic gravity and does not perform 
full multi-angle transport.  Though \fornax\ does follow the multi-group vector fluxes, 
it is currently too expensive to attempt to calculate the full angular specific-intensity 
distributions in 3D for a simulation of reasonable physical duration 
\citep{2014ApJS..214...16N,2017ApJS..229...42N,2019ApJ...878..160N}.

In keeping with the philosophy behind this comprehensive 3D study spanning such a unprecedentedly 
wide progenitor mass range, we start our calculations with a uniform set of progenitors taken from 
\citet{2016ApJ...821...38S}.  The only exception is the 25-M$_{\odot}$ model, taken 
from \citet{2018ApJ...860...93S}.  This 3D full-physics model set includes 9-, 10-, 
11-, 12-, 13-, 14-, 15-, 16-, 17-, 18-, 19-, 20-, 25-, and 60-M$_{\odot}$ massive-star 
progenitors. In addition, we explore for a subset of progenitors the multipole/monopole, 
Horowitz/no-Horowitz, low versus high angular resolution differences for a 
small subset of progenitors, namely the 19-M$_{\odot}$ and 11-M$_{\odot}$ models.
We explored in a previous paper \citep{2019arXiv190503786N} other aspects of the same resolution 
study. In toto, this comprises nineteen 3D simulations with what is thought to be the necessary physical realism.

Unless otherwise indicated, our default spatial resolution is 678$\times$128$\times$256 
(r$\times\theta\times\phi$), we use 12 energy groups, and the outer radius is at 20,000 kilometers (km).
The radial zone width from the center out to $\sim$20 km is 0.5 km, after which the zone width grows logarithmically to
this outer boundary.  The energy groups are logarithmically distributed from 1 MeV to 300 MeV (for $\nu_e$) or 100 MeV (for all 
other species).  To seed instabilities, very modest initial perturbations to the velocity field of 
amplitude 100 km s$^{-1}$ and with $\ell = 10$, $m = 1$, and $n=4$, using the prescription of 
\citet{2015MNRAS.448.2141M}, were imposed 10 milliseconds (ms) after bounce to the 3D model 
that was mapped from the 1D model followed to collapse.  {This is to be compared
to the pre-explosion speeds in front of the bounce shock of $\sim$50,000 km s$^{-1}$
and the immediate post-bounce speeds from $\sim$8000 km s$^{-1}$ to $\sim$4000 km s$^{-1}$.  It is expected
that these perturbations will grow on infall \citep{2000ApJ...535..402L,2014ApJ...794..162T}, 
but not achieve comparable speeds.} All models were non-rotating.
We have attempted to standardize all model runs to ensure our model-to-model comparisons
are as direct as possible.  In this way, one can hope to better ascertain true systematic 
differences in the context of state-of-the-art 3D simulations over this wide progenitor panorama.

{We emphasize that all our 3D models are calculated using exactly the same specifications and setup,
including our admittedly-crude method of initial model perturbation. This is to enable direct comparisons and, thereby,
to extract systematic variations along the progenitor continuum.  It may be that models for which we don't
witness explosions (and vice versa) might explode with rotation, updated physics, higher resolution, or
an improved code, etc.  However, we assert that the {\it relative} tendency to explode, or not to explode,
is captured by our study and will serve as an important theoretical context going forward.}

%% -------------------------------------------------------
\section{Results}
\label{results}
%% -------------------------------------------------------

\subsection{Overview}
\label{overview}

\begin{figure}
  \includegraphics[width=\columnwidth]{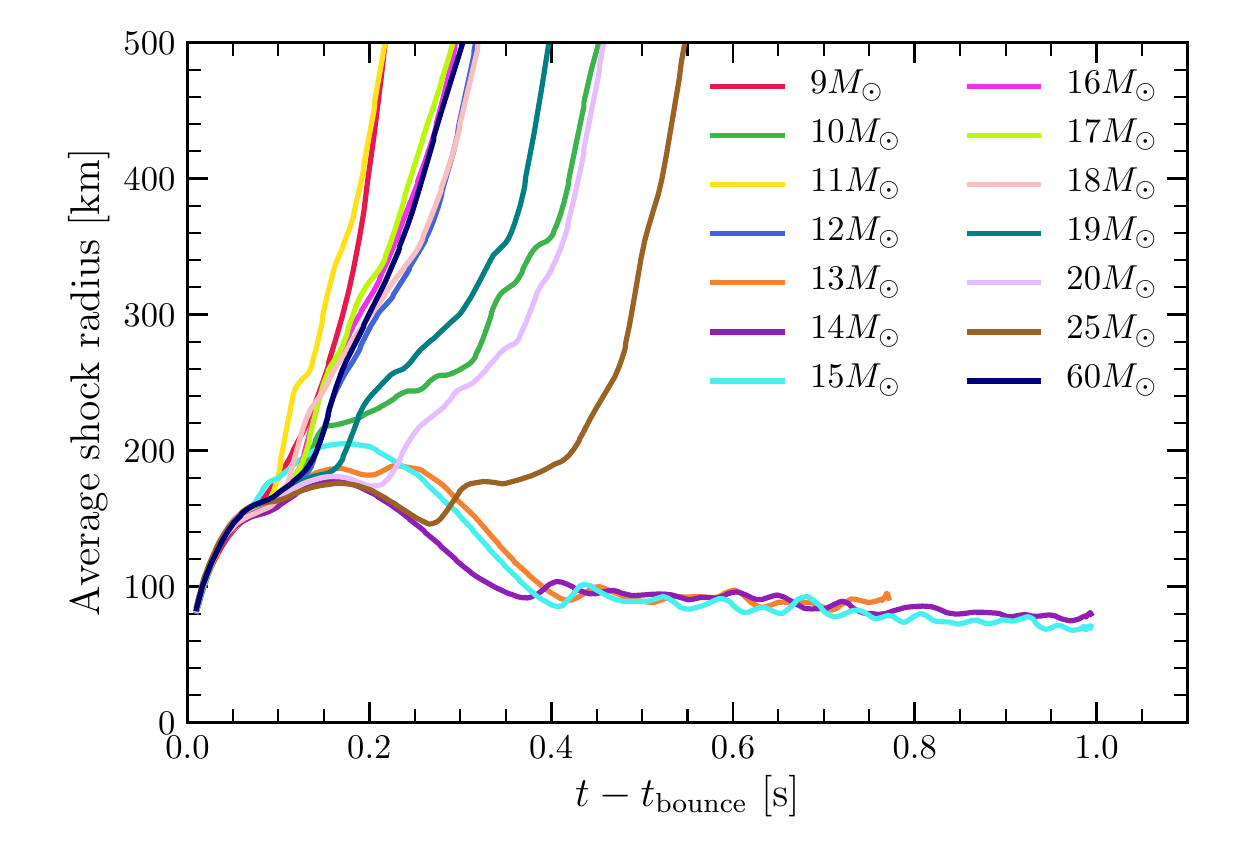}
  \caption{Average shock radii. Our models span a wide range in terms of
  explosion delay times with shock revival occurring from ${\sim}0.1$ to
  ${\sim}0.5$ seconds after bounce. Among the progenitors we consider,
  the 13-M$_\odot$, 14-M$_\odot$ and 15-M$_\odot$ models fail to explode
  within the timeframe we simulate.}
  \label{fig:rshock}
\end{figure}

\begin{figure}
  \includegraphics[width=\columnwidth]{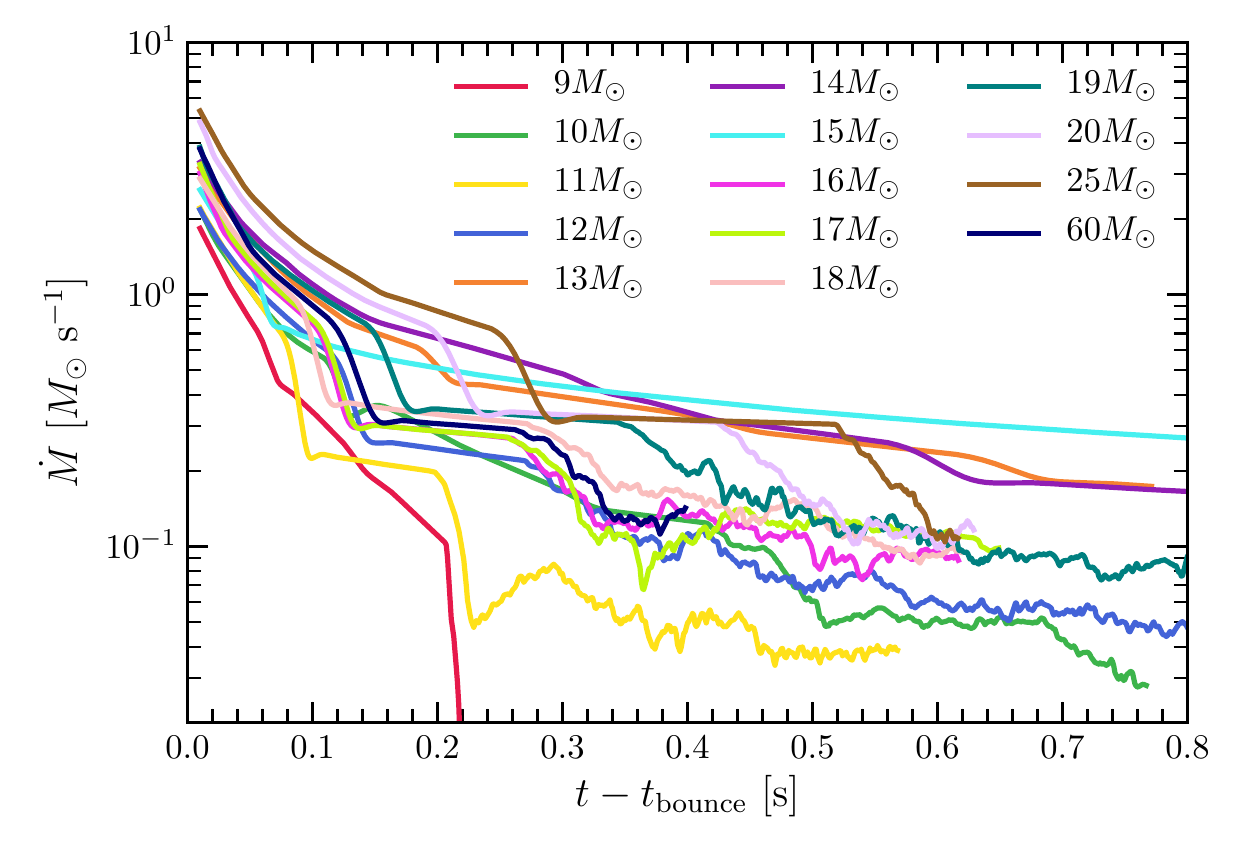}
  \caption{Mass accretion rate at 500~km. All exploding models display a
  sharp drop in the accretion rate corresponding to the infall of the
  Si/O interface. All models, with the exception of the 9-$M_\odot$
  progenitor, show an overall positive net accretion rate onto the inner core even after
  explosion sets in.}
  \label{fig:mdot}
\end{figure}

At this stage in the theoretical development of progenitor models, it should 
not be assumed that the mapping between mass and profile is accurately known. There 
is still much churn in that complicated field, and the effects of multi-dimensional stellar evolution 
\citep{2015ApJ...808L..21C,2016A&A...593A..72J,2016ApJ...822...61C,muller2017,2019MNRAS.484.3307M,2019A&A...622A..74J,2019arXiv190307811Y} 
and binarity \citep{2019MNRAS.484.3307M}, to name only two, have not yet been fully assimilated.  However,
it is reasonable to suggest that the range of possible structures is well-captured by the 
range depicted in Figure \ref{density}.  It is in this spirit that we present our 3D explosion
results and suggest that the general range of outcomes has been approximately corralled.

\begin{figure}
  \includegraphics[width=0.49\textwidth]{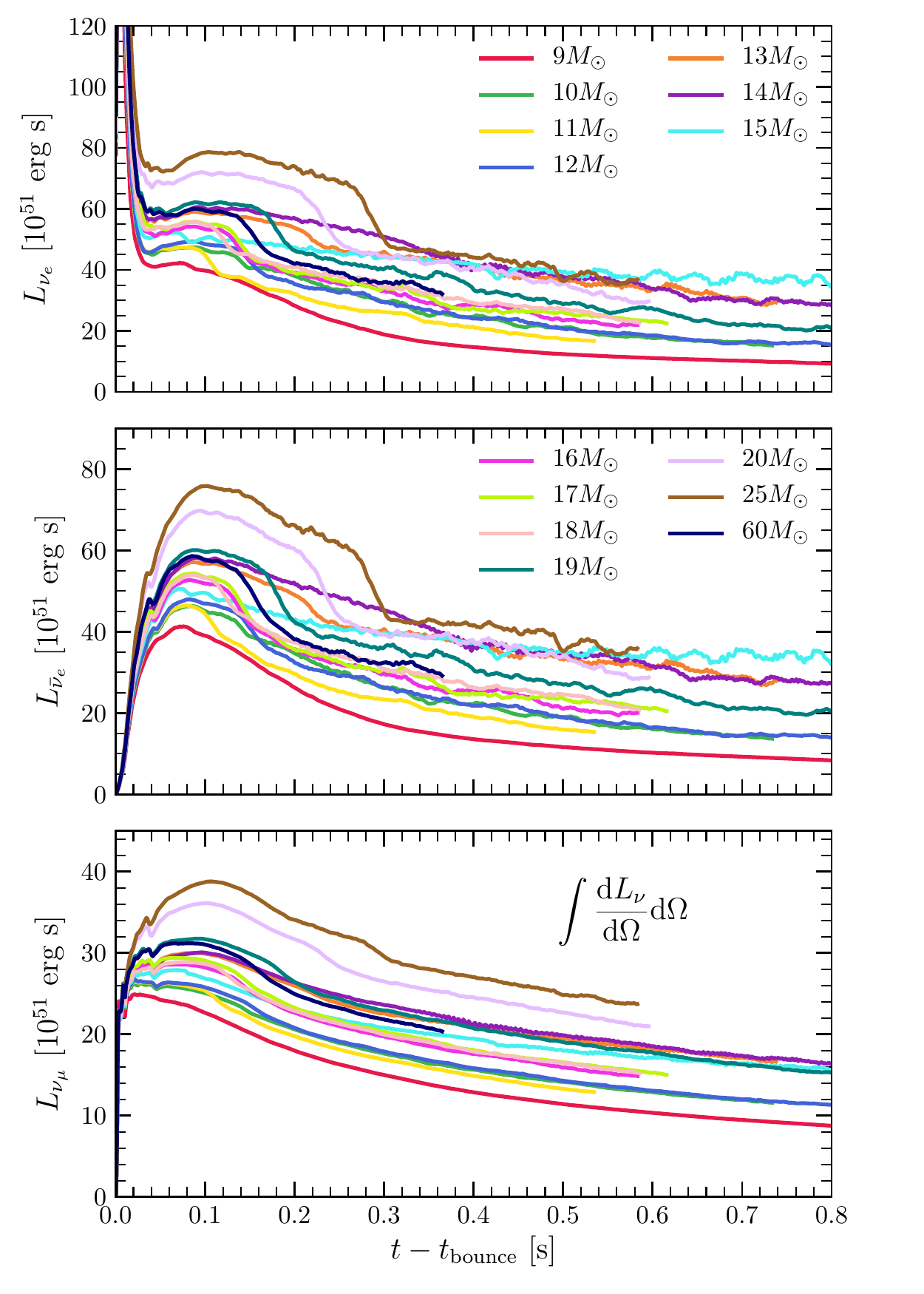}
  \hfill
  \includegraphics[width=0.49\textwidth]{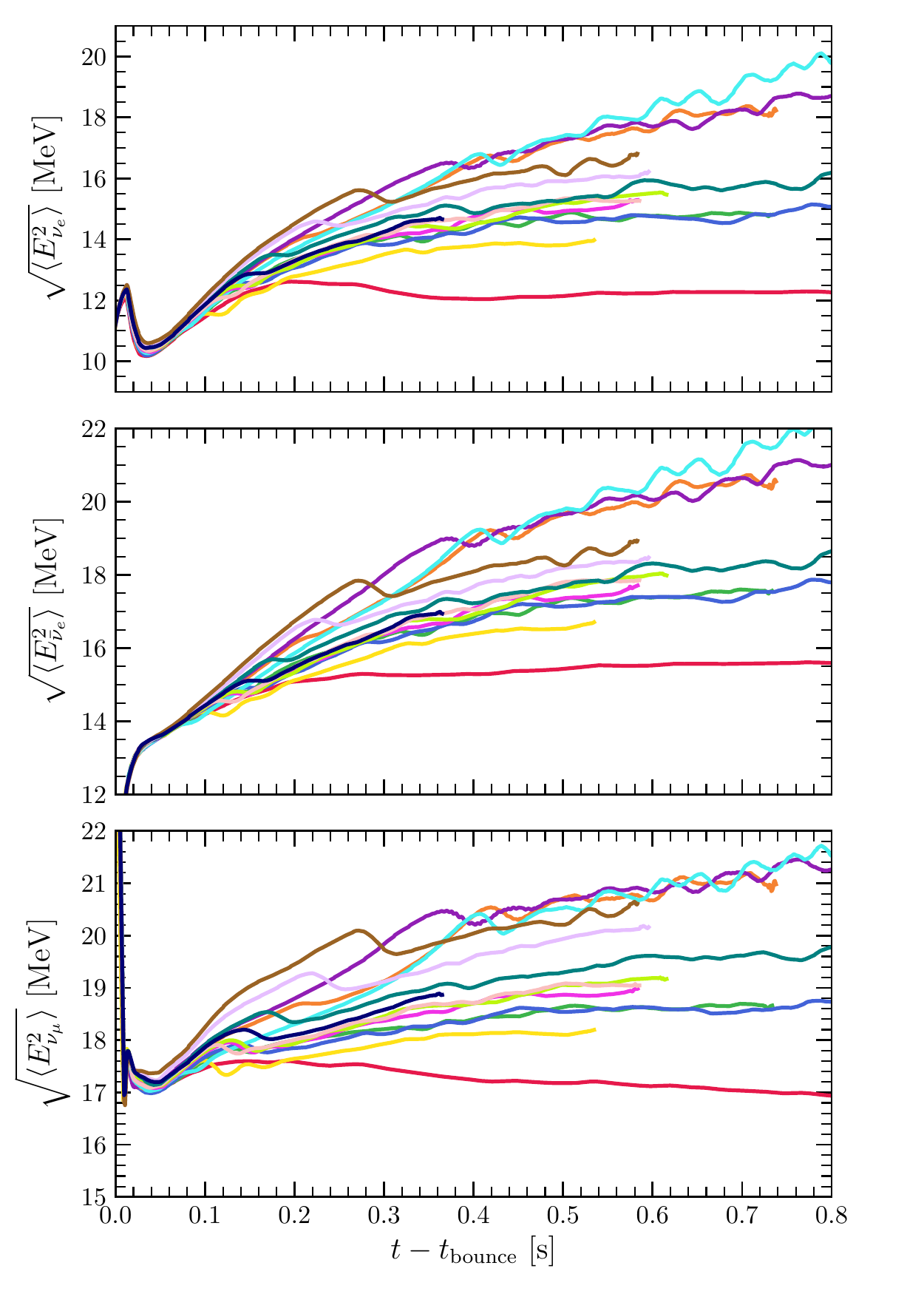}
  \caption{Neutrino luminosities (\emph{left panel}) and rms energies
  (\emph{right panel}) measured at 10,000~km. There is significant
  spread between the different models. The exploding models, with the
  exception of the 9-M$_\odot$ progenitor, show drops in the
  electron-type neutrino luminosities at the time of the accretion of the Si/O interface,
  which coincides with the explosion time. The explosion time is also
  imprinted in the neutrino rms energies, which exhibit a similar drop.}
  \label{fig:nulums}
\end{figure}

\begin{figure}
  \includegraphics[width=\columnwidth]{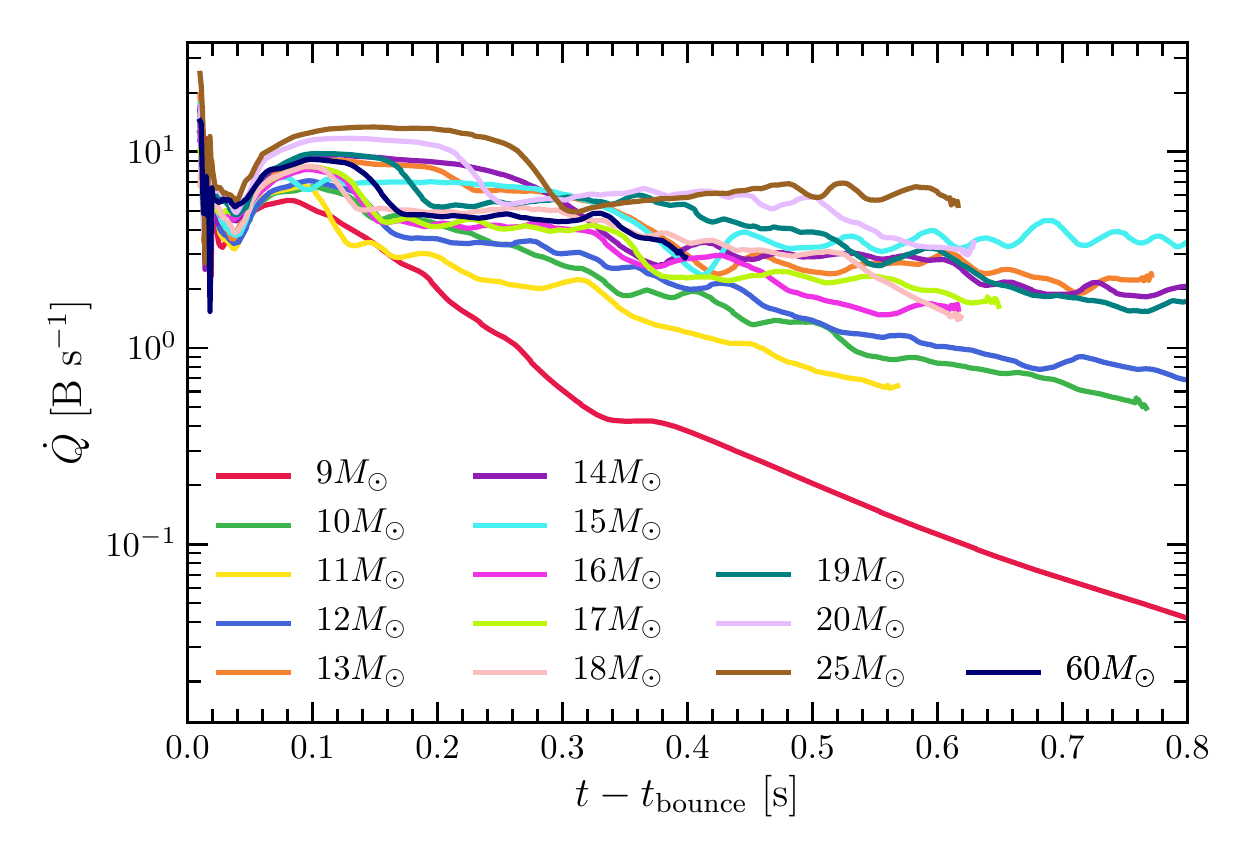}
  \caption{Net neutrino energy deposition rate. More compact progenitors
  typically show higher neutrino luminosities and higher net heating
  rates. With the exception of the 9-M$_\odot$ progenitor, strong
  heating is maintained for the entire duration of our simulations.}
  \label{fig:net_heat}
\end{figure}

Figure \ref{density} depicts the mass density profiles of the suite of models upon which
we focus in this paper.  The range of model slopes exterior to $\sim$1.2 M$_{\odot}$
is quite wide and covers most of the model space historically found in the literature.
The lowest mass representative, the 9-M$_{\odot}$ progenitor, boasts the steepest profile
and the 25-M$_{\odot}$ progenitor the shallowest, and any measure of average declivity
would be a monotonic function of ZAMS mass.  However, as the calculated compactness given 
in Table \ref{sn_tab} demonstrates, the models are not perfectly nested monotonically,
and this is thought to reflect real physical effects 
\citep{2007PhR...442..269W,2016ApJ...821...38S,2018ApJ...860...93S}. Moreover, due 
to significant mass loss, the 60-M$_{\odot}$ of \citet{2016ApJ...821...38S} we employ 
in this paper resides in the middle of the pack. For all the models, the compactness
and shallowness are inversely related to the central density, which helps 
determine the time to bounce.  It should be noticed that most of the models have pronounced
density cliffs at the silicon/oxygen interface, and it has been shown that the accretion of 
such features can itself jump a model into explosion \citep{2018MNRAS.477.3091V,2018SSRv..214...33B,2019MNRAS.485.3153B}.
However, not all progenitors share this feature, with the 13-, 14-, and 15-M$_{\odot}$ models
evincing some of the most modest jumps {of $\sim$1.2 - 1.4}.  
As Figure \ref{fig:rshock} of the post-bounce evolution of 
the mean shock radius demonstrates, these are the models that do not explode, 
and this is one reason.  All our other models explode, with the post-bounce explosion 
times generally shorter for the lower-mass progenitors and longer for the higher-mass progenitors.
{Most of these exploding models have mass density jumps at this interface of $\sim$2.0-2.3.}
{Here, we define the time of explosion rather loosely as the approximate time
the mean shock radius experiences an upward inflexion and is seen to continue its climb.}
In fact, the 19-, 20-, and 25-M$_{\odot}$ stars explode later than most, and the 9- and 11-M$_{\odot}$
models the earliest, with the 10-M$_{\odot}$ model a bit sluggish, perhaps due to the less pronounced
silicon/oxygen ledge and its (seemingly anomalous) shallower density profile. 
However, the general separation of the early-exploding lower-mass branch from the later 
exploding higher-mass branch seems to hold.  The delay of the higher-mass models
seems connected with the larger early mass accretion rate (Figure \ref{fig:mdot}) and 
higher associated ram pressure.  However, when these models do explode they do so more energetically $-$
the higher accretion rates are maintained to translate into higher driving neutrino luminosities
(Figure \ref{fig:nulums}, left) and RMS neutrino energies (Figure \ref{fig:nulums}, right) absorbed 
on a consequently thicker column of mass in the gain region, resulting in a higher neutrino 
power deposition (Figure \ref{fig:net_heat}). As we discuss in \S\ref{energy}, this results in 
a higher accumulation rate of net explosion energy, and likely into higher asymptotic explosion energies.
Nevertheless, we still find that there are models, currently in the middle of the progenitor
continuum, that do not explode, but are bracketed in compactness and other general parameters 
by those that do.  This reiterates the strong conclusion that low compactness is not a 
necessary nor sufficient condition for explodability \citep{2018SSRv..214...33B}.

\onecolumn

\twocolumn

Figure \ref{fig:mdot} renders the evolution of the integrated mass accretion rate ($\dot{M}$, inward) through
a radius of 500 km as a function of time after bounce.   $\dot{M}$ follows the corresponding mass 
density profile (Figure \ref{density}) closely, with the effects of the accretion of the silicon/oxygen 
interface clearly shown.  The post-bounce time of the accretion of this interface is correlated for 
many models with the onset time of explosion (modulo the accretion time from 500 km to the shock). 
$\dot{M}$ for the 9-M$_{\odot}$ model drops precipitously, and accretion effectively ceases
around $\sim$0.2 seconds.  Not unexpectedly, $\dot{M}$ for the non-exploding models (13-, 14-, 
and 15-M$_{\odot}$) continues and eventually (after $\sim$0.6 seconds) supersedes that of any exploding model.
However, apart from the 9-M$_{\odot}$ model, even for the exploding models accretion continues for quite 
some time.  This is due to the fact that in 3D there simultaneously can be accretion in one direction, 
while the star explodes in another. This feature enables accretion to maintain the driving neutrino 
luminosities beyond the onset of explosion at a higher level than would be possible with 
core neutrino diffusion alone, and is clearly pronounced for the 25-M$_{\odot}$ model\footnote{This natural 
facet of CCSN theory in the multi-dimensional context has been discussed 
before, for example in \citet{2016ApJ...818..123B}, \citet{2018SSRv..214...33B}, 
\citet{2019MNRAS.482..351V}, and \citet{2019MNRAS.485.3153B}.}. For the exploding 
models, we note that with time $\dot{M}$ at 500 km begins very weakly fluctuating behavior
on timescales of $\sim$10 milliseconds. This is due to the summed effect of the clumpiness and swirling behavior of the ejected 
material at a radius of 500 km after the exploding turbulent shock has reached and passed it with 
the dominant accretion component associated with the material that continues to infall despite explosion 
at other angles \citep{muller2017,2019MNRAS.482..351V}. 

The post-bounce evolutions of the angle-integrated neutrino luminosities and RMS neutrino energies
are given in Figure \ref{fig:nulums}.  The hierarchy of values expected from the systematics
with progenitor with $\dot{M}$ depicted in Figure \ref{fig:mdot} is continued in these plots.  The lower-mass
progenitors generally achieve lower luminosities, with the plateaus/peaks in the $\nu_e$ (post-breakout) 
and $\bar{\nu}_e$ luminosities ranging by a factor of $\sim$2 and in the $\nu_{\mu}$ luminosities by
$\sim$60\%. There is a similar systematic behavior for the mean and RMS neutrino energies, with higher
neutrino energies generally for the more massive exploding progenitors.  Since the neutrino energy deposition
rate in the gain region behind the shock goes as the absorption cross section, which is quadratic in 
the RMS energy, the more massive models experience the double effect of both high luminosity 
and high neutrino energy. This result underpins the more rapid rise in explosion energy shown 
in Figure \ref{fig:explene} for the 20- and 25-M$_{\odot}$ models.  However, as Figure \ref{fig:nulums} 
indicates, at later times the neutrino energies for the non-exploding models continue to rise to achieve 
the highest values. This is also the case for their late-time luminosities. Therefore, despite the high
values for the non-exploding models of the product of the luminosity and square of the RMS energy, they still
need to explode for them to take advantage of this high product. 

We note that the neutrino energies reached by all models are still significantly lower than
those published by Wilson in his early, pioneering studies \citep{1985ApJ...295...14B,1987ApJ...318..288M}.
This is due to subsequent improvements in the neutrino-matter interaction rates and is reflected
broadly in the modern literature \citep{2016ApJ...818..123B,2017hsn..book.1095J,2018JPhG...45j4001O,2019ApJ...873...45G}.

The heating rates (minus those due to inelastic scattering) in the gain region 
behind the shock are portrayed in Figure \ref{fig:net_heat} and recapitulate
the trends seen in Figures \ref{fig:mdot} and \ref{fig:nulums}.  The high deposition rates of a few to $\sim$10 
Bethes (10$^{51}$ ergs) per second should not lead one to infer that an asymptotic explosion energy of a Bethe
is quickly achieved.  Before explosion, this energy is completely reradiated and after the onset of explosion much
of this power goes into lifting the ejecta out of a deep potential well.

Table \ref{sn_speed} provides the mean shock radius and mean shock speed at the end of each of our baseline 
simulations.  {In general, soon after the shock is launched its mean speed stays roughly constant.}
The 9-M$_{\odot}$ model proceeded the furthest after bounce, at which point its explosion shock 
achieved a mean radius of $\sim$12,400 km and a mean shock speed of $\sim$1.6$\times{10}^9$ cm s$^{-1}$.  The other exploding models achieved mean speeds of $\sim$5$-$8$\times{10}^9$ cm s$^{-1}$, while, as Table \ref{sn_speed} clearly indicates, the 13-, 14-, and 15-M$_{\odot}$ do not explode.

\subsection{Explosion Energies}
\label{energy}

\begin{figure}
  \includegraphics[width=\columnwidth]{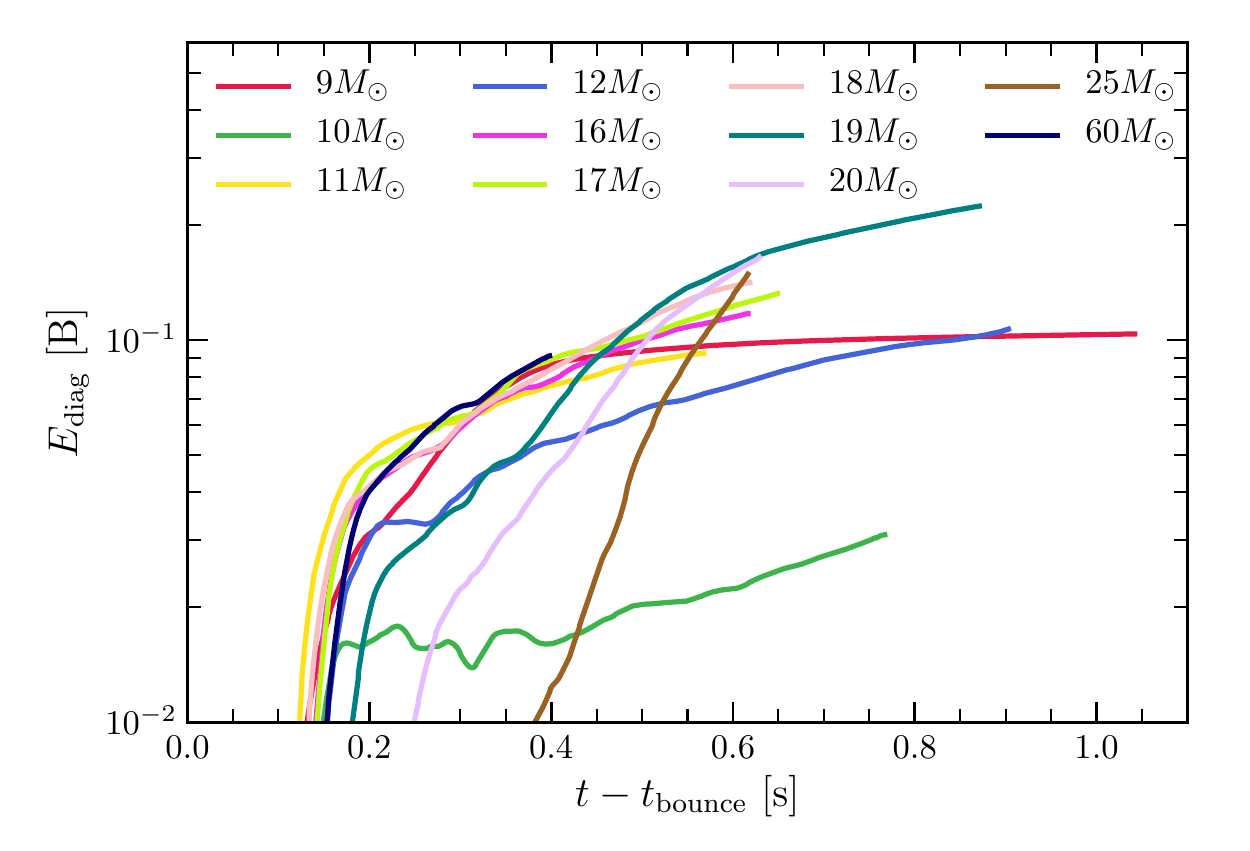}
  \caption{Diagnostic explosion energies. With the exception of the 9-M$_\odot$
  progenitor, the explosion energies have not yet reached their
  asymptotic value. However, most of the models appear to be approaching
  diagnostic explosion energies near a $\mathrm{few}\times 10^{50}\ {\rm erg}$, except 
  for the more massive models which are poised to
  achieve at later times even higher explosion energies. 
  The total ``diagnostic" supernova explosion energy is the sum of
  the internal thermal, kinetic, and gravitational energies of the ejecta, defined as the matter
  with positive Bernoulli integral.  The gravitational term is the largest and much of the
  deposited neutrino power goes into work against it.  We include in the diagnostic energy
  the ``reassociation energy" of the debris into nuclei and the gravitational binding energy
  of the matter exterior to the explosion shock, but interior to the 20,000-km boundary.
  Note that the total diagnostic energy must still be corrected for the binding energy exterior to the outer computational
  boundary, provided in Table \protect\ref{sn_tab}, to obtain the total explosion energy.  
  This correction is small, except for the more massive models,
  where it can approach one Bethe.}
  \label{fig:explene}
\end{figure}

  Though many of our models were carried out to post-bounce times that would be considered 
late in the context of most other published 3D models, we find that the 
majority of our simulations still need to be carried out even further to asymptote to their 
final explosion energies. In a resource-constrained computational environment, deciding to be wider in 
progenitor space naturally translates into being shorter in mean duration.  Nevertheless, ours is still
by far the largest number of total physical-seconds explored in 3D.  As Figure \ref{fig:explene}
shows, though the 11- and 12-M$_{\odot}$ simulations {seem, by the curvature of their
energy curves,} to have reached within greater than $\sim$50\% of their asymptotic explosion 
energies at simulation end, the only model in our set to actually asymptote to its 
final explosion energy is the 9-M$_{\odot}$ model \citep{2019MNRAS.485.3153B}. 
It achieves an explosion energy of $\sim$10$^{50}$ ergs ($\sim$0.1 Bethe) after $\sim$0.5 seconds 
and was continued to $\sim$1.0 seconds.  

Importantly, the higher-mass progenitors explode late (Figure \ref{fig:rshock}), but, 
as stated in \S\ref{overview}, accumulate total energy at a more rapid rate (Figure \ref{fig:explene}).  For 
the 25-M$_{\odot}$ model, that rate is $\sim$1 Bethe per second and for the 20-M$_{\odot}$ model it is only a bit less,
implying that, carried for another few seconds, these models would achieve what are considered to 
be ``canonical" supernova energies of one Bethe or more.  A caveat is that the total binding energy 
of the mantle exterior to our computational boundary at 20,000 km must be paid.  As Table \ref{sn_tab} indicates,
though this number is quite small for the low-mass progenitors, it is approaching one Bethe 
for the 25-M$_{\odot}$ star, necessitating a longer energy ramp at the rate witnessed in 
Figure \ref{fig:explene} to achieve a kinetic energy at infinity of 
order one Bethe.  This longer time for the more massive stars is in keeping with the results 
of \citet{bmuller_2015}, who concluded the same using a simpler computational infrastructure.
Hence, our results suggest that the more massive models that explode a bit later, likely ramp up more quickly
to larger explosion energies after a longer evolution. For some massive models, perhaps the 25-M$_{\odot}$ model, the mantle binding 
energy penalty may be too high and a black hole may result\footnote{Whether a weaker supernova could still emerge 
in this scenario is an interesting possibility for future study.}.  We note that since we have 
neglected nuclear burning, it is for the 25-M$_{\odot}$ model that this neglect may be most relevant.  
As we see in \S\ref{ejecta}, the amount of core material ejecta for this model is large and 
a fraction of this mass (to be determined) may burn to boost this explosion even further. In this context 
it should be remembered that the burning of one solar mass of oxygen yields approximately 
a Bethe of energy. 

The lower-mass progenitors explode, when they do, 
earlier after bounce, but achieve lower asymptotic explosion energies. This is the systematics in explosion energy
with progenitor structure/mass that we infer from the results of this 3D progenitor model set.  Importantly, this
is also consistent with what is emerging from the progenitor-mass/explosion-energy correlation inferred
in recent analyses of Type IIp light curves\citep{2018ApJ...858...15M,2019arXiv190801828M,2019arXiv190807762E,2013MNRAS.436.3224P}\footnote{See, in particular, 
Figure 6 in \citet{2018ApJ...858...15M} and Figure 5 in \citet{2019arXiv190801828M}.}.
Clearly, future 3D simulations should push to longer post-bounce physical times.
Moreover, the chaos in the convective turbulence will naturally introduce a degree of stochasticity in the 
outcomes and their parameters, including explosion energy. Therefore, determining the distribution 
functions in these observables, even for a given progenitor, will be an interesting long-term challenge for theory.

%\onecolumn

\begin{table}
  \center{
  \begin{tabular}{llll}
    \hline
            Progenitor    &  t(final) & Shock Radius & Shock Speed \\

            (M$_{\odot}$) & (seconds) & (1000 km)        & (1000 km s$^{-1}$)  \\
    \hline
s9.0  &    1.042   &        12.419    &                16.287   \\
s10.0 &    0.767   &        1.963     &                6.647    \\
s11.0 &    0.568   &        2.754     &                7.996    \\
s12.0 &    0.903   &        4.088     &                6.944    \\
s13.0 &    0.771   &        0.090     &                0.078    \\
s14.0 &    0.994   &        0.077     &                0.044    \\
s15.0 &    0.994   &        0.069     &                0.072    \\
s16.0 &    0.617   &        2.265     &                6.717    \\
s17.0 &    0.649   &        2.527     &                6.621    \\
s18.0 &    0.619   &        2.122     &                7.870    \\
s19.0 &    0.871   &        3.879     &                7.848    \\
s20.0 &    0.629   &        1.415     &                7.330    \\
s25.0 &    0.616   &        0.735     &                6.594    \\
s60.0 &    0.398   &        0.808     &                5.233    \\
    \hline
  \end{tabular}
  \caption{A table of the mean shock radius and mean shock speed at the end of each baseline 3D simulation.
The simulation end time is given in seconds, the mean shock radius is given in units of 1000 km, and the mean shock
speed is given in units of 1000 km s$^{-1}$.  Note that the non-exploding models (13-, 14-, and 15-M$_{\odot}$)
have correspondingly low values for both quantities.
  \label{sn_speed}
  }}
\end{table}

\begin{table}
  \center{
  \begin{tabular}{llll}
    \hline
            Progenitor    &  t(final) & Baryon Mass & Grav. Mass \\

            (M$_{\odot}$) & (seconds) &  (M$_{\odot}$) & (M$_{\odot}$)  \\
    \hline
s9.0  &    1.042    &             1.342    &           1.233   \\
s10.0 &    0.767    &             1.495    &           1.358   \\
s11.0 &    0.568    &             1.444    &           1.317   \\
s12.0 &    0.903    &             1.517    &           1.377   \\
s13.0 &    0.771    &             1.769    &           1.577   \\
s14.0 &    0.994    &             1.824    &           1.619   \\
s15.0 &    0.994    &             1.774    &           1.580   \\
s16.0 &    0.617    &             1.585    &           1.431   \\
s17.0 &    0.649    &             1.615    &           1.455   \\
s18.0 &    0.619    &             1.606    &           1.448   \\
s19.0 &    0.871    &             1.757    &           1.567   \\
s20.0 &    0.629    &             1.887    &           1.667   \\
s25.0 &    0.616    &             1.993    &           1.747   \\
s60.0 &    0.398    &             1.647    &           1.481   \\
    \hline
  \end{tabular}
  \caption{At the final time after bounce for each simulation (in seconds), 
the baryonic mass of the PNS (in M$_{\odot}$) and the gravitational mass of the residual neutron star (in M$_{\odot}$)
for all the models of this study (except the extra 3D models associated with the sensitivity study of \S\ref{sensitivity}).
All models, except the 13-, 14-, and 15-M$_{\odot}$ models, explode.
  \label{sn_tab2}
  }}
\end{table}

%\twocolumn

\subsection{Proto-neutron Star Masses}
\label{pns}

\begin{figure}
  \includegraphics[width=\columnwidth]{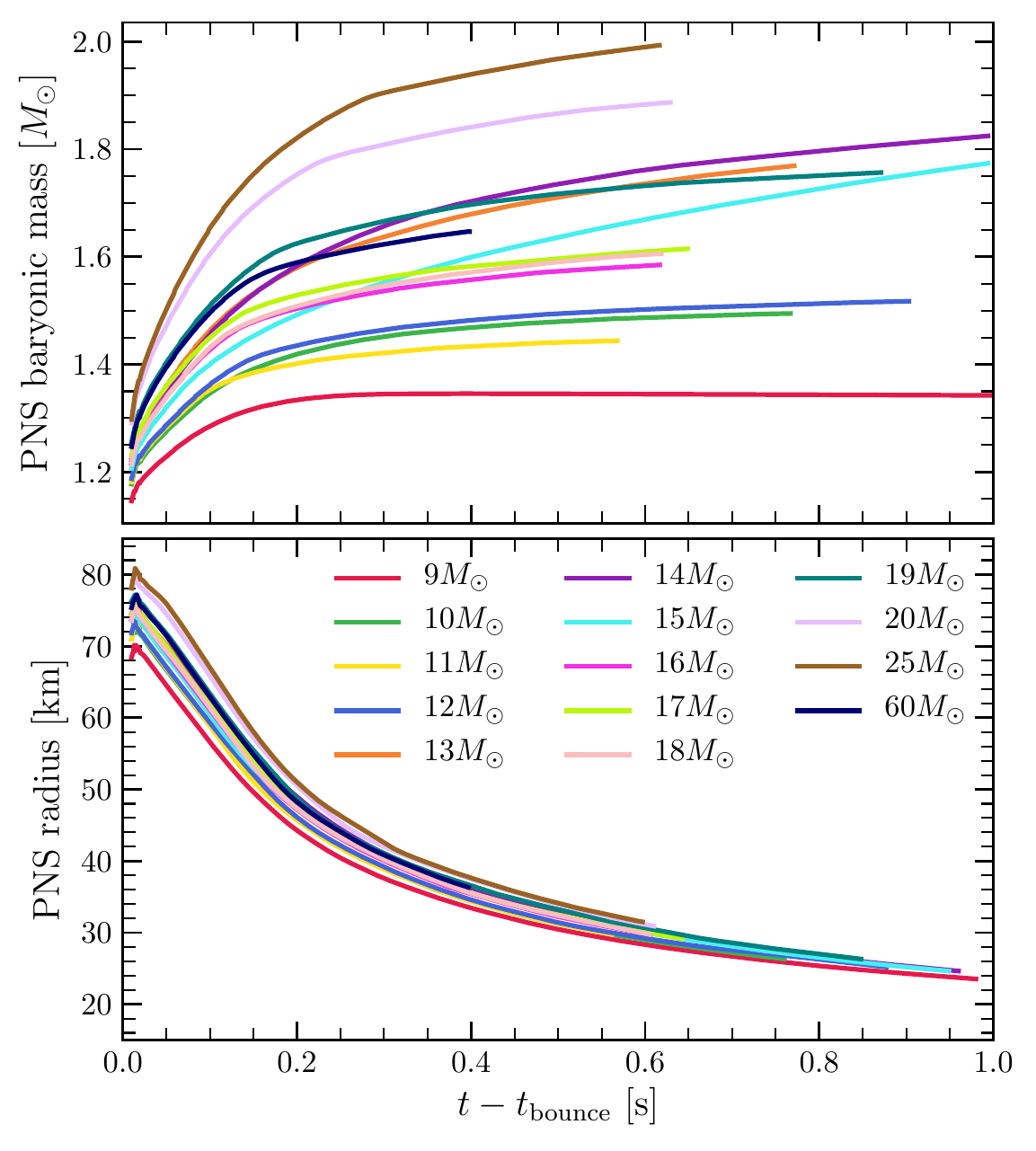}
  \caption{PNS baryonic masses (in M$_{\odot}$) (\emph{upper panel}) and radii (in km)
  (\emph{lower panel}). For most of the models the growth rate of the
  PNS has dropped substantially by the time we terminate our
  simulations. The PNS of the 25-M$_\odot$ progenitor, which explodes
  very energetically ${\sim}0.5$ seconds after bounce, is still growing
  rapidly at the end of the simulation.}
  \label{fig:pns}
\end{figure}

Figure \ref{fig:pns} shows the baryon mass accumulated within an isodensity surface
of mass density 10$^{11}$ g cm$^{-3}$ for all the simulations of this investigation.  This PNS mass ranges
from a low of $\sim$1.3 M$_{\odot}$ for the 9-M$_{\odot}$ model to a high near $\sim$2.0 M$_{\odot}$
for the 25-M$_{\odot}$ progenitor.  In Table \ref{sn_tab2}, we tabulate the baryon and gravitational PNS masses
at the end of each simulation.  {The latter is the gravitational mass for the cold 
neutron star in beta equilibrium, using the SFHo EOS.} Except for the 9-M$_{\odot}$ simulation,
for which the PNS mass has asymptoted, the PNS masses for the other models are still growing at a 
rate bounded by $\sim$5\% per second at the termination of each run. While the PNS radius 
(defined as the $10^{11}$ g cm$^{-3}$ radius) for all models varies
from one model to the next by no more than $\sim$15\%, the gravitational mass varies by $\sim$40\% from 1.233 
to 1.747 M$_{\odot}$, a range in keeping with general expectations for neutron stars in the galaxy 
\citep{2007PhR...442..109L}.  However, it does not extend to the highest values measured to date 
($\sim$2.1 M$_{\odot}$, \citet{2013Sci...340..448A}; $\sim$1.97 M$_{\odot}$, \citet{2010Natur.467.1081D};
$\sim$2.14 M$_{\odot}$, \citet{Cromartie}).
Nevertheless, we see a clear trend among the exploding models from low neutron star masses for low-mass progenitors
to higher mass neutron stars for the high-mass progenitors, with the 10-M$_{\odot}$ out of order (as discussed). 
Given the non-monotonicity of the initial mass-density profiles (Figure \ref{density}) with progenitor,
we would not expect a monotonic correspondence between progenitor mass and residual PNS gravitational mass.
Again, the non-exploding models in the middle of the compactness continuum are ``out of sequence." Of 
course, if they never explode in the first seconds after bounce, they should birth black holes. 

The PNS radii depicted in Figure \ref{fig:pns} achieve values of only $\sim$25 km by the end of the simulations.
This is not the standard 10-12 km because the PNS is still lepton-rich and hot.  It will require many more seconds to one 
minute to cool and deleptonize into the ``cold, catylyzed" state of a galactic neutron star \citep{1986ApJ...307..178B}.

\subsection{Explosion Morphology}
\label{morphology}

\begin{figure*}
  \includegraphics[width=0.40\textwidth]{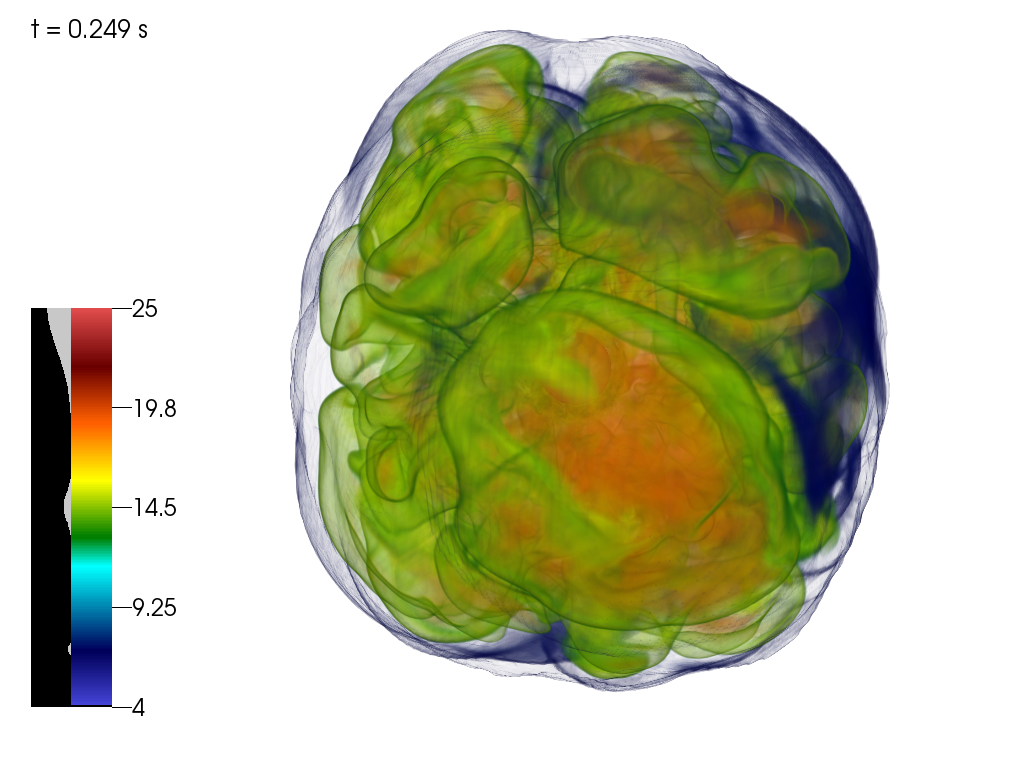}
  \hfill
  \includegraphics[width=0.40\textwidth]{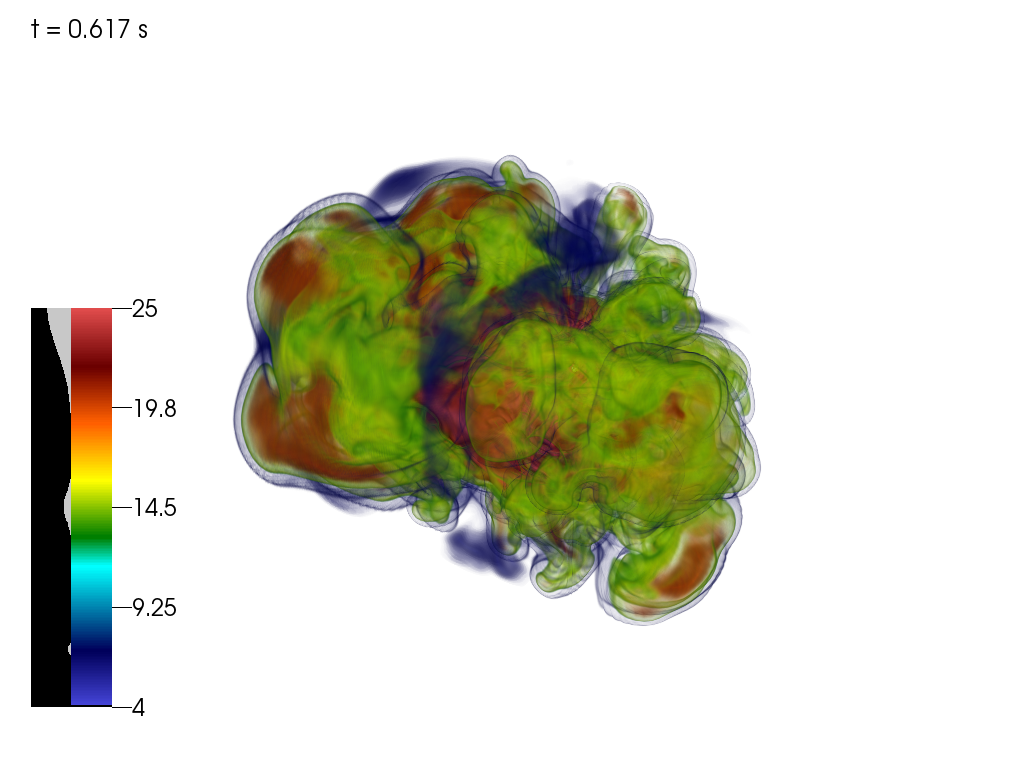}
  \hfill
  \includegraphics[width=0.40\textwidth]{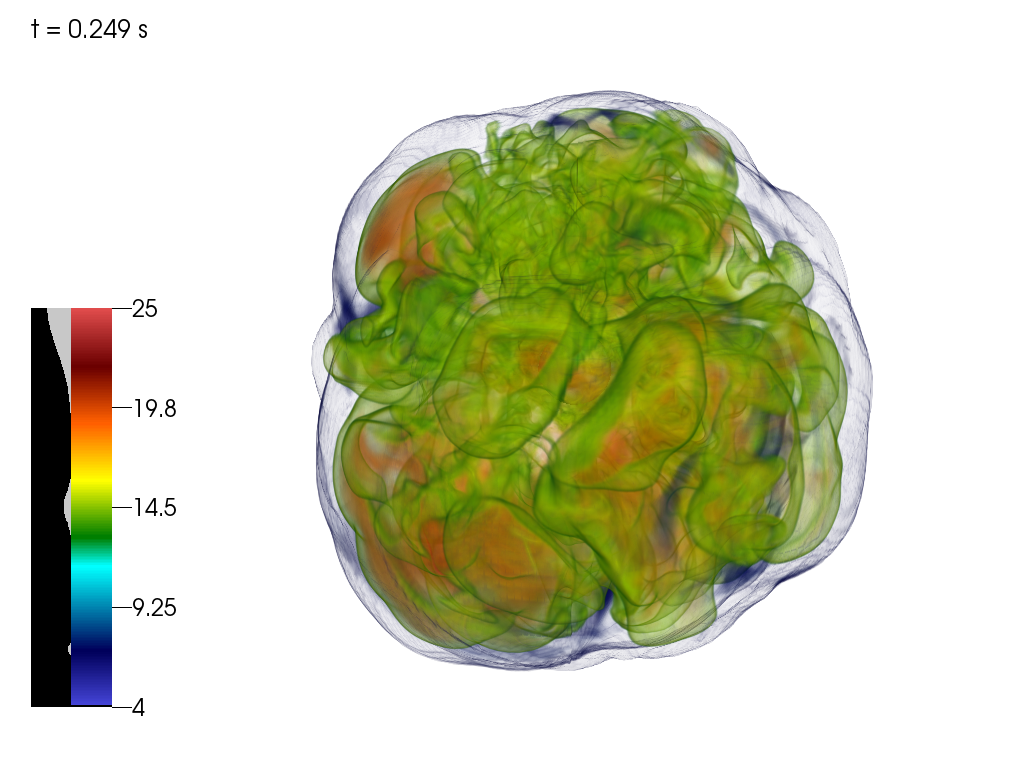}
  \hfill
  \includegraphics[width=0.40\textwidth]{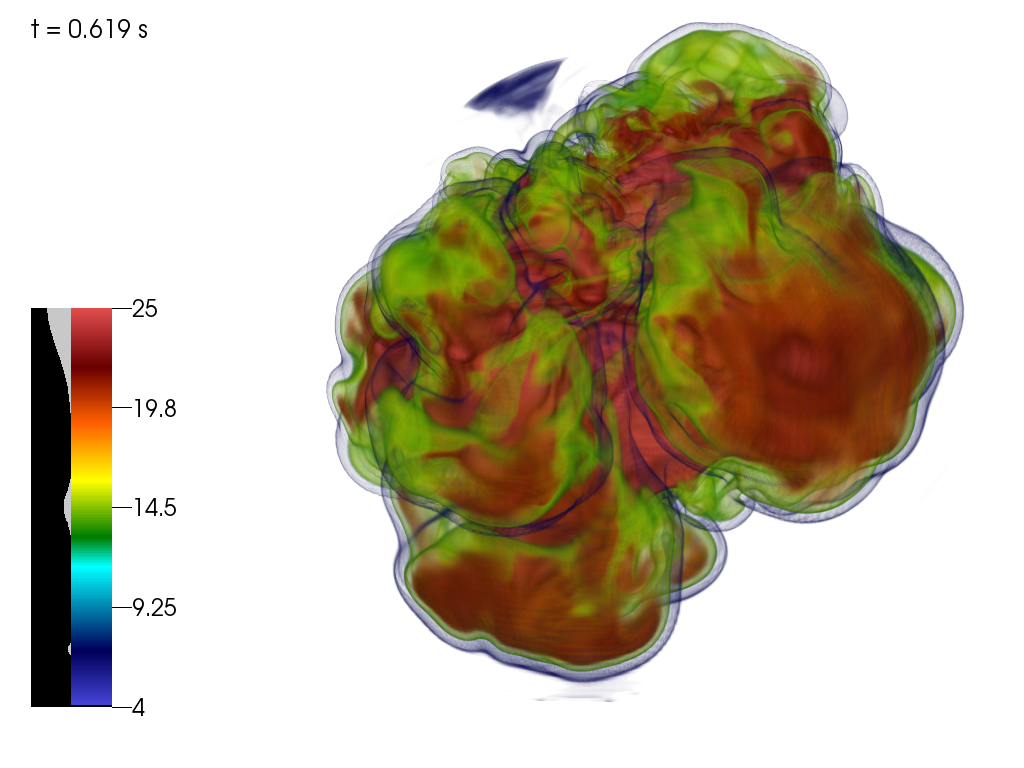}
  \hfill
  \includegraphics[width=0.40\textwidth]{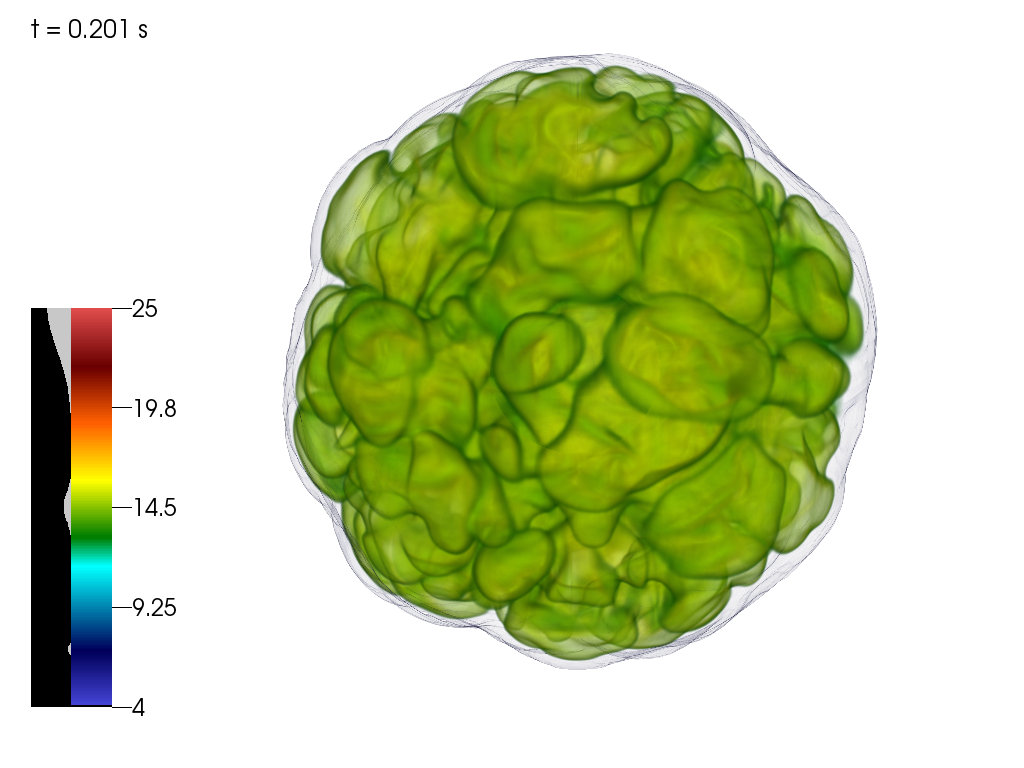}
  \hfill
  \includegraphics[width=0.40\textwidth]{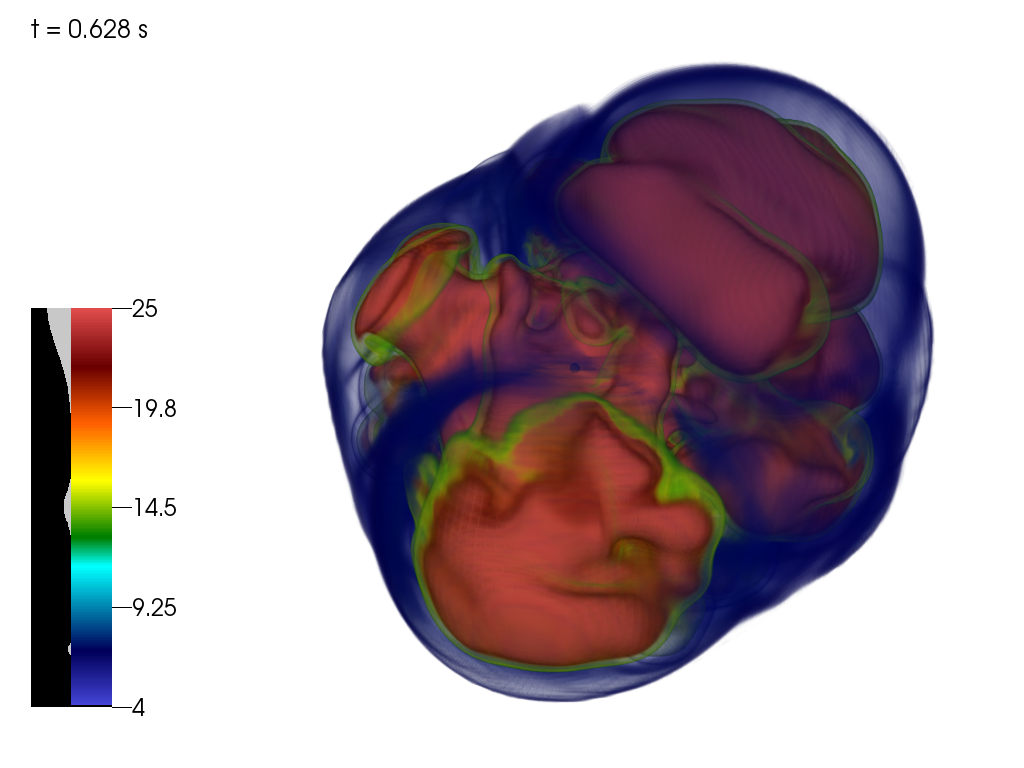}
  \hfill
  \includegraphics[width=0.40\textwidth]{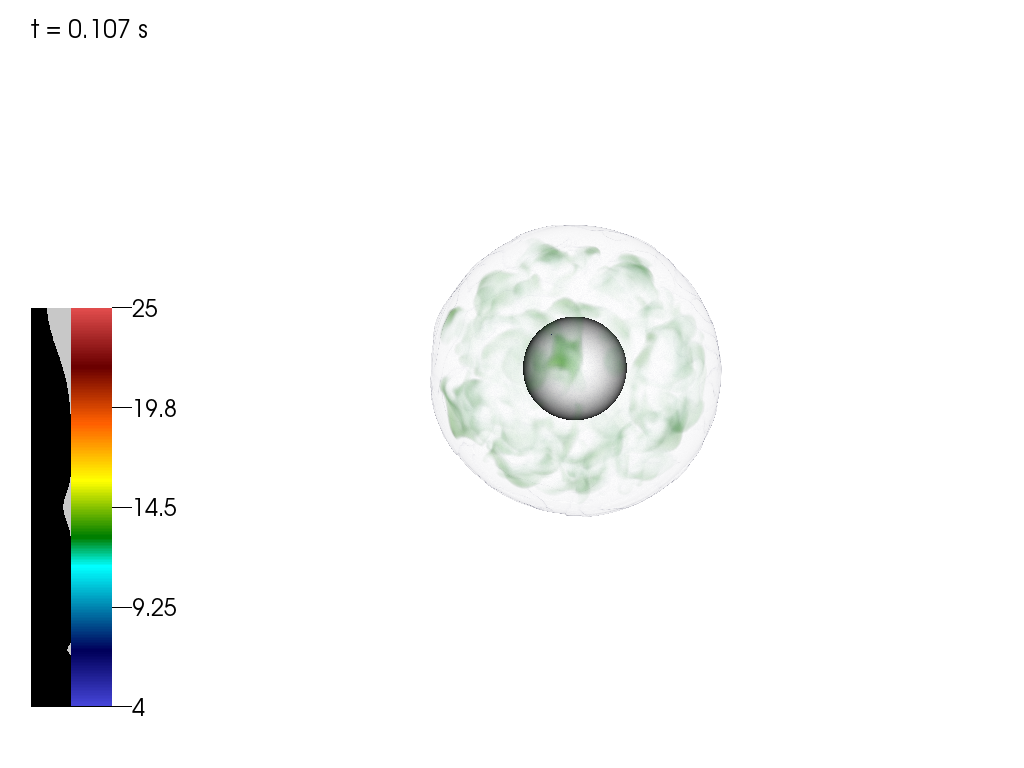}
  \hfill
  \includegraphics[width=0.40\textwidth]{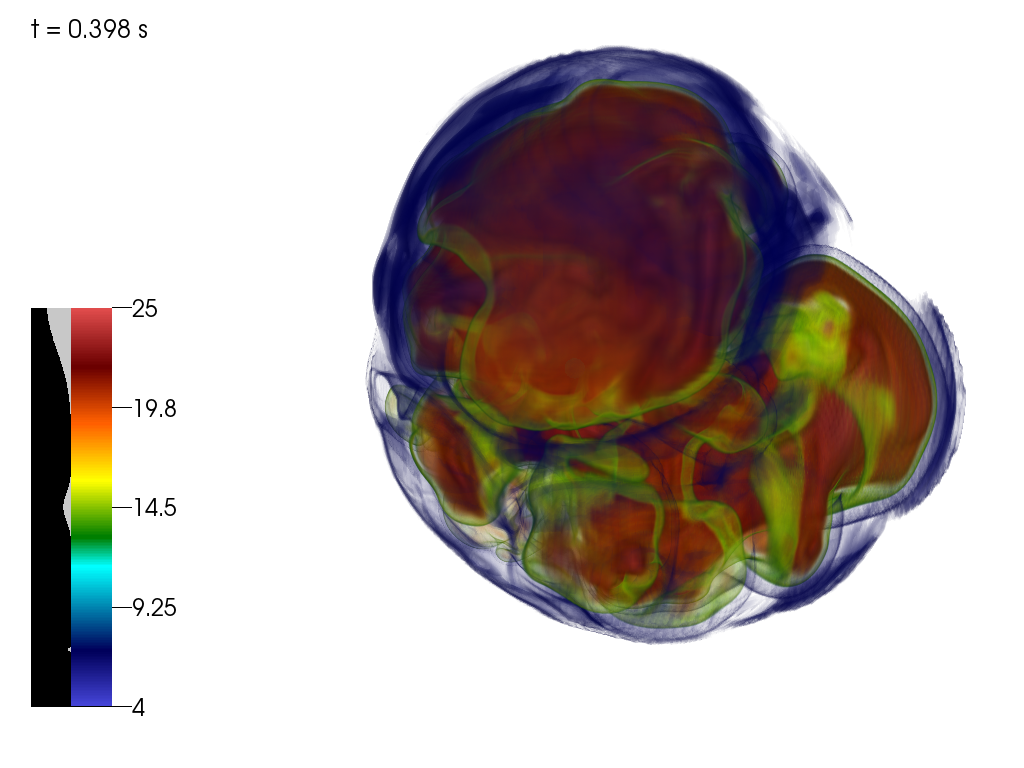}
  \caption{Volume renderings at two different times of four representative 3D models 
   that explode. The 16-, 18-, 20-, and 60-M$_{\odot}$ models are portrayed from top to bottom.
   For these 16-, 18-, 20-, and 60-M$_{\odot}$ models, the times for the stills on the left
   250, 250, 200, and 100 ms after bounce, respectively, and on the right are 637, 639, 651, 
   and 400 ms after bounce, respectively. The blue veil on each plot roughly traces the position of the shock. 
   Also shown are the entropy (per baryon per Boltzmann's constant) colormaps for each
   model and time. Note the transition to redder colors at the later times, 
   indicative of the higher entropies generated through ongoing neutrino heating.
   The physical scales are (left plot, right plot) = ($\pm$320 km/$\pm$2240 km; $\pm$320 km/$\pm$2000 km;
   $\pm$160 km/$\pm$1600 km; and $\pm$160 km/$\pm$960 km), respectively.
}
  \label{morph}
\end{figure*}

\begin{figure*}
  \includegraphics[width=0.40\textwidth]{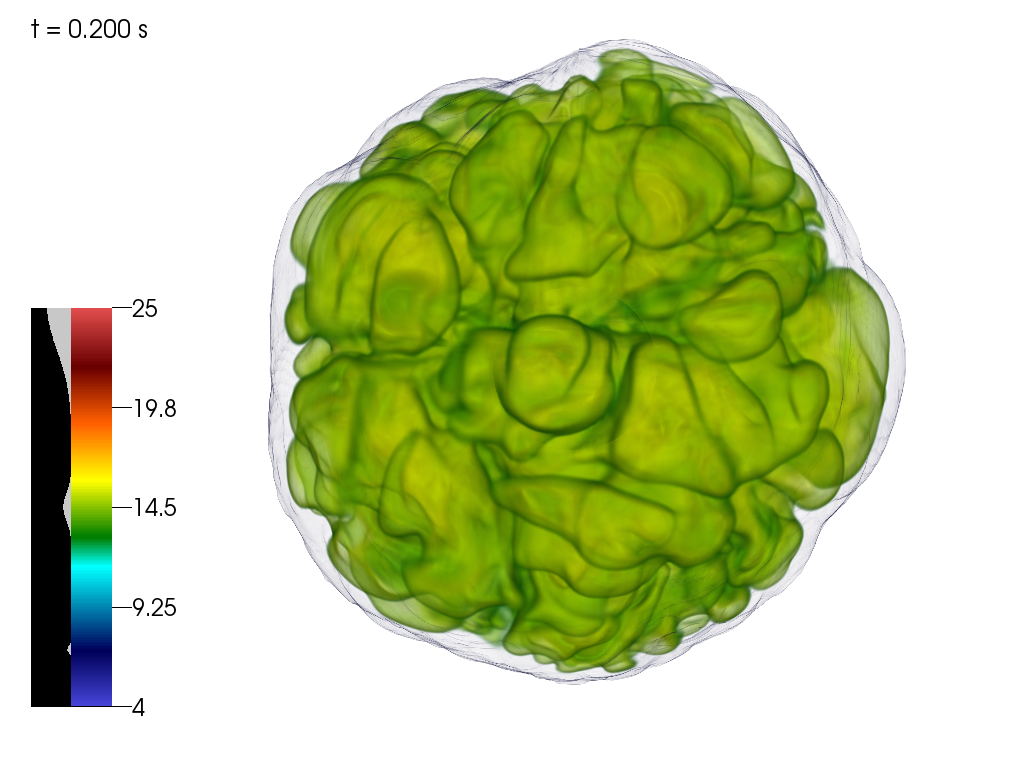}
  \hfill
  \includegraphics[width=0.40\textwidth]{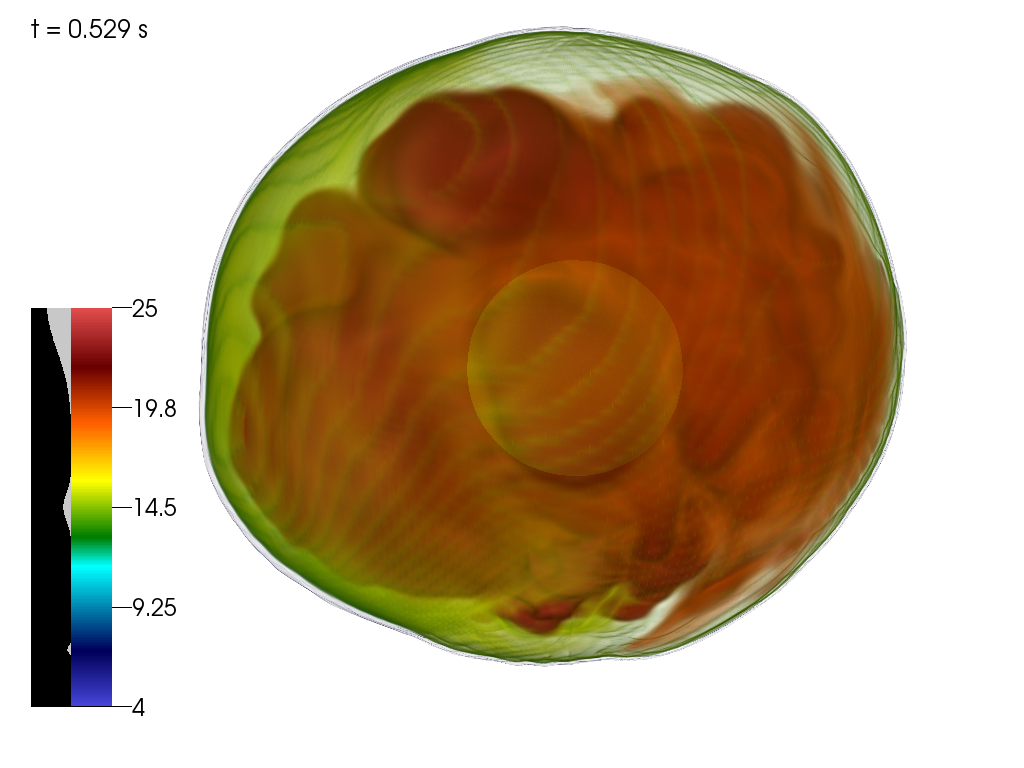}
  \hfill
  \includegraphics[width=0.40\textwidth]{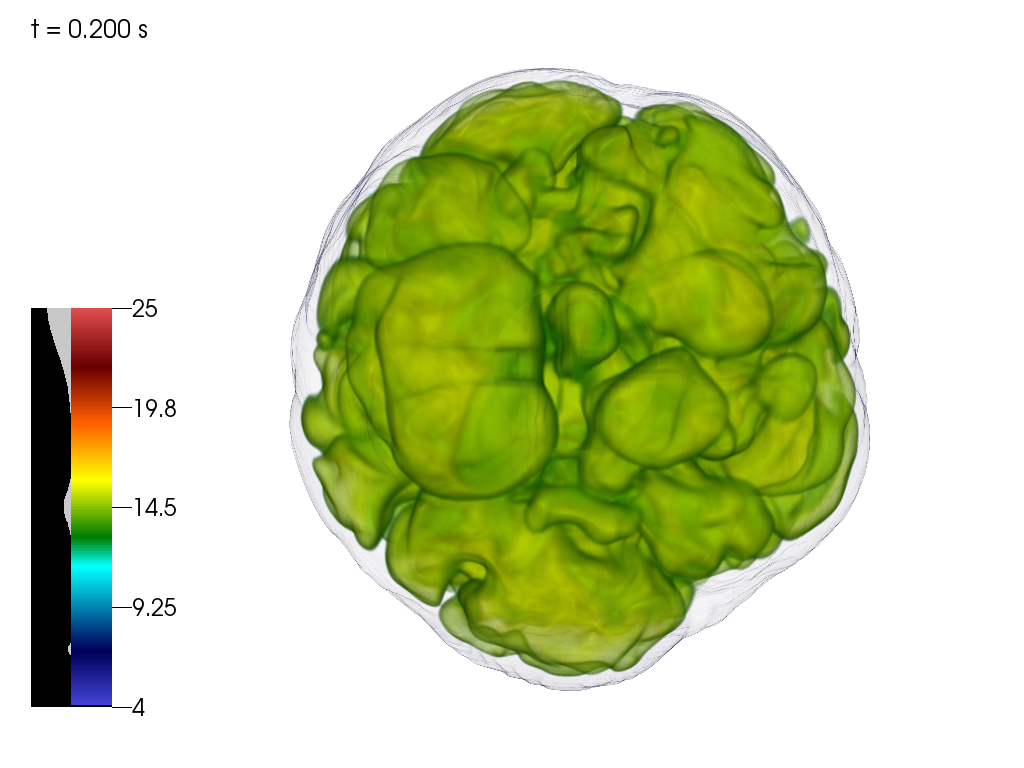}
  \hfill
  \includegraphics[width=0.40\textwidth]{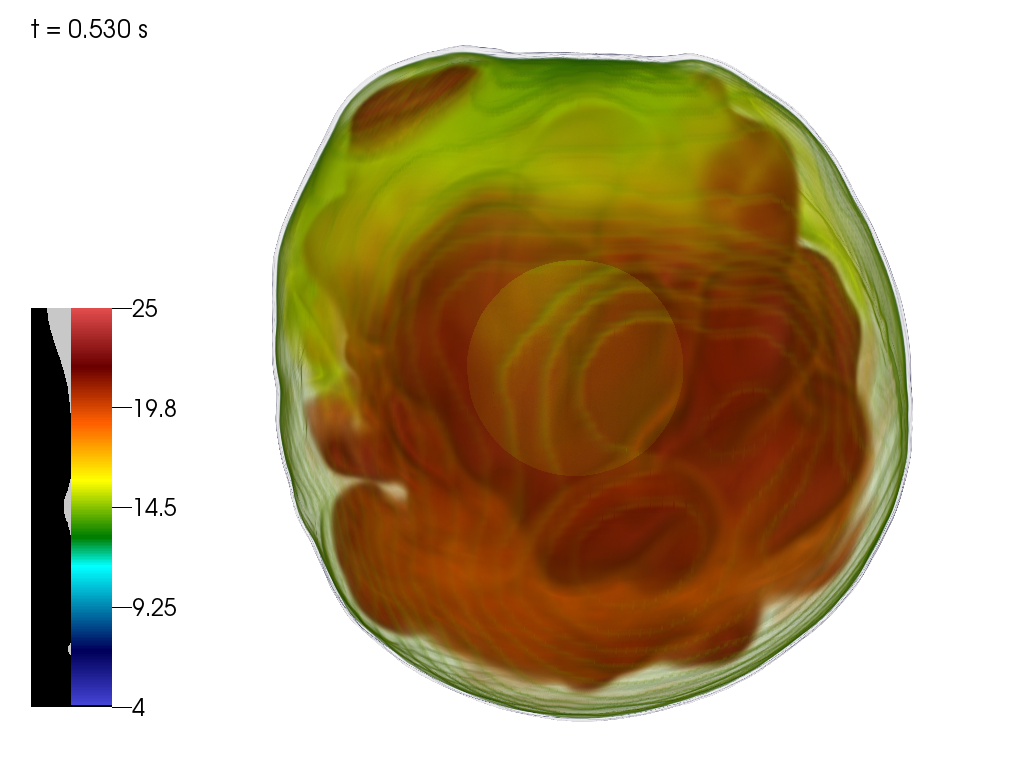}
  \hfill
  \includegraphics[width=0.40\textwidth]{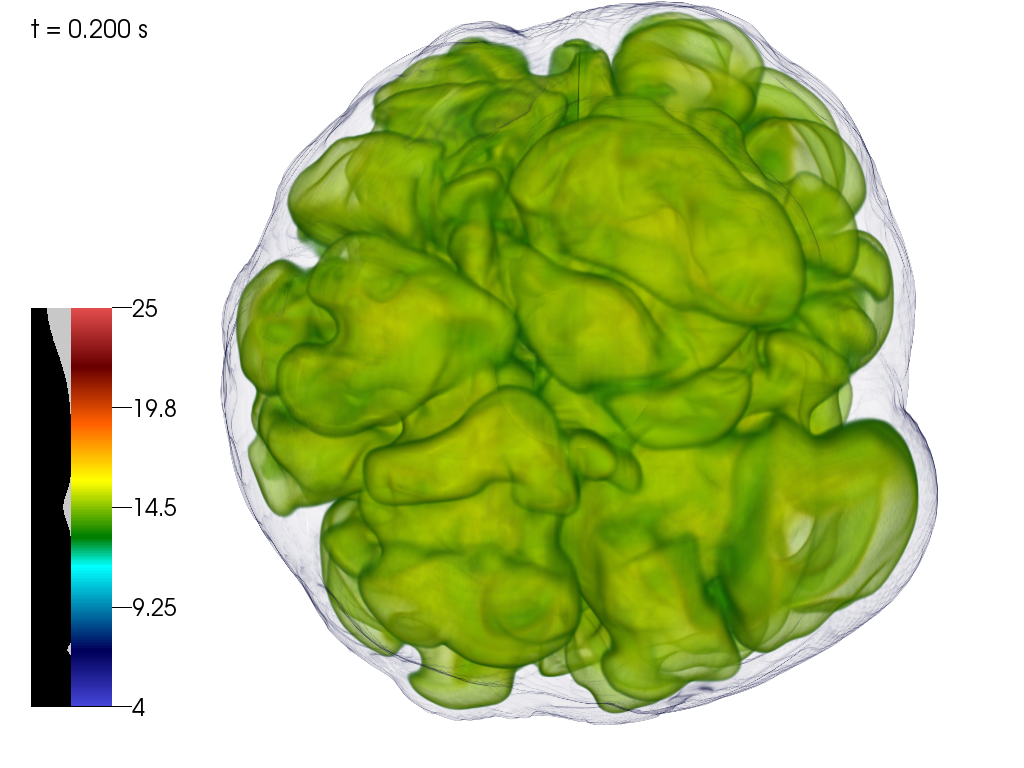}
  \hfill
  \includegraphics[width=0.40\textwidth]{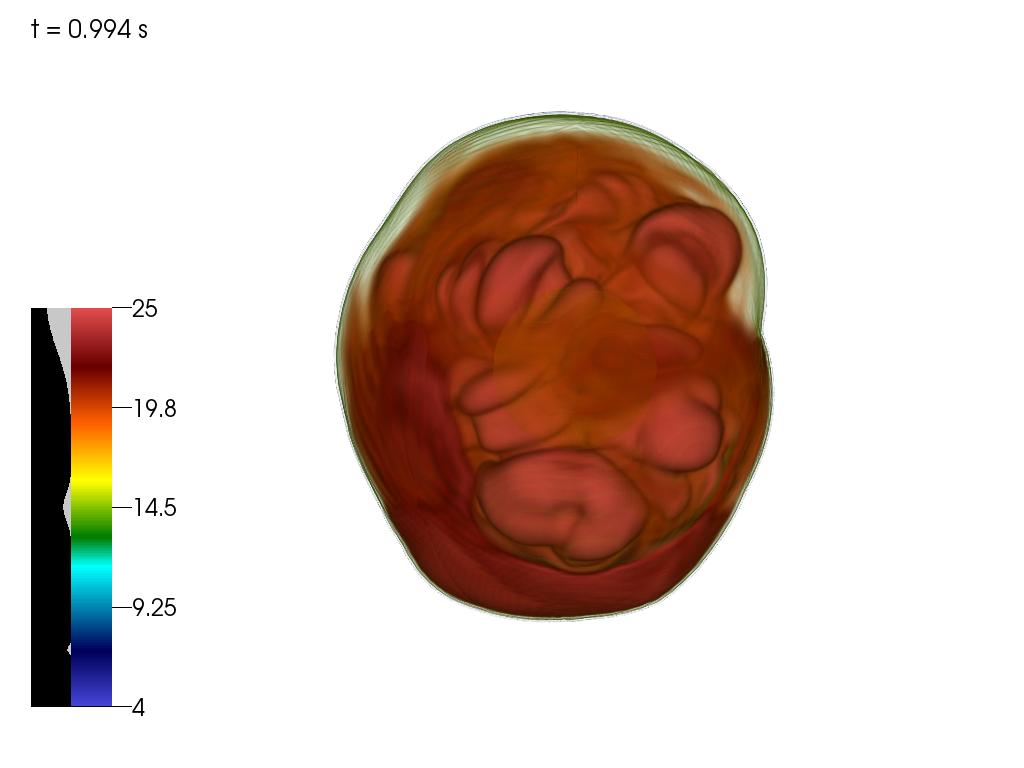}
  \caption{Same as Figure \ref{morph}, but for the three default 3D models that do not explode.
   The 13-, 14-, and 15-M$_{\odot}$ models are portrayed from top to bottom.
   For these 13-, 14-, and 15-M$_{\odot}$ models, the times for all the stills on the left
   are all 200 ms after bounce, and on the right are 550, 550, and 1048 ms after bounce, 
   respectively. Despite the transition to redder colors at the later times,
   indicative of the higher entropies generated through ongoing neutrino heating, 
   these models do not explode. Note the presence of the spheres in the centers on the right.
   This is the proto-neutron star (PNS), here given as a 10$^{11}$ gm cm$^{-3}$ mass-density 
   isosurface. The physical scales are $\pm$160 km/$\pm$80 km (left/right) for all three model plots.
   See text for a discussion.
}
  \label{morph2}
\end{figure*}

Figure \ref{morph} summarizes the evolution before and after explosion of a subset of our
exploding models.  At early times before or near the onset of explosion the rough shape of 
the shock is spherical, with significant bubble structures and a range of scales.  The green color
that dominates on the left (early phase) indicates modest entropies generated due 
to ongoing neutrino heating (Figure \ref{fig:net_heat}). With time and as the explosion progresses,
these colors turn progressively more red as the entropies rise and higher-entropy bubbles drive 
the explosion. A feature of many of our exploding models is the slight pinching near the middle
of the exploding structures \citep{2019MNRAS.482..351V,2019MNRAS.485.3153B} seen on the 
right panels of Figure \ref{morph}.  This is due to accretion of matter in a belt while the 
rest of the mantle is exploding.  The direction of explosion and the positions of this 
``wasp waist" emerge randomly for our non-rotating models. Such accretion helps maintain
a respectable neutrino luminosity during explosion that continues to drive by neutrino heating
that same explosion.  Spherical models can not accomplish this and this feature is one positive aspect
of explosions in the multi-dimensional context that naturally emerges in most of our 3D explosion models. 

Figure \ref{morph2} depicts similar transitions from early to late, but for our non-exploding 13-, 14-, and 
15-M$_{\odot}$ models. For these models, despite the increase in entropy behind the shock no explosion is seen 
and the shock shrinks in radius.  Moreover, there emerges a spiral SASI (Standing-Accretion-Shock-Instability 
\citep{2007ApJ...656..366B,2015PASA...32....9F}) mode that assumes a quite regular wobbling motion.  This quasi-periodic
spiral arm motion may be generic of failed non-rotating core collapse and has distinctive 
gravitational-wave and neutrino signatures \cite{2019MNRAS.tmp.2235V}.  The central sphere
seen at late times in the centers of these stills is the PNS (proto-black-hole?), whose 
radius relative to that of the shock demonstrates the degree to which the shock radius 
has receded at late times for these non-exploding models.

\onecolumn

\begin{figure}
  \includegraphics[width=\columnwidth]{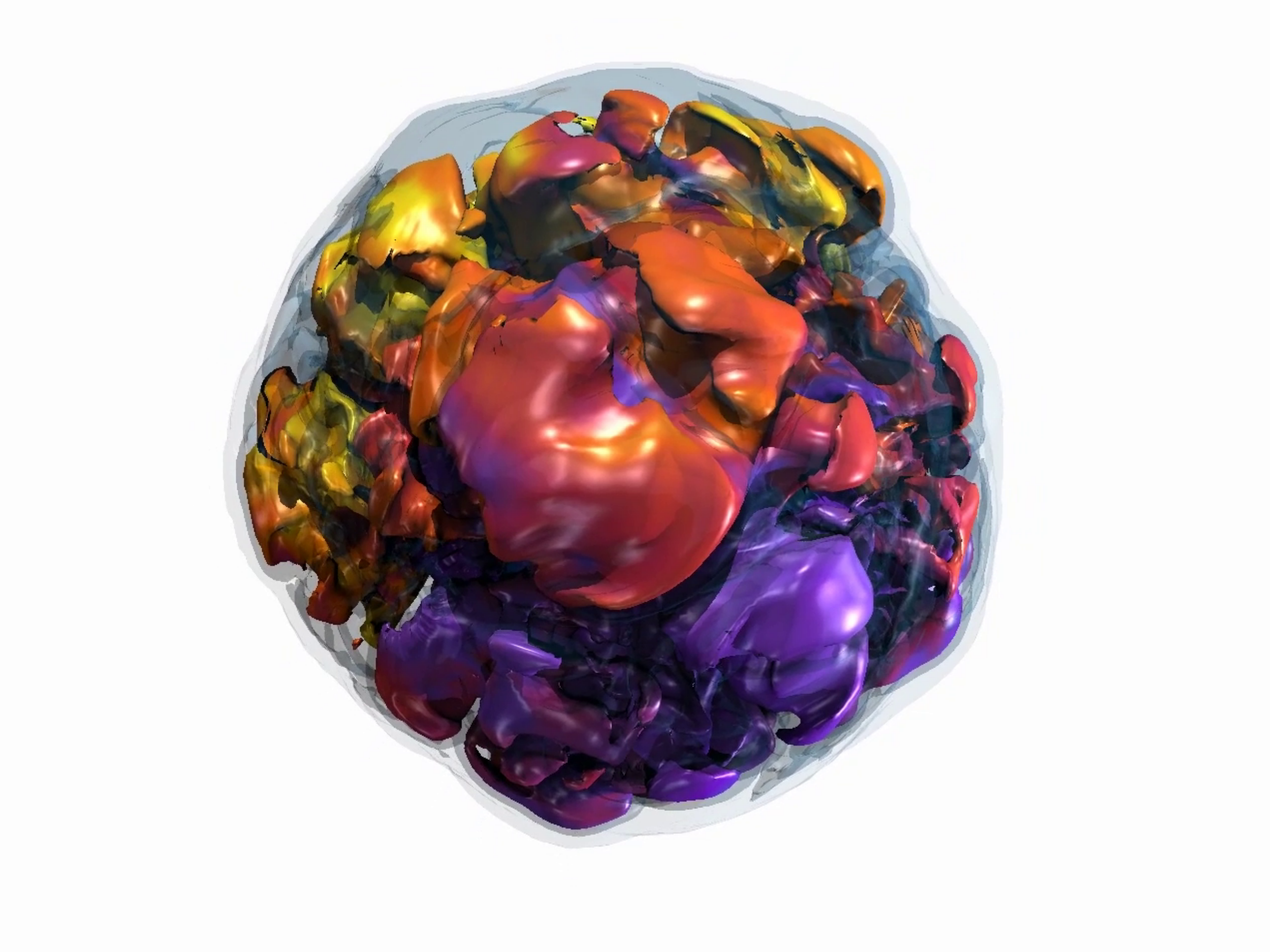}
  \caption{Isoentropy surfaces of the 25-M$_{\odot}$ model, painted by Y$_e$,
  at 390 milliseconds after bounce.  The outer surface is at an entropy per 
  baryon per Boltzmann's constant of 20.  The spatial scale is 500 km on a side. 
  The Y$_e$ colors range from purple ($\sim$0.45), through yellow ($\sim$0.5), 
  to red ($\sim$0.55).  This is just after the onset of explosion
  and portrays the neutrino-driven turbulent structures of the matter behind the 
  shock wave.  The top-bottom contrast in the color indicates that there is a hemispheric
  dependence of the electron-fraction distribution.  This is indicative of the LESA
  phenomenon and is a feature seen in all our exploding models after $\sim$200 milliseconds
  after bounce.
}
  \label{sn25_fig}
\end{figure}

\twocolumn

Figure \ref{sn25_fig} is a representative depiction of the neutrino-driven bubble structures
in our 3D models near the onset of explosion, in this case for the 25-M$_{\odot}$ model.
Shown are isoentropy surfaces painted by Y$_e$.  A range of scales are visible, with larger
scales starting to dominate.  Note that there is a distinct color contrast between the top and bottom
hemispheres in this model at this time.  This Y$_e$ asymmetry is an indirect signature of the LESA
phenomenon \citep{2014ApJ...792...96T} and we go into this in more detail in \citet{2019MNRAS.tmp.2235V}.

\subsection{Shock Shape and Structure}

\begin{figure}
  \includegraphics[width=\columnwidth]{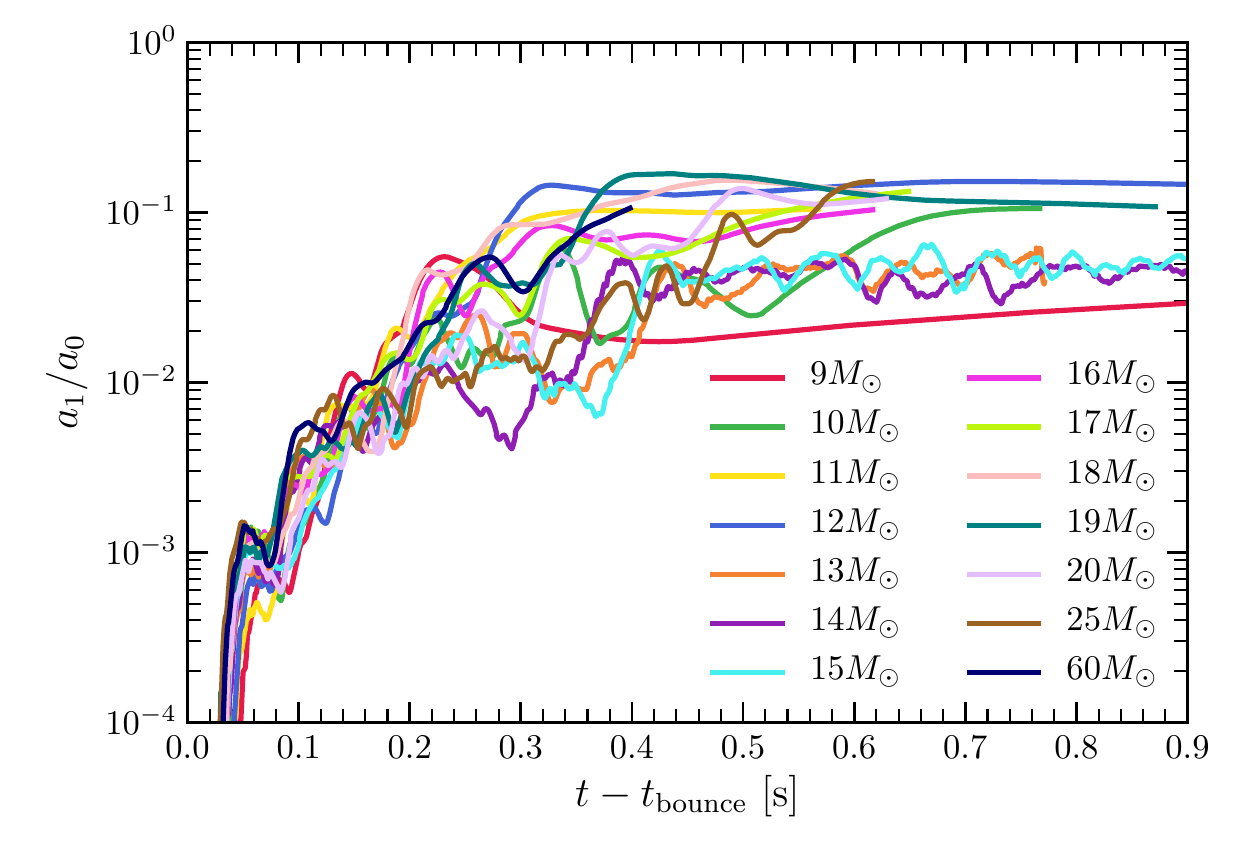}
  \caption{Normalized shock dipole moment. All models show fast growing
  deformations of the shock. The shock dipole saturation level for
  exploding models is a factor of a few larger than for failed
  explosions despite the appearance of the spiral SASI mode for the latter. The
  9-M$_\odot$ progenitor explodes almost spherically and is an exception to
  this general trend.}
  \label{fig:rshock.l1}
\end{figure}

One way to characterize and depict the asphericities of the pre- and post-explosion 
hydrodynamics is to study the spherical harmonic decomposition of the shock wave surface \citep{burrows2012}.
The monopole is the solid-angle-averaged shock radius and the higher-order multipoles
reflect the degree of corrugation of this surface in response to turbulent upwelling 
and distortion during its propagation. It has been shown in the past \citep{dolence:13,2019MNRAS.482..351V} 
that the explosion monopole is accompanied by a dominant dipole, but the exploration of this
decomposition for a large, uniform suite of 3D models has not been possible until now.
Figure \ref{fig:rshock.l1} depicts the magnitude of the monopole-normalized dipole (N.B., the dipole is a vector)
for the fourteen fiducial models of this paper.  There are a few notable aspects of this collection
to emphasize.  First is that, with individual variation on timescales of tens of milliseconds,
all the dipole magnitudes experience a quasi-exponential linear growth phase during the first $\sim$200 ms after bounce,
with a time constant (e-folding time) of $\sim$20 ms.  This time scales with the advection and sound travel times
between the stalled shock at $\sim$150 km and the inner core.  Second, the 9-M$_{\odot}$ model (red)
never achieves a signifcant dipole after explosion, though early in its explosion it seems to be on
a trajectory to achieving one.  This reflects the near spherical behavior of this explosion, which 
resembles more a wind driven by the quasi-spherical diffusive flux from the inner core, unaided
by much accretion-fueled neutrino power \citep{2019MNRAS.485.3153B,1995ApJ...450..830B}.  Third, the shock waves of the non-exploding 
13-, 14-, and 15-M$_{\odot}$ models retain a much more spherical character than most of the other exploding models, 
with their dipoles limited to only a $\sim$0.5\% effect.  The exploding models (except the 9-M$_{\odot}$) 
all achieve a dipole whose magnitude is as much as an order of magnitude larger than seen for the non-exploding models.  
This distinction between non-exploding and exploding models, except in the case of 
the 9-M$_{\odot}$ model, seems striking and may be robust.  Once the non-linear turbulence sets in
and vigorous convection is manifest, the subsequent neutrino- and turbulent-pressure-aided explosion 
almost always assumes a pronounced dipolar component.  

We note that the onset of the non-linear turbulent 
phase will likely depend upon the magnitude and character of the seed perturbations (velocity and thermal fields) 
in the progenitor.  Had we used a different approach to seeding the flow (\S\ref{methods}), the details 
of the developments just described could well have been quantitatively, though probably not qualitatively, 
different.

\begin{figure}
  \includegraphics[width=0.49\textwidth]{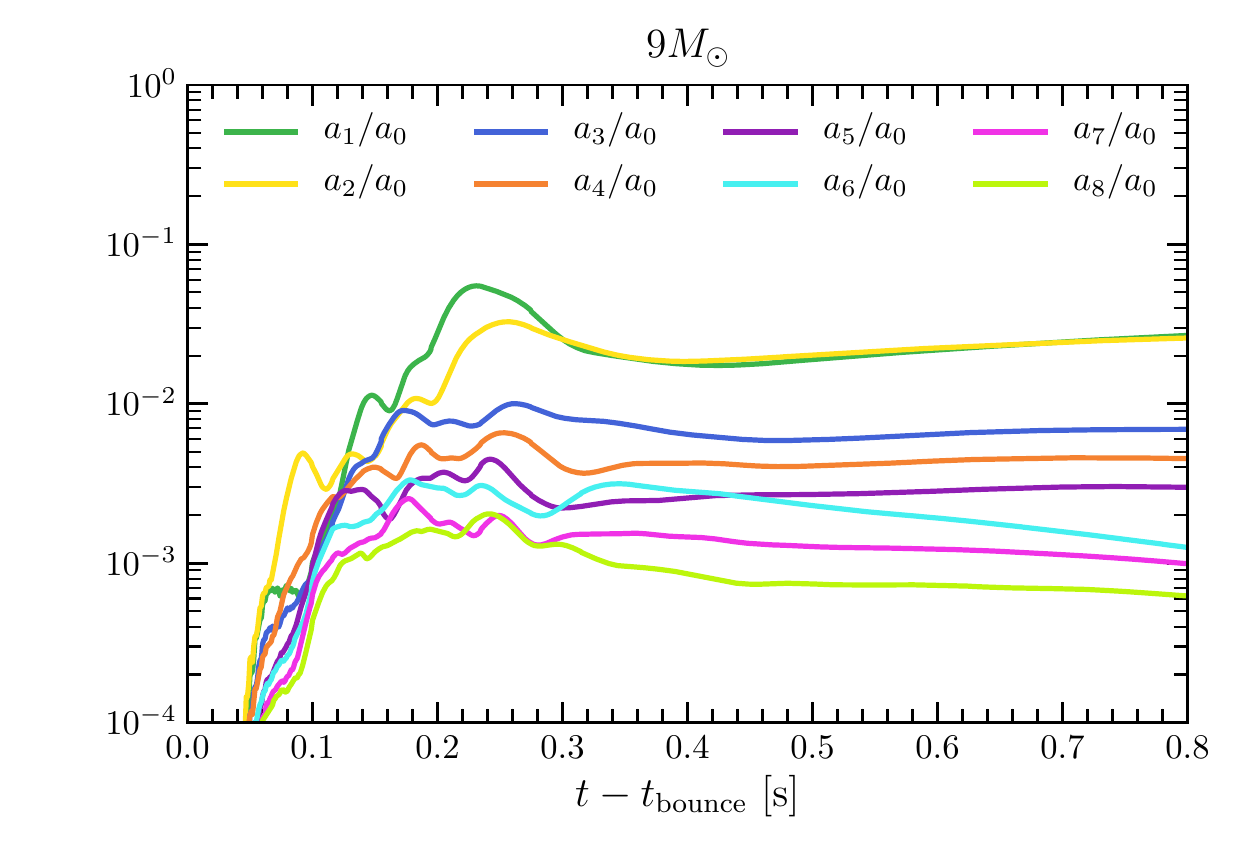}
  \hfill
  \includegraphics[width=0.49\textwidth]{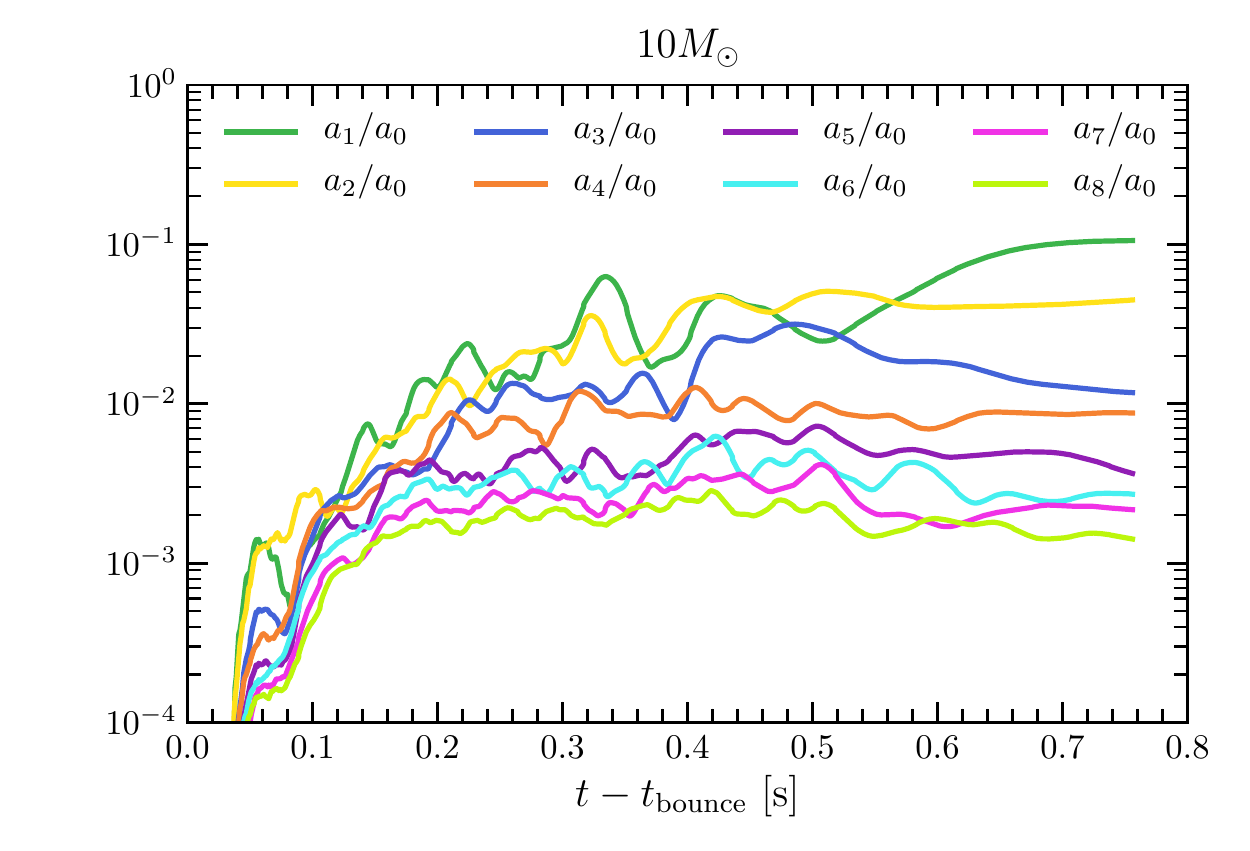}
  \includegraphics[width=0.49\textwidth]{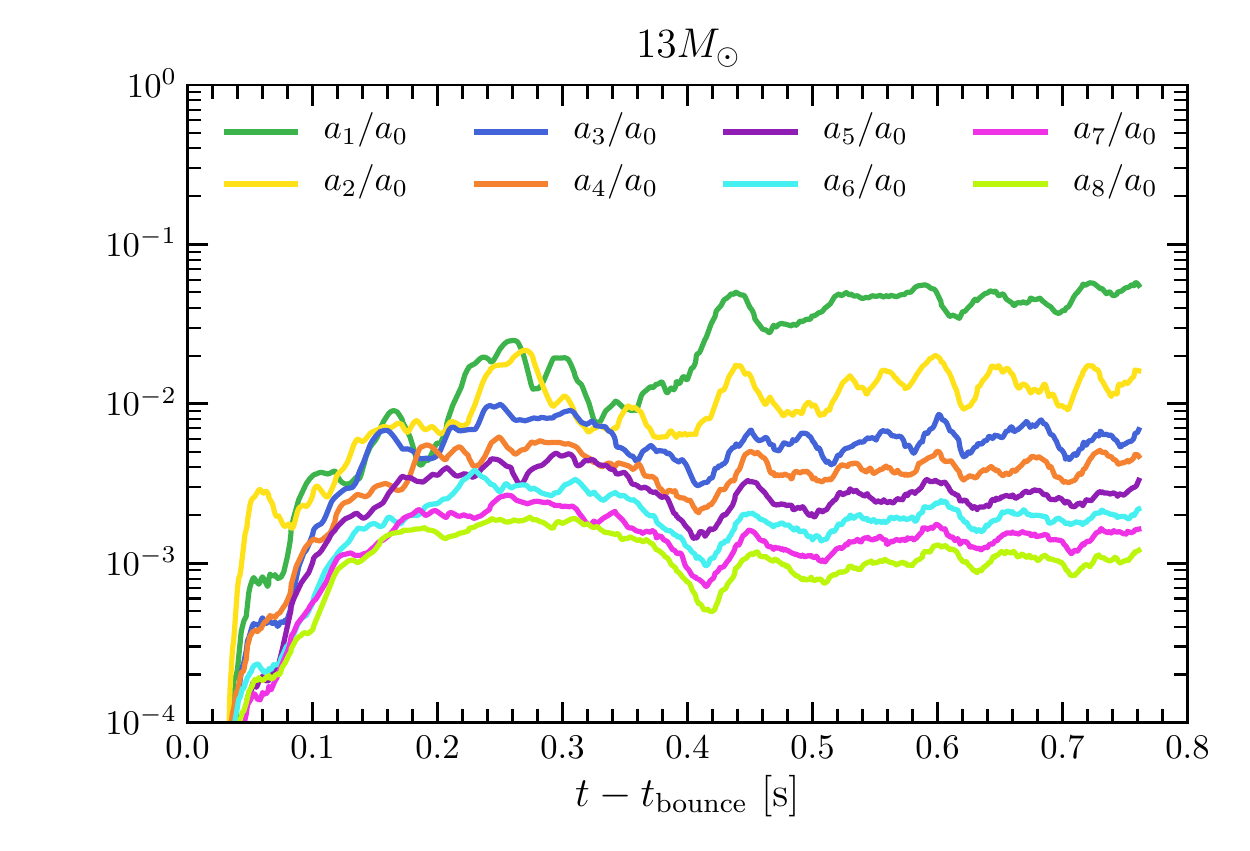}
  \hfill
  \includegraphics[width=0.49\textwidth]{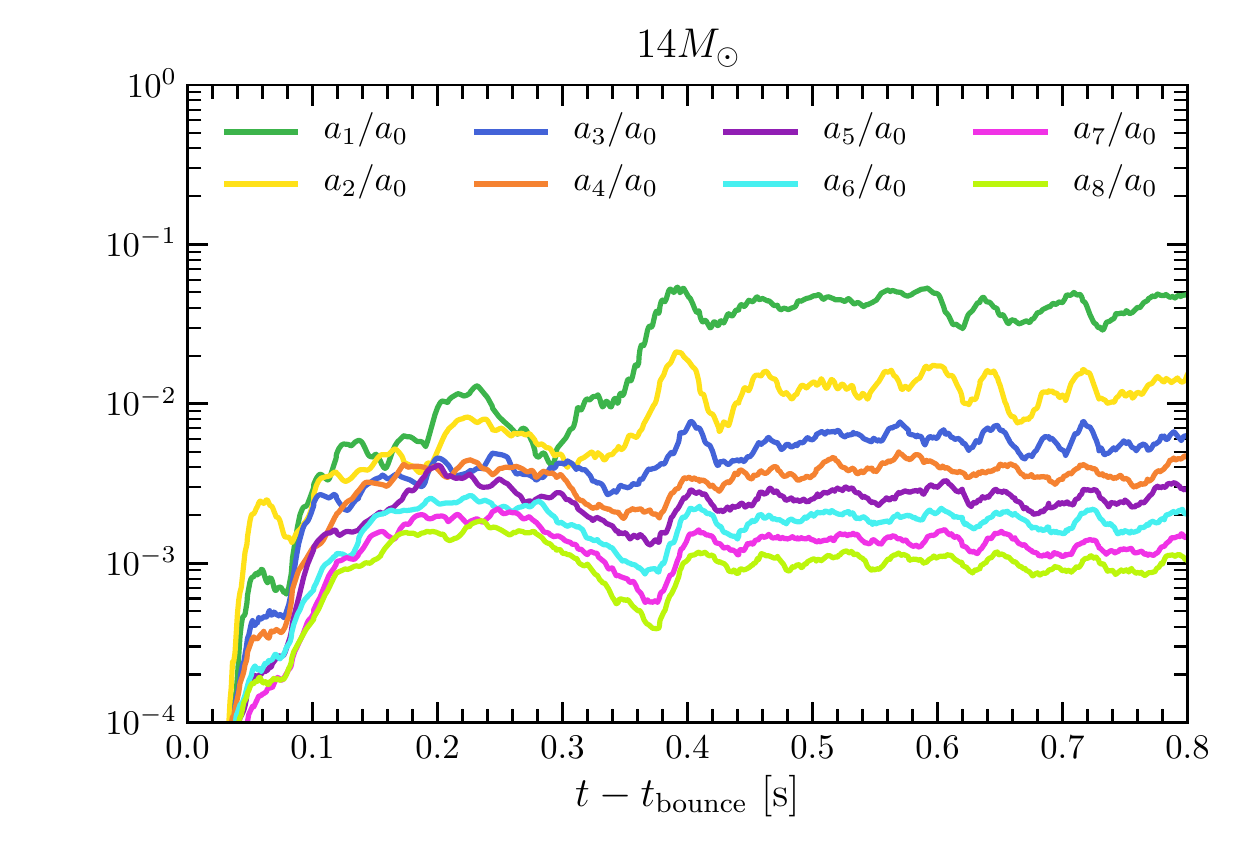}
  \includegraphics[width=0.49\textwidth]{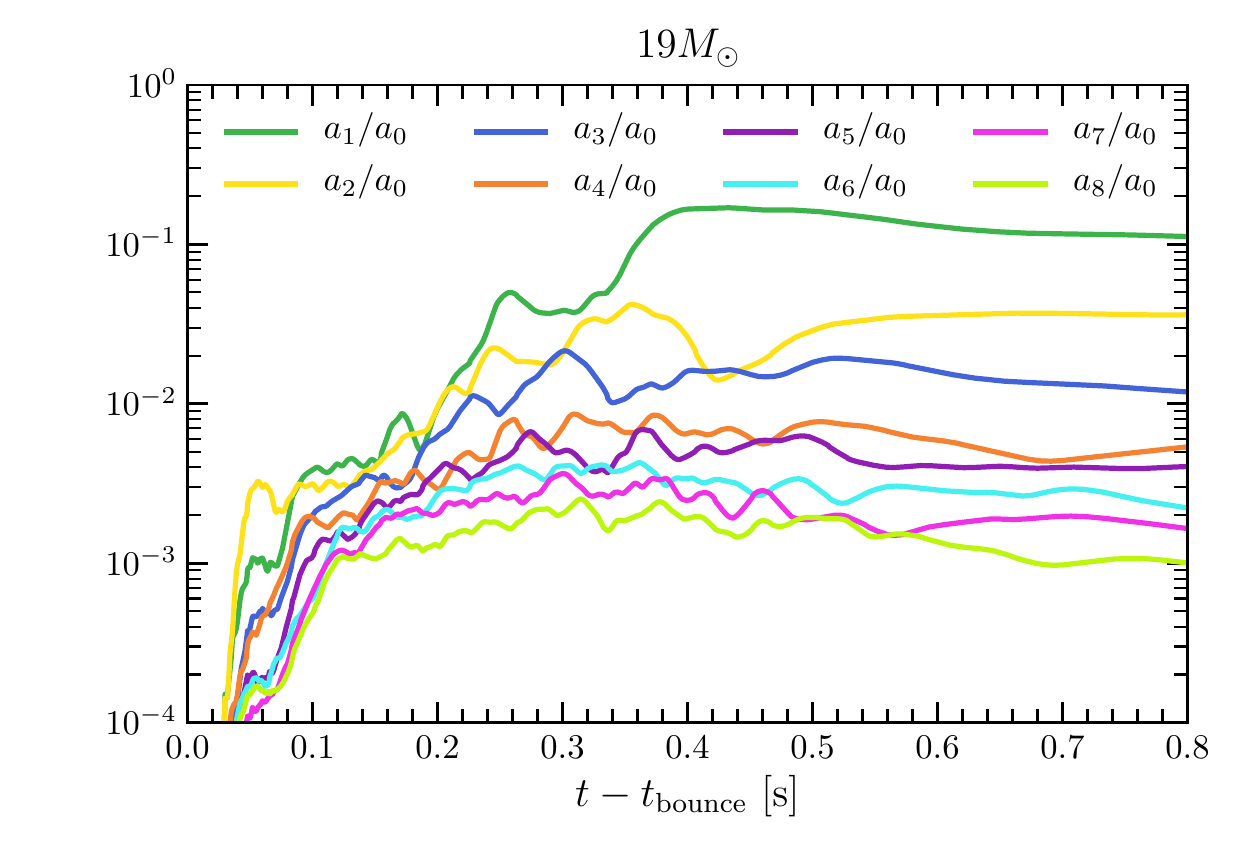}
  \hfill
  \includegraphics[width=0.49\textwidth]{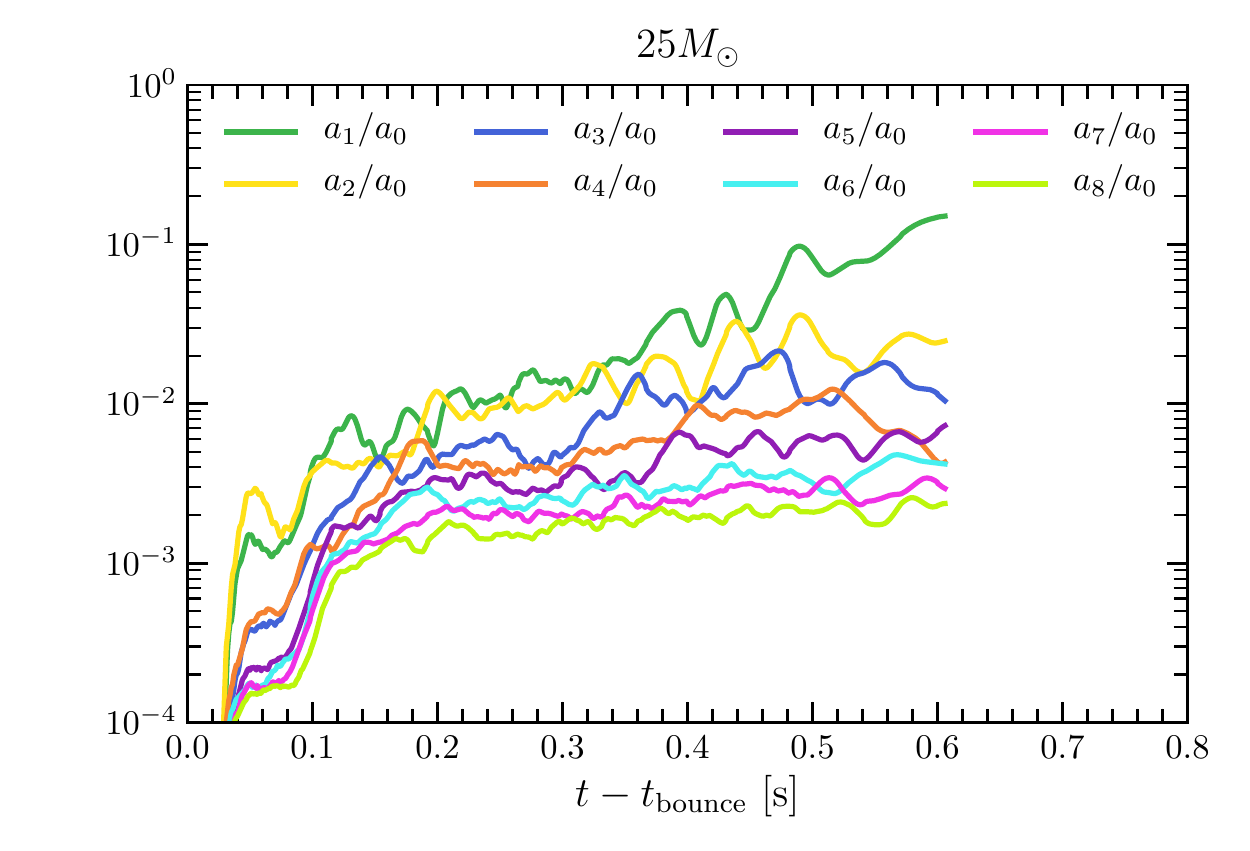}
  \caption{Spherical harmonics decomposition of the shock radius for
  selected models. All our models show significant deformations of the shock
  predominantly of dipolar character. The deformations are typically
  larger for the exploding models, with the exception of the 9-M$_\odot$
  progenitor where the shock front remains close to spherical.}
  \label{fig:rshock.lm}
\end{figure}

Figure \ref{fig:rshock.lm} depicts the hierarchy of normalized shock multipoles for a representative subset
of progenitors.  The square root of the sum of the squares of the $m$ subcomponents for each $\ell$, a rotational
invariant, is what is plotted. Not only is the prominence of the dipole clearly reinforced, but the various multipoles 
seem to be nested (by and large) one over the other as a function of spherical harmonic order $\ell$, the highest-order $\ell$
having the lowest magnitude. It is almost as if explosion is correlated with the onset of the growth (as opposed to damped oscillation) of
various harmonic mode perturbations of the shock surface, with the growth rates of the $\ell$ modes being a monotonically 
decreasing function of $\ell$. The $\ell = 0$ and $\ell = 1$ modes have the greatest rates of growth, and we can see this 
in the morphologies of the blasts (\S\ref{ejecta}). An intriguing future project would be to explore whether 
this speculation concerning a modal analysis has quantitative merit.

\onecolumn

\twocolumn

\begin{figure}
  \includegraphics[width=\columnwidth]{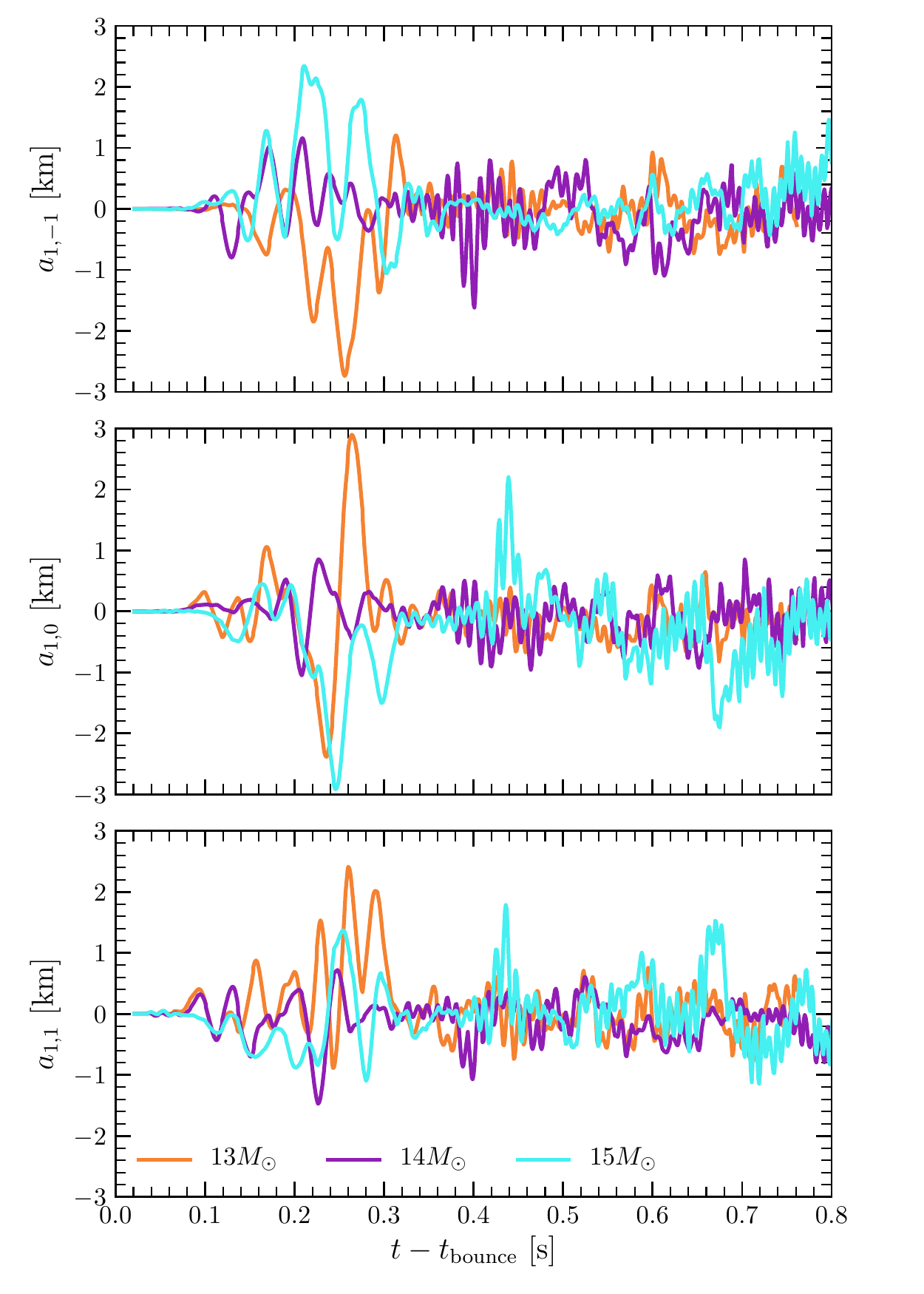}
  \caption{Shock dipole moments for the models that fail to explode.
  Large spiral-SASI-like oscillations are observed at late times for all the
  models that do not achieve shock revival.}
  \label{fig:sasi}
\end{figure}

Another intriguing trend is shown in Figure \ref{fig:sasi}.  This figure portrays the dipole
subcomponents of the shock surface for the three non-exploding progenitors as a function of time after bounce.
The qualitative behavior does not differ from one model to the next. However, there are intriguing
features of these plots and models that bear mentioning. The vector amplitudes of all three grow with 
time in the first $\sim$0.3 seconds, during which the mean shock radius also increases, but after which 
the magnitude of the dipole subsides.  The latter phase marks the shrinkage of the mean shock radius and
is near when it becomes clear the model will not explode (at least as the others did). The characteristic 
pulsation time is of order $\sim$20-25 ms, again comparable to the sound and advection times between the 
shock and the inner core when the shock is at its greatest extent.  Afterwards, with the shrinkage of 
the shock radius, the characteristic timescale of the subsequent oscillations diminishes to $\sim$10-15 ms. 
However, the oscillation frequency becomes a bit more regular.  What is not obvious from these plots is
that the phases of the component oscillations are such that we are witnessing a spiral mode, with the timescale
of variation the timescale of the rotation of the mode.  This is likely the spiral SASI 
\citep{2007ApJ...656..366B,2015PASA...32....9F}, and has distinctive gravitational-wave 
and neutrino-emission signatures \citep{2016ApJ...829L..14K,2019MNRAS.tmp.2235V}.  
Otherwise, only for a short interval during the early post-bounce phase of the 25-M$_{\odot}$ model do 
we see the original SASI mode \citep{blondin2003,2019ApJ...876L...9R}. Moreover, neither the original nor the spiral 
SASI are in evidence during any of the exploding phases. At least in our calculations, all variants of SASI seem 
to show up clearly only in the most compact shock configurations. However, whether the spiral SASI leads to 
a more dynamical later phase \citep{2016MNRAS.461L.112T}, beyond the horizon of our simulations, 
seems unlikely, but is yet to be determined.

\begin{figure}
  \includegraphics[width=\columnwidth]{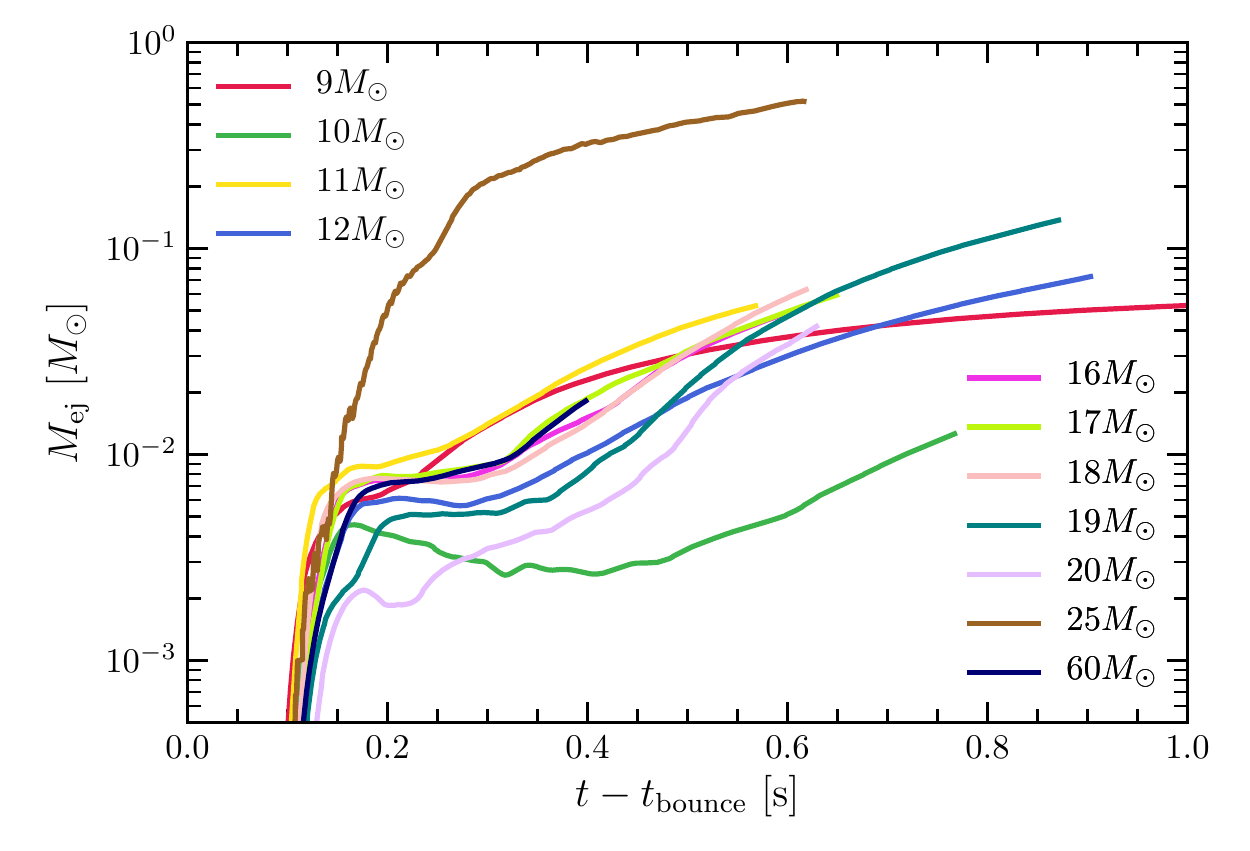}
  \caption{Mass of the material that is not gravitationally bound. Note
  that for non-exploding models some material is formally unbound for a 
  brief period of time. However, as the shock recedes deeper into the
  potential well of the PNS, all of the matter behind the shock becomes
  bound.}
  \label{fig:mej}
\end{figure}

\begin{figure}
  \includegraphics[width=\columnwidth]{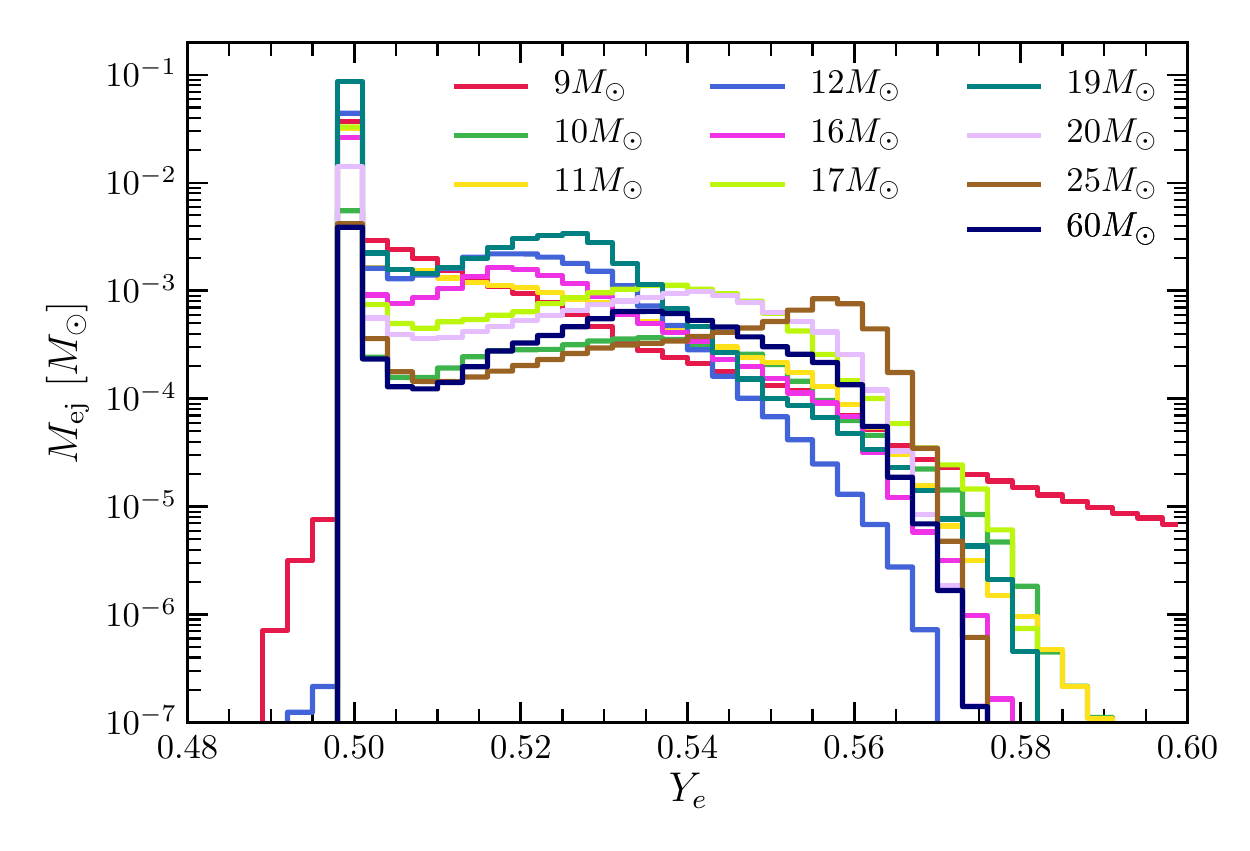}
  \caption{Electron fraction distribution of the ejecta. All models
  display a peak at $Y_e = 0.5$ associated with the production of
  ${}^{56}{\rm Ni}$. Additional abundance peaks are found at different
  $Y_e$'s for different simulations, suggesting that the nucleosynthesis
  yields might depend on the structure of the imploding stellar cores
  and show differences between low-compactness and high-compactness
  progenitors.}
  \label{fig:hist.ye}
\end{figure}

\begin{figure*}
  \includegraphics[width=0.49\textwidth]{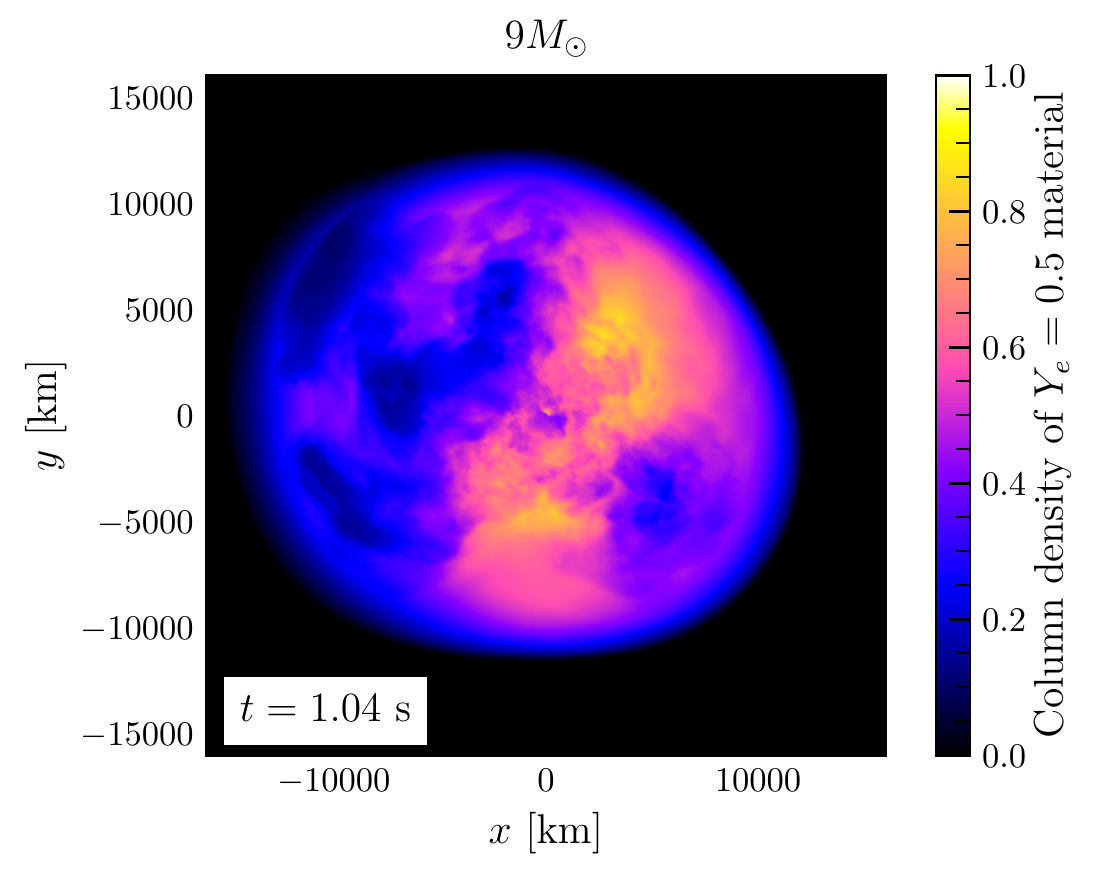}
  \hfill
  \includegraphics[width=0.49\textwidth]{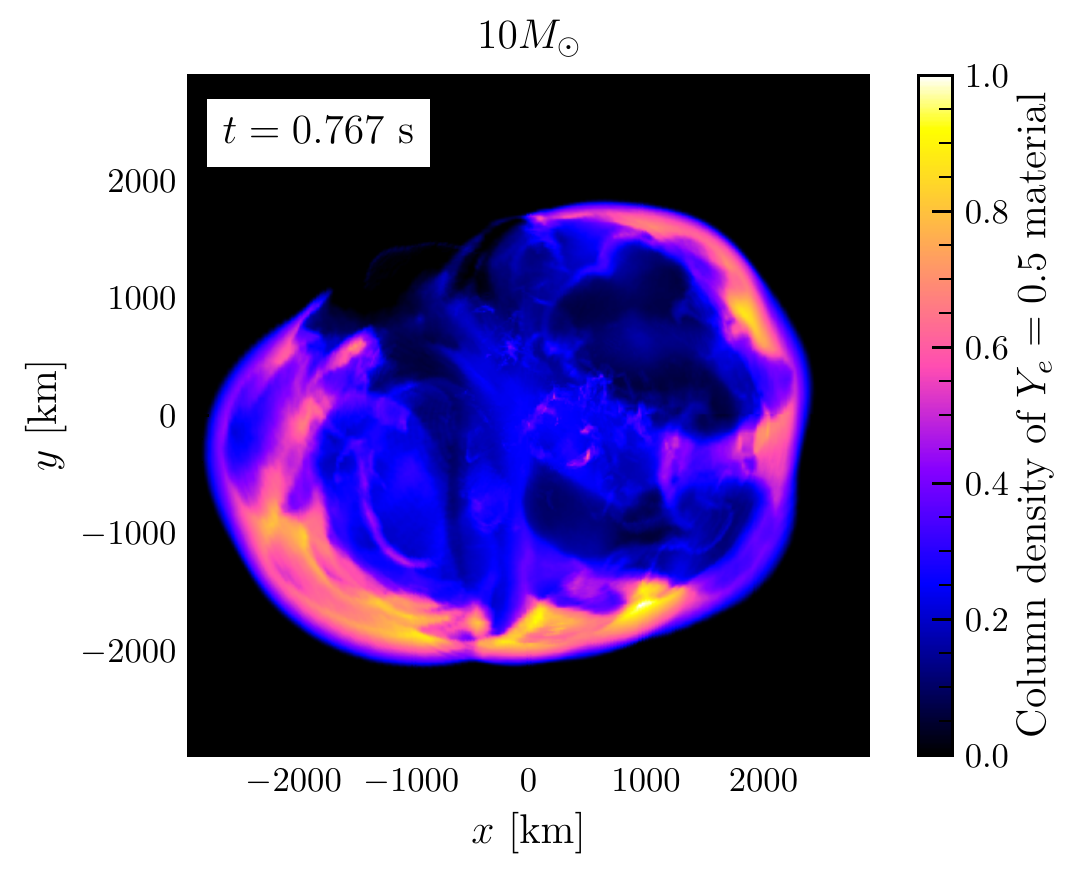}
  \includegraphics[width=0.49\textwidth]{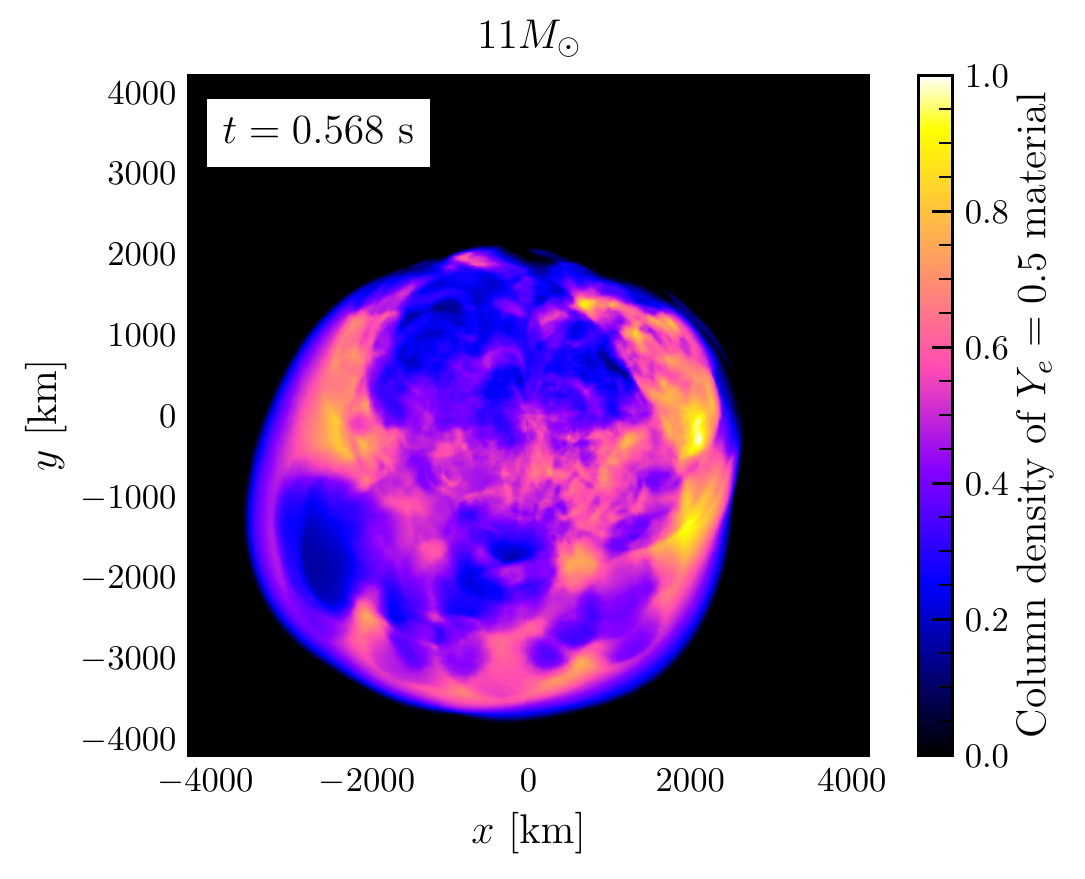}
  \hfill
  \includegraphics[width=0.49\textwidth]{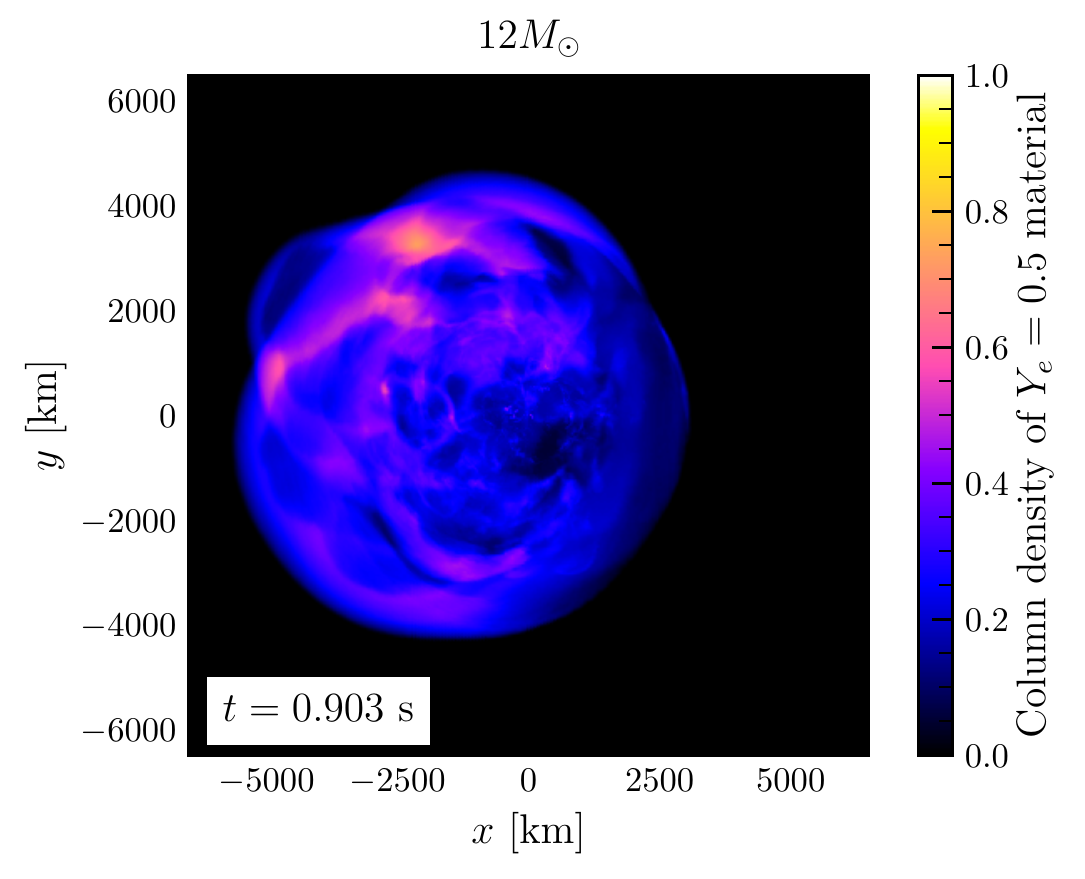}
  \includegraphics[width=0.49\textwidth]{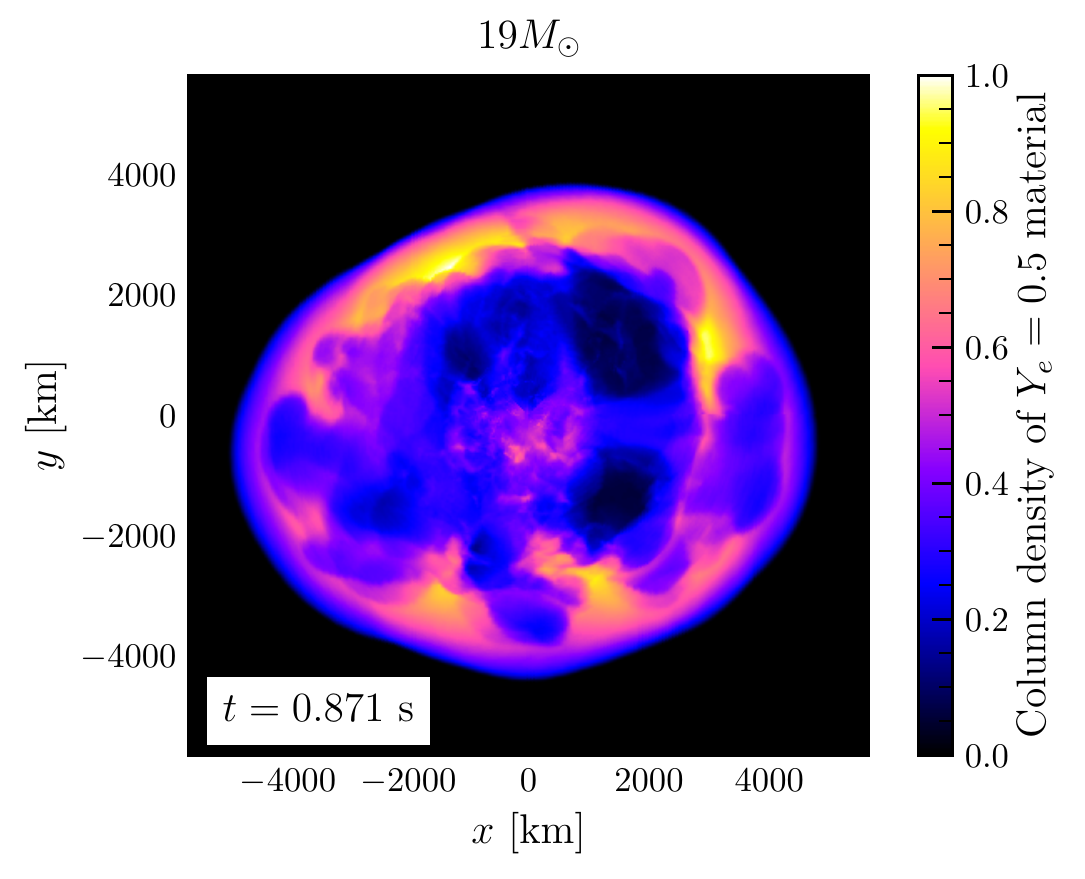}
  \hfill
  \includegraphics[width=0.49\textwidth]{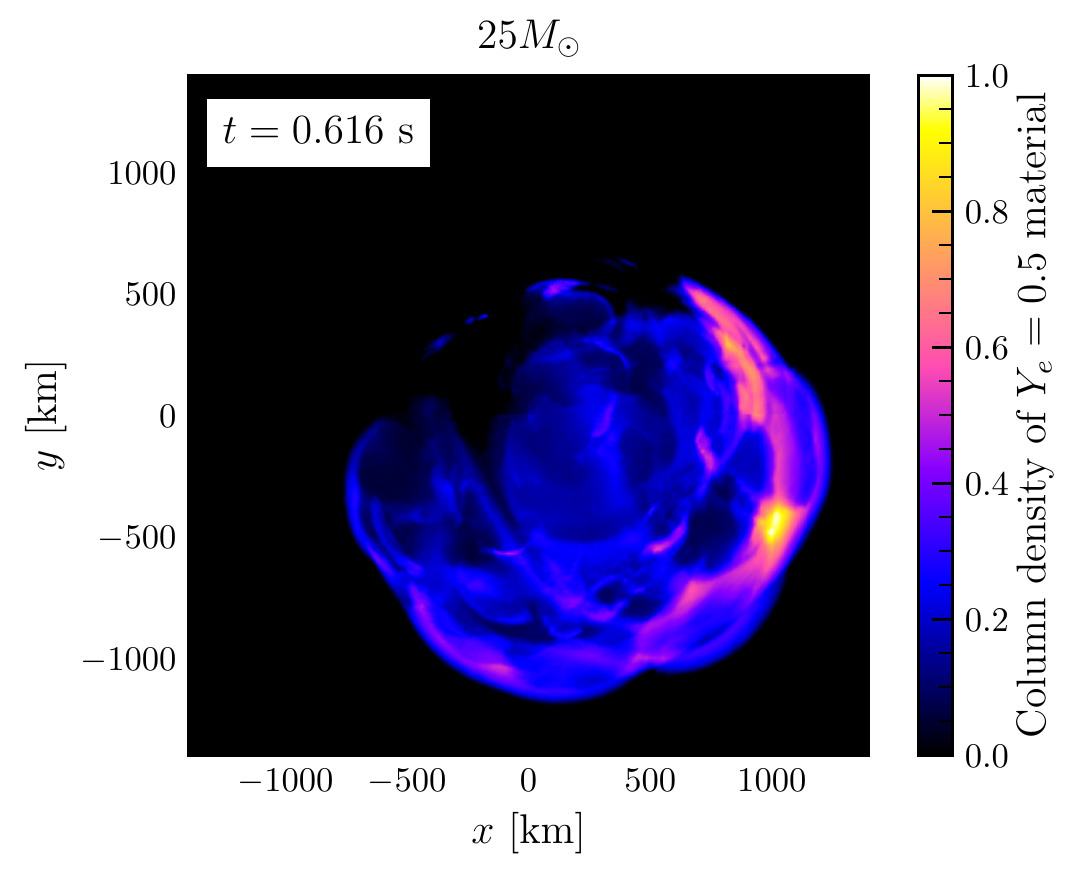}
  \caption{False-color maps of the column density of {the Y$_e$ = 0.5 ejecta, wherein 
   ${}^{56}{\rm Ni}$ may reside} at the end of
   the 9-, 10-, 11-, 12-, 19-, and 25-M$_{\odot}$ simulations. {Note that
   since we have not performed proper nucleosynthetic calculations these maps
   are merely notional indications of the approximate positions of whatever 
   ${}^{56}{\rm Ni}$ might be ejected.} The ejecta have a complex morphology reminiscent of elemental
   abundance maps of supernova remnants. However, much longer simulations with nuclear
   burning and nucleosynthesis turned on are necessary to establish whether
   or not such similarities are fortuitous.}
  \label{fig:ejecta1}
\end{figure*}

\begin{figure*}
  \includegraphics[width=0.49\textwidth]{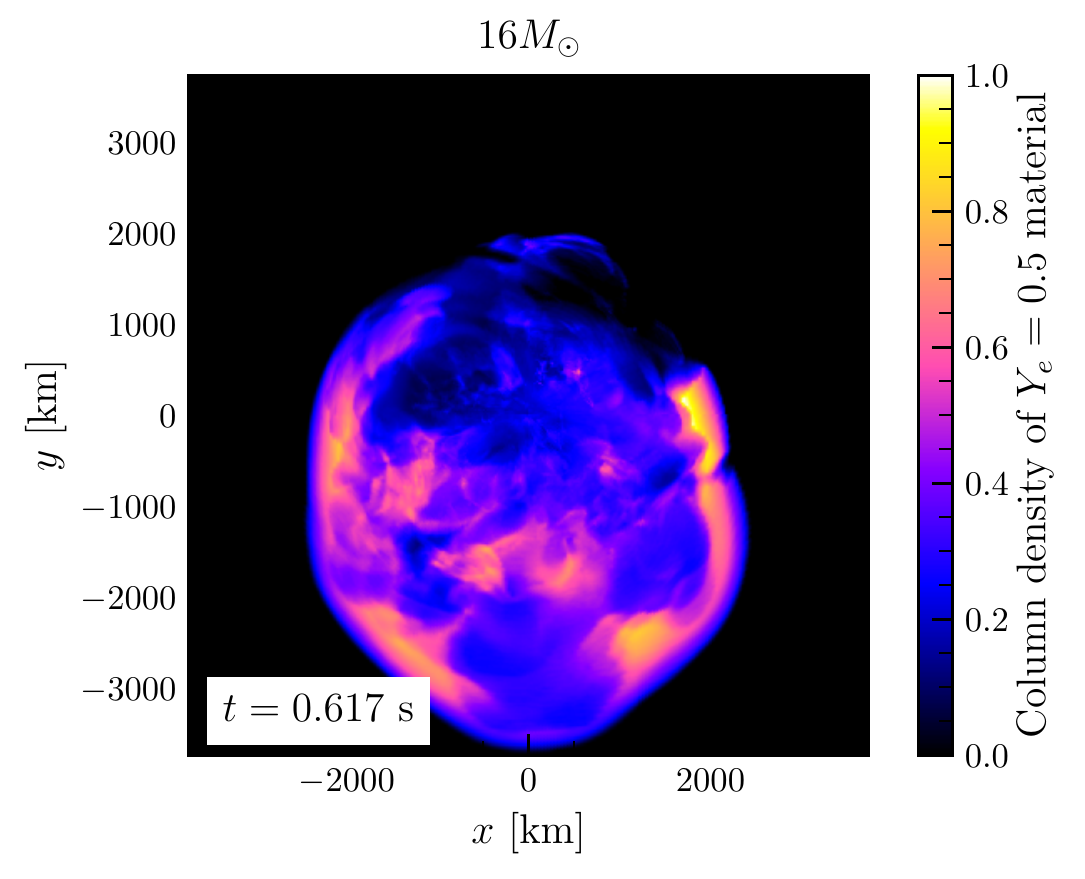}
  \hfill
  \includegraphics[width=0.49\textwidth]{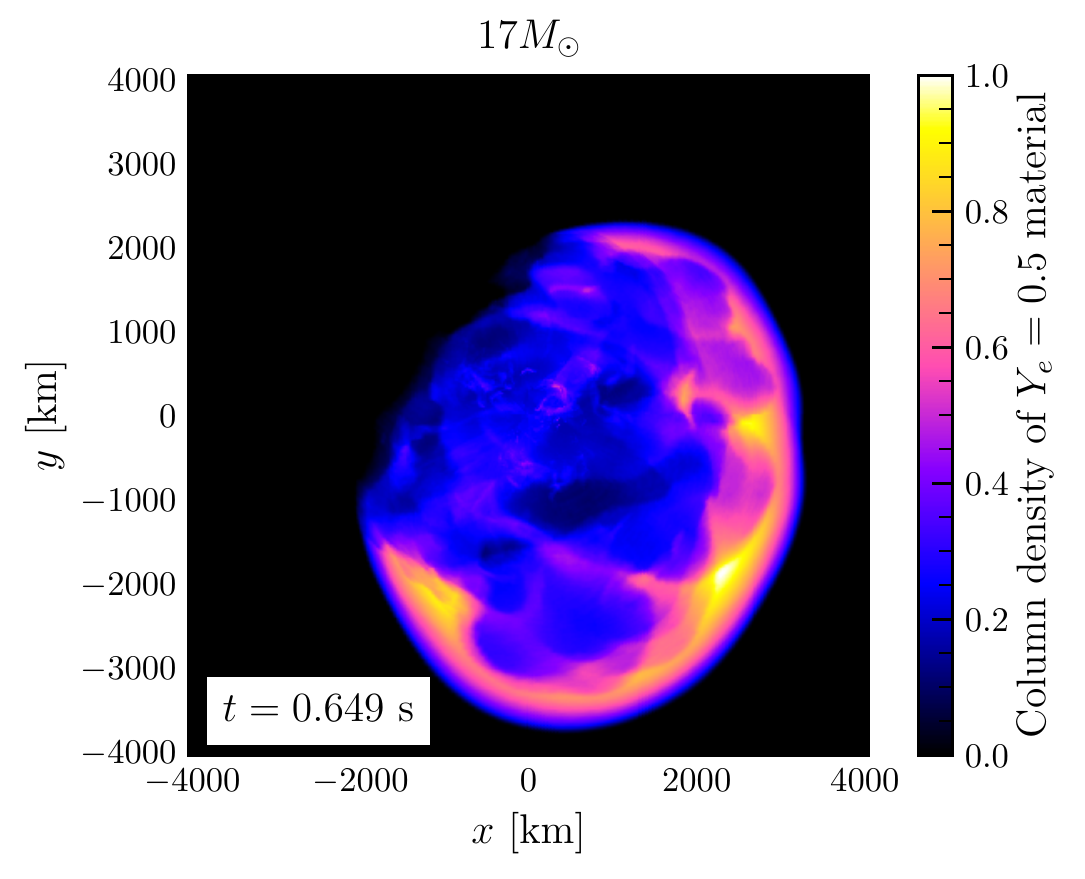}
  \includegraphics[width=0.49\textwidth]{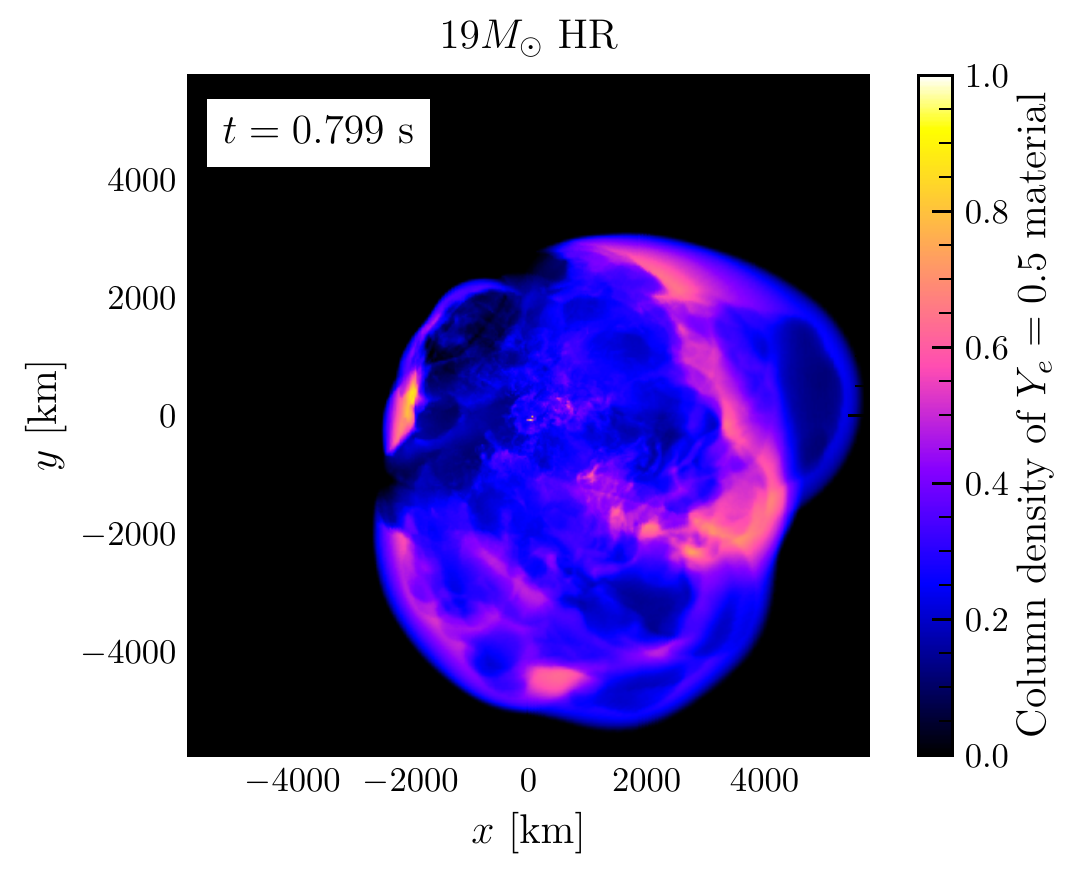}
  \hfill
  \includegraphics[width=0.49\textwidth]{Ni56_s19_0.pdf}
  \includegraphics[width=0.49\textwidth]{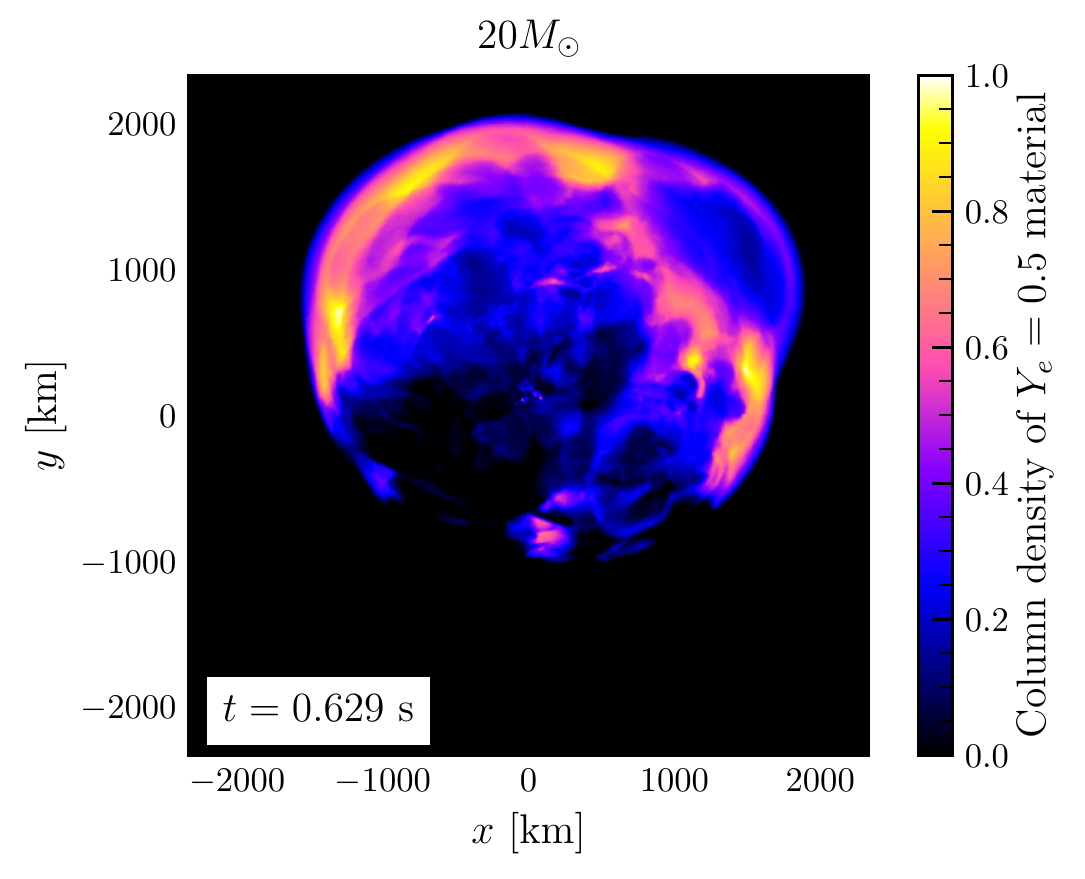}
  \hfill
  \includegraphics[width=0.49\textwidth]{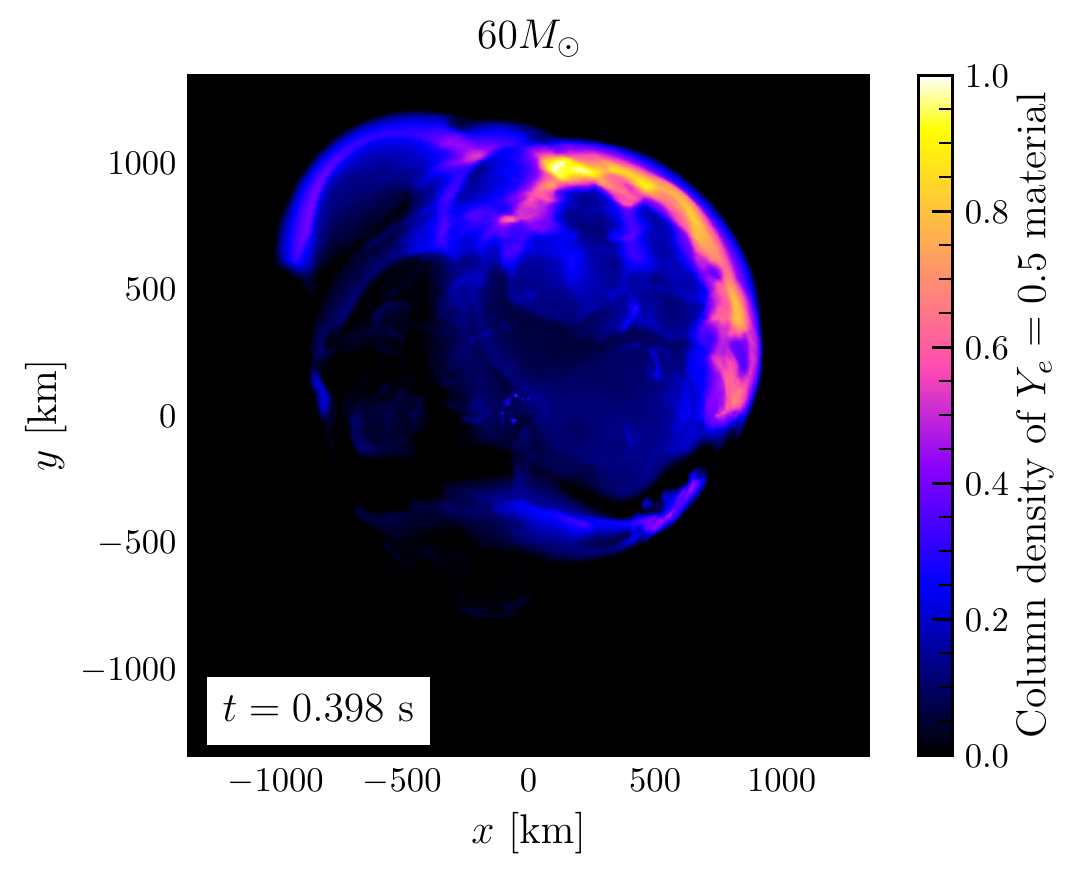}
  \caption{The same as Figure \ref{fig:ejecta1}, but for
   the 16-, 17-, 19 (HR)-, 19-, 20-, and 60-M$_{\odot}$ simulations.  The default 
   19-M$_{\odot}$ model is repeated here for the purpose of direct comparison with its
   high-resolution variant.}
  \label{fig:ejecta2}
\end{figure*}

\subsection{Ejecta}
\label{ejecta}

A calculation should be continued to the end of the explosive phase, at which point
the mass cut between the ejecta and the residual neutron star (we surmise, the majority 
of the time) or black hole would be determined.  The ejecta mass and total explosion energy 
will have then asymptoted (\S\ref{energy}) to their final values.  However, our simulations, except
for that for the 9-M$_{\odot}$ progenitor, were truncated before asymptoting.  Nevertheless,
one can still derive general trends in the ejecta mass from the high-density inner region.
Recall, that our computational grid extends to 20,000 km.  
{We designate the matter on the grid whose Bernoulli constant (``$\epsilon
+ P/\rho +v^2/2 -GM/r$", where ``M" is the mass interior to the given radius) is 
positive as the ejecta mass (M$_{ej}$) and track this quantity with time after bounce until the end of a run.}
{Here, $\epsilon$ is the specific
internal thermal energy.} This number should be a good measure of the outer material 
of what was the white-dwarf-like progenitor blown off the compact residue in the explosion.  
It includes the freshly-minted iron-peak material and almost all material processed during
explosion by the emerging $\nu_e$ and $\bar{\nu}_e$ neutrinos.  Figure \ref{fig:mej} depicts
the evolution with time after bounce of M$_{ej}$ for all the eleven exploding baseline
progenitors in this paper.  The ejecta masses at the end of each respective simulation range from about a percent of 
a solar mass to many tenths of a solar mass for the 25-M$_{\odot}$ model\footnote{The large 
inner ejecta mass for the 25-M$_{\odot}$ model 
suggests nuclear burning would boost its explosion energy even further.}. 
The 9-M$_{\odot}$ model has asymptoted to $\sim$5$\times$10$^{-2}$ solar masses.  The 10-M$_{\odot}$ model takes a while
to reach 10$^{-2}$ M$_{\odot}$, while the 11-M$_{\odot}$ model jumped to $\sim$5$\times$10$^{-2}$ M$_{\odot}$ early on.
The 19-M$_{\odot}$ model achieves nearly 0.15 M$_{\odot}$ by the end of its run.  
The 20- and 25-M$_{\odot}$ models, the latter taking the most time, eventually experience phases 
of rapid growth in M$_{ej}$. This is due to their shallower initial density profiles (Figure \ref{density}) $-$ there
is a lot of mass to work with, though this also delays explosion.  Similar, but to a lesser degree, 
are the 16- and 18-M$_{\odot}$ models, with M$_{ej}\sim 0.05$ and $\sim$0.06 M$_{\odot}$, respectively.

The electron fraction (Y$_e$) distributions of these ejecta are given in Figure \ref{fig:hist.ye}.  Most of the ejecta
have Y$_e = 0.5$ and this material will likely result in radioactive $^{56}$Ni (and then stable $^{56}$Fe).  However,
most of the rest of these inner ejecta have higher Y$_e$s on the proton-rich side.  This is the result of net $\nu_e$
absorption that exceeds in effect $\bar{\nu}_e$ absorption to elevate Y$_e$s above that of symmetric matter.  Importantly, 
this effect depends upon the speed with which the ejecta leave the core.  For models, such as the 9-M$_{\odot}$ {and 12-M$_{\odot}$} 
models, which explode and eject matter quickly, some of the ejecta will not have had time to transition from neutron-rich 
to proton-rich through the agency of net $\nu_e$ neutrino absorption.  These models will be a source of 
{some} neutron-rich material.  {At the end of each simulation, these neutron-rich ejecta 
have reached entropies per baryon per Boltzmann's constant around $\sim$35 for the  9-M$_{\odot}$ model and around 
$\sim$20 for the 12-M$_{\odot}$ model.}  Those other models that explode later and initially more slowly seem to have enough time  
to elevate the Y$_e$ of more of their ejecta to proton-rich values.  {Apart from the disassembly-speed dependence, the 
fundamental tendency to eject proton-rich material in CCSNe is related to the electron-lepton excess of the PNS.  
Due to electron-neutrino trapping on infall, there will naturally be a net electron-lepton 
excess in the emitted neutrinos, which, if given enough time, will push the ejecta Y$_e$ upward.  
However, the possible effects on this conclusion of neutrino oscillations, not addressed in this study, have yet to be ascertained.} 

{Nevertheless}, this progenitor- and explosion-speed dependence is an intriguing conclusion that has important nucleosynthetic 
consequences.  In any case, all our 3D models show a preference for proton-rich ejecta and this, if true, has a consequence for 
the isotope yields and nucleosynthesis of core-collapse supernovae. Note that this conclusion is contingent upon the proper 
handling of neutrino transport and is provisional, but is highly suggestive. Similar effects have been seen 
by \citet{2018ApJ...855..135B} and \citet{2018ApJ...866..105B} for PNS winds, and something like a wind component 
is contributing here. A note of caution, however, is in order.  Two-dimensional results for the same progenitors 
can show different ejecta Y$_e$ distributions that have more neutron-rich ejecta \citep{2018MNRAS.477.3091V} 
(though the ejecta are still predominantly proton-rich). This is mostly a consequence of the different trajectory 
histories of individual matter parcels during the early explosive phases $-$ 2D and 3D dynamics are not the same, though
the neutrino emissions can be similar (see Figure \ref{fig:nulums} and \S\ref{overview}). 

As noted, however, most of these inner ejecta have a Y$_e$ of 0.5 and we expect this material 
to {include} $^{56}$Ni.  {We emphasize, however, that we have not 
performed the necessary nucleosynthesis calculations, nor incorporated tracer particles
to facilitate such calculations, and have left this to future work.}  
Figures \ref{fig:ejecta1} and \ref{fig:ejecta2} show column density maps of 
the {``Y$_e$ = 0.5" regions wherein we expect whatever $^{56}$Ni 
that is created for a sample of our exploding models to reside.  
These figures should not be interpreted to indicate perfectly our  
produced $^{56}$Ni, but merely to indicate in a rough fashion where we expect it to reside.}
The distributions vary significantly from model to model, some having roughly symmetric angular distributions
and others very asymmetrical angular distributions.  All, however, have shells of $^{56}$Ni and are not fully 
filled in. Both these observations may have consequences for the observed distributions of the iron-peak 
elements in supernova light curves, spectra (via line profiles), and remnants.

\section{Brief Sensitivity Investigation}
\label{sensitivity}

It is important to gauge the sensitivity of simulation results to variations 
in input physics and methodologies; it is in this way that the qualitative import of 
uncertainties in the relevant physics and of approximations in the numerical schemes 
can be ascertained.  To this end, there is already a large literature spanning decades wherein 
the dependence of CCSN dynamics on changes in the physics has been inspected.
However, given the complexity of the overall core-collapse supernova simulation 
enterprise, this is not something easily determined nor quantified.  Nevertheless, 
such efforts are ongoing and modern codes incorporate many of the lessons learned.

There have not, however, been many such studies to determine the consequences 
of such alterations in the context of full 3D simulations.  Until recently, 
this would have been prohibitively expensive.  In this section, we provide a few such 
comparisons, altering only a few aspects of 3D simulations.  These include 
the angular spatial resolution, the effect of the \citet{2017PhRvC..95b5801H} many-body correction
to the axial-vector term in the neutrino-nucleon scattering rates, and the use of a monopole
versus a multipole \citep{1995CoPhC..89...45M} gravity expansion.  \citet{2019arXiv190503786N}, 
to which the reader is referred, have provided more details and interpretation for the 
angular resolution study of the 19-M$_{\odot}$ model, but here we augment that study with a few 
additional observations. In addition, we contrast the behavior of 3D models with and 
without the many-body correction for both the same 19-M$_{\odot}$ progenitor and 
our 11-M$_{\odot}$ progenitor, at our standard resolution (\S\ref{methods}). 
The 19-M$_{\odot}$ progenitor is also the context of our single multipole/monopole comparison.

\begin{figure}
  \includegraphics[width=\columnwidth]{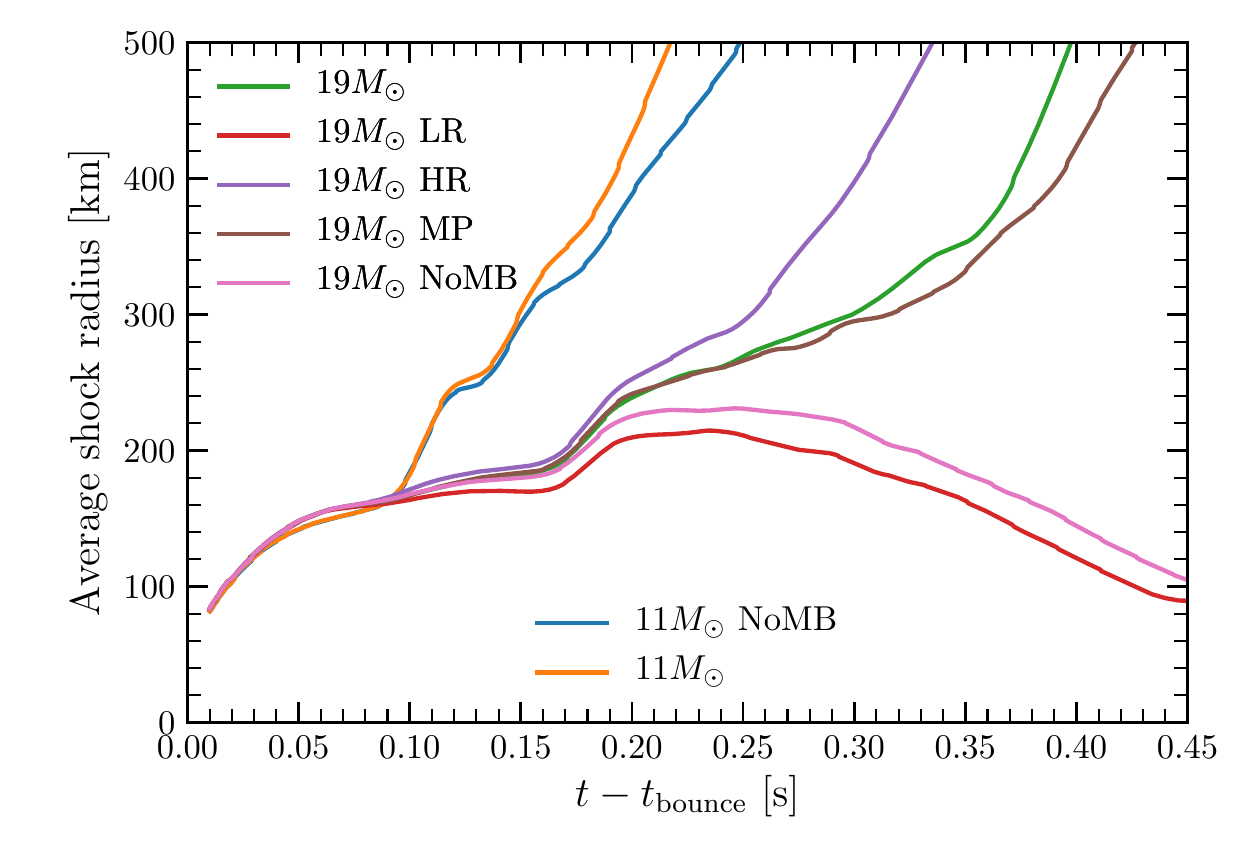}
  \caption{Shock radius evolution for the 11-M$_\odot$ and 19-M$_\odot$
  progenitors, with and without many-body corrections to the neutrino-nucleon
  scattering cross section. The inclusion of many-body effects
  is crucial for the explosion of the 19-M$_\odot$ progenitor, and
  strengthens the explosion of the 11-M$_\odot$ progenitor.}
  \label{fig:sens.rshock}
\end{figure}

\begin{figure}
  \includegraphics[width=\columnwidth]{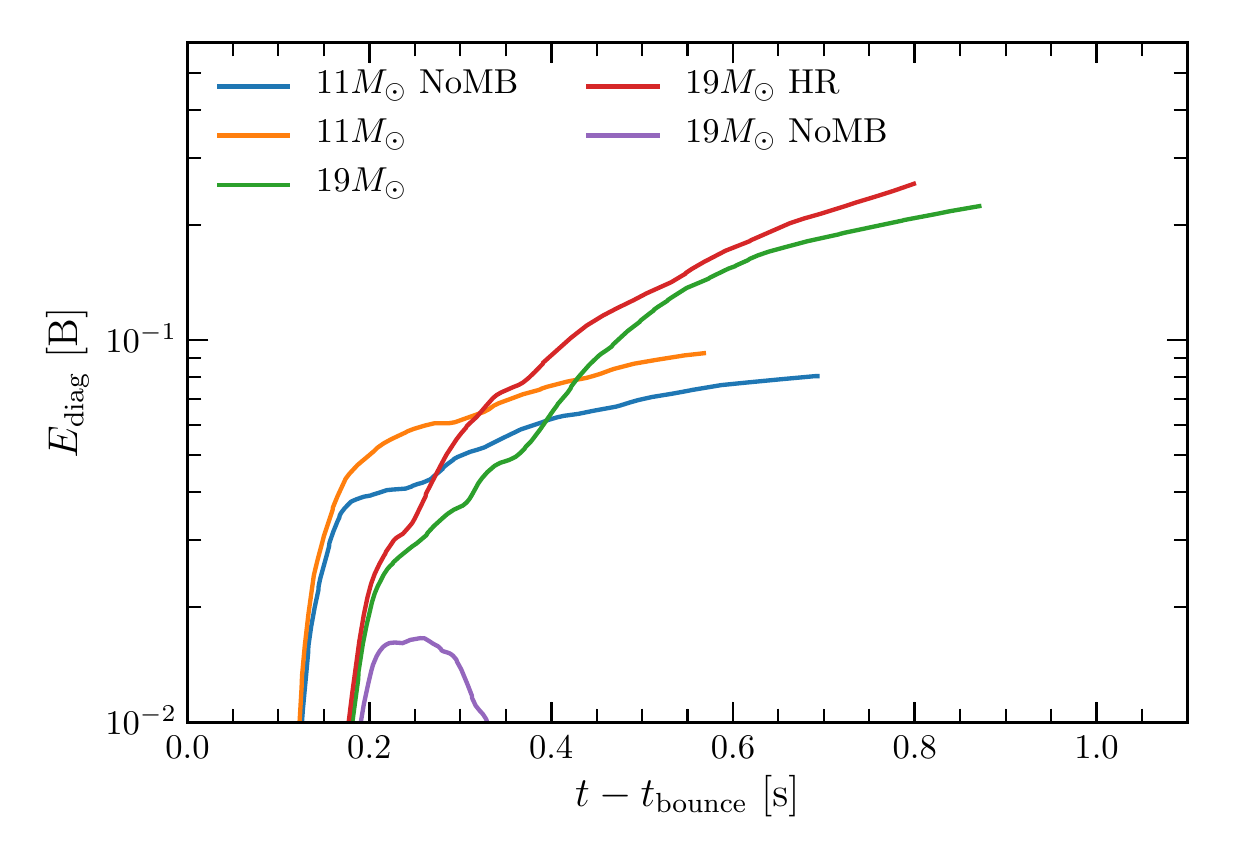}
  \caption{Diagnostic explosion energy for the 11-M$_\odot$ and 19-M$_\odot$
  progenitors, with and without many-body corrections to the neutrino
  nucleon scattering cross-section. See the text for a discussion.}
  \label{fig:sens.explene}
\end{figure}

\begin{figure}
  \includegraphics[width=\columnwidth]{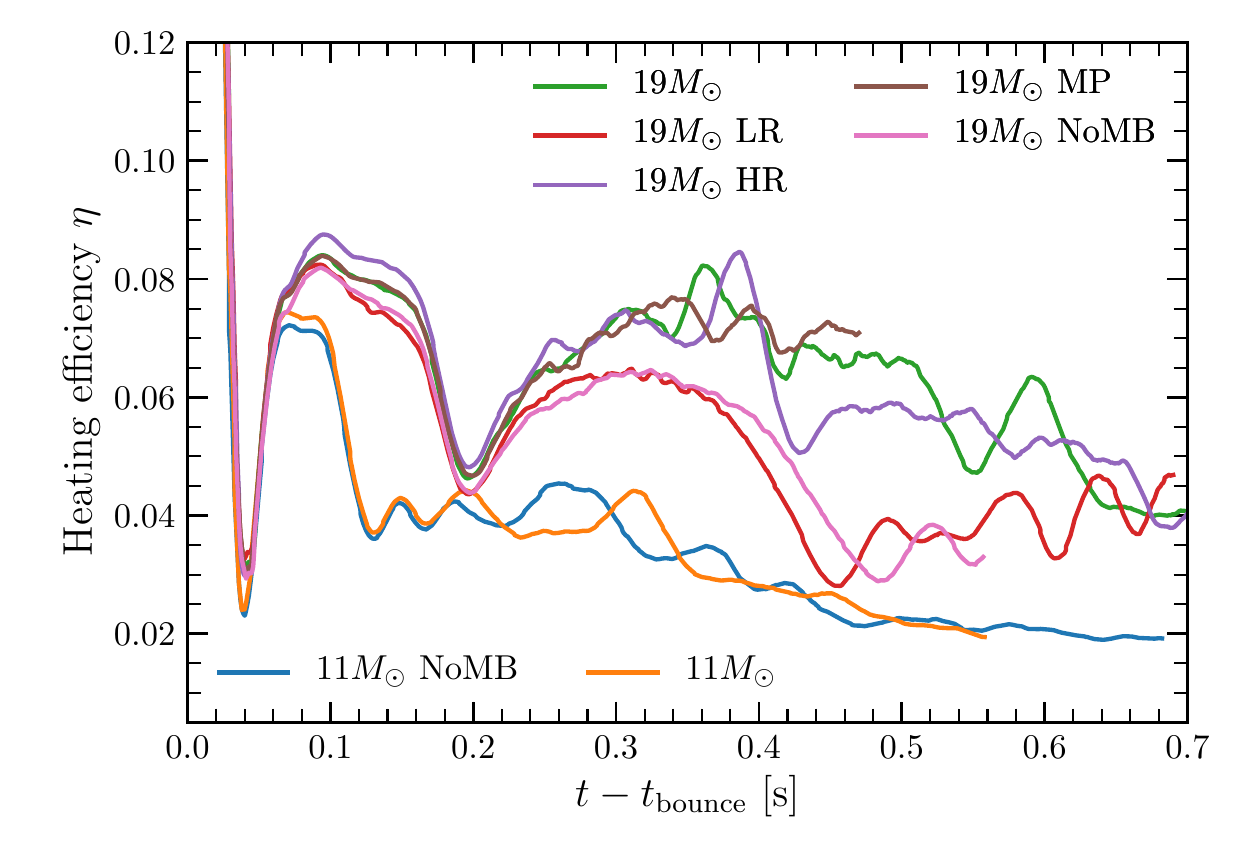}
  \caption{Heating efficiencies for the 11-M$_\odot$ and 19-M$_\odot$
  progenitors, with and without many-body corrections to the neutrino-nucleon
  scattering cross section. The inclusion of many-body effects
  leads to a more rapid contraction of the PNS, resulting in slightly
  higher neutrino rms energies, and, consequently, higher heating
  efficiencies. See text for a discussion.}
  \label{fig:sens.eta}
\end{figure}

Figure \ref{fig:sens.rshock} provides a comparison of the evolution of the mean shock 
radius with time after bounce for all the models in our modest sensitivity study.  
From a comparison of models 19-M$_{\odot}$ (default: 678$\times$128$\times$256, green), 
19-M$_{\odot}$-LR (low-resolution: 678$\times$64$\times$128, red), and 19-M$_{\odot}$-HR 
(high-resolution: 678$\times$256$\times$512, purple) we see that if the resolution is too low a model
that otherwise explodes will not.  This is, of course, a qualitative difference and 
is explained and analyzed in more detail in \citet{2019arXiv190503786N}.  The increased numerical
viscosity at lower resolution inhibits the turbulent pressure important in almost all 
neutrino-driven models of explosion.  We also see that the higher resolution model explodes 
earlier.  This result puts a premium on spatial resolution as a factor in the interpretation of
model results in the literature.  We note that this 19-M$_{\odot}$-HR model is one of the highest 
resolution 3D supernova models ever performed using a spherical grid. 

From Figure \ref{fig:sens.rshock}, we learn that, whereas the many-body correction makes little
qualitative difference for the 11-M$_{\odot}$ progenitor (11-M$_{\odot}$ versus 11-M$_{\odot}$-NoMB), 
without it (19-M$_{\odot}$-NoMB, magenta) our otherwise default 3D 19-M$_{\odot}$ model does not explode.  
The density profile of the 11-M$_{\odot}$ progenitor all but ensures explosion for a range of microphysics, 
but to get the 19-M$_{\odot}$ model (and, presumably, other more massive progenitors) to explode the many-body
correction, as we have currently implemented it \citep{2017PhRvC..95b5801H}, has proven supportive.
{The many-body effect decreases slightly the neutrino-nucleon scattering rate, thereby accelerating
the shrinkage of the core.  This raises by the resulting compression the temperatures around the $\nu_e$ and $\bar{\nu}_e$
neutrinospheres and, as a result, the heating rates due to absorption on nucleons near the stalled shock wave.
This facilitates explosion.} What the effect may be of anticipated improvements down the road in this class of corrections is 
yet to be determined \citep{1998PhRvC..58..554B,sawyer1999,roberts2012,roberts_reddy2017}.

Also on Figure \ref{fig:sens.rshock}, we find that there is little difference between models using 
the full multipole gravitational expansion (19-M$_{\odot}$-MP) and those that retain only the monopole.  
This is due to the strong central concentration of the generic core-collapse structure 
and the fact that all our initial models are non-rotating.  

Figure \ref{fig:sens.explene} plots the evolution with time after bounce of the diagnostic energy 
of exploding models.  We see that the many-body correction increases the explosion energy of 
the 11-M$_{\odot}$ progenitor by $\sim$20\% and that higher resolution does the same (at least 
in this comparison study) for the 19-M$_{\odot}$ model. These are not qualitative differences, but 
important ones, as we attempt to determine, or at least bracket, the salient quantities 
of theoretical CCSN explosions.

Figure \ref{fig:sens.eta} displays the heating efficiencies ($\eta$) for all our sensitivity calculations.  The 
efficiency is defined as the ratio of the neutrino power deposition rate by $\nu_e$ and $\bar{\nu}_e$ absorption
in the gain region behind the shock wave and the sum of the angle- and group-integrated $\nu_e$ and $\bar{\nu}_e$
luminosities.  This number does not include the subdominant heating rate due to inelastic scattering, though the simulations 
do. $\eta$ is approximately a measure of the ``optical depth" to neutrino absorption and ranges from $\sim$4\% 
to $\sim$8\%.  Core-collapse supernovae are a ``5$-$10-\%" effect, not the ``$\sim$1\%" effect often quoted.  
We see that during the first $\sim$0.2 seconds there is little difference between the various 
models with the same progenitor mass.  The high-resolution 19-M$_{\odot}$ model does have a slightly
higher energy deposition rate than the default model, and higher still than the low-resolution realization.
This is one of the reasons for the qualitative difference in the outcomes (HR versus LR) \citep{2019arXiv190503786N}.  
In addition, the default 11-M$_{\odot}$ model with the many-body correction has a 
$\sim$3\% higher heating rate early on, but in a time-averaged sense
is not much different after explosion.  Not unexpectedly, the comparison between the two models
with and without the higher-order multipole gravity terms reveals no appreciable differences.  One thing we do notice
is that the efficiency is not a good predictor or index of explosion, since models such as 19-M$_{\odot}$-NoMB
and 19-M$_{\odot}$-LR have approximately the same $\eta$ history, but only one explodes, and the $\eta$
values for the two non-exploding 19-M$_{\odot}$ models are higher than those for the exploding 11-M$_{\odot}$ models.

The small set of 3D simulations we have performed here to address some sensitivity issues can not in any way be construed
as definitive, nor adequate to the general task of exploring the important dependences of CCSN theory on
the outstanding ambiguities concerning progenitors, microphysics, resolution, and numerical technique.  This
is the ongoing program for the community of supernova theorists.  However, ours are some of the first to be performed 
in 3D with a state-of-the-art supernova code, and, as such, are meant in part to indicate what is now possible.

%% -------------------------------------------------------
\section{Conclusions}
\label{conclusions}

For this paper, we have conducted and assembled for analysis nineteen state-of-the-art 
3D core-collapse supernova simulations spanning a broad range of progenitor masses and structures.  
This is, we believe, the largest such collection of sophisticated 3D supernova simulations ever performed.
A goal was to determine the behavior of the family of CCSN progenitors, not just one
model at a time, but collectively, and to determine overarching trends vis \`a vis explodability
and outcomes.  We have found that while the majority of this suite explode, not all do, and that 
even models in the middle of the available progenitor mass range may be less explodable.
This does not mean that those models for which we did not witness explosion would not explode 
in Nature, but that they are less prone to explosion than others in this cohort. One clear
consequence is that the ``compactness" measure is not a metric for explodability $-$ we,
find as have others \citep{2012ApJ...757...69U,2015ApJ...806..275P,2016ApJ...818..124E},  that models
with both low and high compactness can explode, but that some with an intermediate
value may not. As we have discussed in previous work \citep{2018SSRv..214...33B}, 
since a core-collapse supernova explosion is a critical bifurcation, explodability is 
still sensitive to the detailed microphysics and numerical schema. 
Despite our attempts here to incorporate the necessary realism and address all the major
issues with the latest methods and physics, every feature of our \fornax\ implementation
and simulations should be considered provisional.  The supernova theory community
continues its decades-long investigations into neutrino-matter interactions, the nuclear
equation of state, and massive star evolution and progenitors with the goal of obtaining
a robust understanding of the core-collapse supernova phenomenon. This paper, though it 
contains an unprecedentedly large set of 3D CCSN simulations, is but one contribution
to this ongoing collective effort.    

We found that a preponderance of lower-mass massive star progenitors likely experience lower-energy explosions,
while the higher-mass massive stars likely experience higher-energy explosions. The latter explode 
a bit latter after bounce than the former, so time of explosion seems weakly correlated or anti-correlated with 
explosion vigor. However, this is a statistical statement and we have not determined the full range of 
possible explosion energies for a given progenitor in the context of chaotic turbulence and chaotic 
initial models. Not unexpectedly, we confirm in 3D that neutrino-driven turbulence behind the stalled 
shock wave is a major factor in the viability of the neutrino-driven mechanism of CCSN.  Moreover, 
as was determined in \citet{2019MNRAS.482..351V} and \citet{2019MNRAS.485.3153B}, most 3D models 
have a dominant dipole morphology, have a pinched, wasp-waist early structure, and experience simultaneous 
accretion and explosion.  Continuing accretion during explosion maintains the neutrino power
during the crucial early launch phase.

Coupled with the earlier calculations of \citet{2017ApJ...850...43R} concerning the sources 
from $\sim$8$-$8.8 M$_{\odot}$ progenitors of the lowest-mass pulsars (down to $\sim$1.17 
M$_{\odot}$ gravitational; \citep{martinez:2015mya}), we have now been able to 
reproduce in a qualitative sense the general range of residual neutron-star 
masses inferred for the galactic neutron-star population. However, the mapping of massive star mass function
to initial neutron star mass function has not been attempted and is likely a job for the future.
One of our most important conclusions is that the most massive progenitor models need to be continued for longer physical 
times, perhaps to many seconds, to asymptote to a final state, in particular vis \`a vis explosion energy.
This seems to be a firm conclusion of our 3D study, and was anticipated by \citet{bmuller_2015}.
Moreover, we find that while the majority of the inner ejecta have Y$_e = 0.5$, there is a substantial proton-rich 
tail.  Those models that explode more lethargically and a bit later after bounce tend not to include
much neutron-rich ejecta, while those that explode more quickly, such as the lowest-mass progenitors 
(e.g., the 9-M$_{\odot}$ model), can ejecta some more neutron-rich matter.  However, in all 
our 3D models, the inner ejecta have a net proton-richness.  If true, this systematic result has important 
consequences for the nucleosynthetic yields as a function of progenitor. 

We find that the non-exploding models eventually evolve into compact inner configurations that experience
a quasi-periodic spiral SASI mode. We otherwise see little evidence of the SASI in the exploding
models, except during a brief period at early post-bounce times for the 25-M$_{\odot}$ model.  For the latter 
model, the slightly smaller initial post-bounce shock radius, by dint of the greater early accretion it experiences,
is likely responsible for this transient phase. 

We are now in a position to articulate the features of a progenitor and physics model supportive of explosion.
Foremost, perhaps, is the initial progenitor mass density profile $-$ all else being equal, such structures
determine the outcomes of collapse.  Associated is the seed perturbation field inherited
from the pre-collapse core.  Jump starting and continuing to seed turbulent convection behind the stalled shock 
wave is necessary for a vigorous outcome, though the detailed character of the requisite seed turbulence
has yet to be determined.  Along with these two aspects of a progenitor is a third, the presence of
a sharp silicon/oxygen interface.  We here and elsewhere \citep{2018MNRAS.477.3091V} confirm that 
explosion is ofttimes inaugurated upon accretion of this interface.  There is a delay of tens of milliseconds 
between accretion through the shock and the response of the emergent neutrino luminosities to the consequent
decrease in accretion rate, with the result that the countervailing effect of the accretion ram pressure
is temporarily diminished.  The upshot is often explosion. 

Though we have not addressed this in this paper, increasing the mean dwell time in the gain region of a
given parcel of newly-shocked matter increases the exposure of that parcel to neutrino heating and facilitates
explosion \citep{Murphy:2008dw}.  Turbulence behind the stalled shock wave does just this, and such an enhancement is one positive
feature of multi-dimensional motions absent in spherical models.  However, as has been made clear in numerous 
publications, the Reynolds stress itself of the neutrino-driven convection behind the shock wave contributes 
centrally to explosion and may be the most important aspect of multi-dimensional turbulent motion.  
Converting some of the accretion gravitational energy into turbulence channels energy 
into a component (turbulence) that, if it were a gas, would have an effective $\gamma$
of $\sim$2 (not $4/3$) and would be anisotropic in the radial direction \citep{Murphy:2008dw}.  
This means that turbulence is an effective means to generate needed outward ``pressure" 
stress \citep{1995ApJ...450..830B,2015ApJ...799....5C,2019arXiv190503786N} behind the shock wave. Turbulence, 
hence, is more effective than fluid pressure for the same energy density.

Of course, central to the neutrino mechanism of core-collapse supernova explosions is the power
deposited by the $\nu_e$ and $\bar{\nu}_e$ neutrinos in the gain region behind the shock $-$ this is the ultimate source 
of the supernova energy when the rotation rate is small. Though charged-current absorption 
on nucleons dominates this rate, inelastic neutrino-electron and neutrino-nucleon scattering 
play positive roles, perhaps in aggregate by as much as $\sim$10$-$15\% \citep{2018SSRv..214...33B}.  
The large energy transfer of neutrino-electron scattering happens at a small rate
and the small energy transfer of neutrino-nucleon scattering occurs at a more rapid rate.  The upshot is
a comparable (to within a factor of $\sim$2) contribution, though neutrino-electron scattering seems generally 
more important.

The many-body correction of \citet{2017PhRvC..95b5801H} to the axial-vector term in neutral-current neutrino-nucleon
scattering is also a factor.  In particular, the resultant decreased interaction cross sections for such scattering lead 
to a more rapid loss of $\nu_{\mu}$, $\bar{\nu}_{\mu}$, $\nu_{\tau}$, and $\bar{\nu}_{\tau}$ neutrinos.  This accelerates the contraction
of the inner core, with the result that the temperatures around the $\nu_e$ and $\bar{\nu}_e$ neutrinospheres
are increased.  Increased temperatures harden their emergent spectra and, since the rate of charged-current absorption 
goes as the square of the neutrino energy, the heating rates in the gain region increase. Therefore, and ironically,
enhanced energy leakage into a less productive channel (the ``$\nu_{\mu}$s") facilitates explosion.  This is similar to the
published effect of general-relativity $-$ despite the redshifting of the emergent neutrinos and the deeper potential well, 
more compact relativistic configurations aid explosion \citep{2001ApJ...560..326B,2012ApJ...756...84M}. However, the full set of
many-body corrections has not yet been calculated \citep{1998PhRvC..58..554B,sawyer1999,roberts2012,roberts_reddy2017} 
nor implemented, so their ultimate effect has yet to be determined.

We also note that PNS convection in the inner core around $\sim$20$\pm5$ km increases the ``$\nu_{\mu}$" loss rate,
performing, though to a lesser degree, a similar function to that of the many-body correction \citep{2017ApJ...850...43R}.
The effect is similar in both 2D and 3D simulations (H. Nagakura et al., in preparation).

The nuclear EOS is a perennial central issue. \citet{2018NatAs...2..980F} have explored the possible 
effects of a quark-hadron phase transition at super-nuclear densities.  \citet{2019arXiv190602009S} have 
shown that a higher effective neutron mass near nuclear density can aid explosion,
through its effect on nuclear specific heats and, hence, on the temperatures achieved during and after
collapse. The EOS we have employed (SFHo) has an effective mass near 0.7$\times$$m_n$, while \citet{2019arXiv190602009S}
find that values close to 1.0$\times$$m_n$, such as are found in the LS220 EOS \citep{1991NuPhA.535..331L}, 
could support greater explodability. However, the LS220 EOS and such a high effective mass currently 
seem incompatible with known nuclear constraints \citep{2017ApJ...848..105T}.  Another physical effect
that may have some bearing on the question of explodablity and that has an indirect EOS connection 
is the electron capture rate during infall \citep{2016ApJ...816...44S,2018JPhG...45a4004T,2019ApJS..240...38N}. 
This rate depends upon the free proton abundance, an 
EOS-dependent quantity that, along with the capture rate on heavy nuclei, determines the 
rate of electron loss.  The loss of electrons translates into a loss of pressure that affects the rate of 
collapse and time to bounce.  Hence, variations in the total capture rate result in variations in
the time to bounce \citep{2012ApJ...747...73L}.  Since alterations in the time to bounce affect the timing of the subsequent
mass accretion of the outer core onto the inner core, and the $\dot{M}$ history factors into the 
explodability, scrutiny of these issues in the future could bear fruit.  In addition, the stiffness of 
the EOS at high density will help determine the size of the inner core, the depth of the gravitational
potential well, and the neutrinopshere temperatures.  These factors influence the work against gravity 
needed to launch the ejecta, as well as the neutrino deposition powers in the gain region. 
So, there remain issues surrounding the nuclear EOS, not addressed in this paper, that could prove illuminating.

Finally, we have not in this paper looked into the possible effects of rotation and/or magnetic fields.
The latter, if the core is not rotating fast, are unlikely to alter our findings to an interesting degree.
Magneto-turbulence will be similar to hydrodynamic turbulence as far as aggregate stress is concerned.
Rapid differential rotation, on the other hand, has the potential to generate large magnetic stresses 
\citep{2007ApJ...664..416B,2014ApJ...785L..29M,2019arXiv190901105O}, with the result that strong jets can emerge.
However, it is thought that most pulsars are not born rotating fast \citep{2006ApJ...643..332F}
and that the majority of CCSNe are not magneto-rotationally powered. Nevertheless, it remains to 
determine whether even slow or modest rotation has a role to play in the overall context of CCSNe.  
An intriguing possibility is that even slow rotation might promote our recalcitrant 13-, 14-, and 15-M$_{\odot}$ 
models into explosion.  

With the advent of \fornax\ and the ongoing development of an international constellation 
of full-physics codes, multiple 3D simulations per year are now the new standard in core-collapse theory.
Not only does this finally ensure an extensive exploration of parameter space in the full 
three dimensions of Nature, but it mitigates the resource penalties of the few inevitable mistakes. 
Though much remains to be done, as a result of this extensive study using \fornax, we can 
now feel confident that a decades-long theoretical challenge is finally yielding many of its secrets.

\section*{Acknowledgements}
We acknowledge support from the U.S. Department of Energy Office of Science
and the Office of Advanced Scientific Computing Research via the
Scientific Discovery through Advanced Computing (SciDAC4) program and
Grant DE-SC0018297 (subaward 00009650). In addition, we gratefully
acknowledge support from the U.S. NSF under Grants AST-1714267 and
PHY-1144374 (the latter via the Max-Planck/Princeton Center (MPPC) for
Plasma Physics). DR cites partial support as a Frank and Peggy Taplin
Fellow at the Institute for Advanced Study. JD acknowledges 
support from the Laboratory Directed Research and Development program 
at the Los Alamos National Laboratory. Help with the equation of state 
(Evan O'Connor), electron capture on heavy nuclei (Gabriel Mart\'inez-Pinedo), 
the initial progenitor models (Tug Sukhbold and Stan Woosley), and inelastic
scattering (Todd Thompson) was provided.  We thank Joe Insley of ALCF
for visualization support. An award of computer time was provided 
by the INCITE program using Theta at the Argonne Leadership
Computing Facility, which is a DOE Office of Science User Facility
supported under Contract DE-AC02-06CH11357. In addition, this overall
research project is part of the Blue Waters sustained-petascale
computing project, which is supported by the National Science Foundation
(awards OCI-0725070 and ACI-1238993) and the state of Illinois. Blue
Waters is a joint effort of the University of Illinois at
Urbana-Champaign and its National Center for Supercomputing
Applications. This general project is also part of the
``Three-Dimensional Simulations of Core-Collapse Supernovae" PRAC
allocation support by the National Science Foundation (under award
\#OAC-1809073). Moreover, access under the local award \#TG-AST170045 to
the resource Stampede2 in the Extreme Science and Engineering Discovery
Environment (XSEDE), which is supported by National Science Foundation
grant number ACI-1548562, was crucial to the completion of this work.
Finally, the authors employed computational resources provided by the
TIGRESS high performance computer center at Princeton University, which
is jointly supported by the Princeton Institute for Computational
Science and Engineering (PICSciE) and the Princeton University Office of
Information Technology, and acknowledge our continuing allocation at the
National Energy Research Scientific Computing Center (NERSC), which is
supported by the Office of Science of the US Department of Energy (DOE)
under contract DE-AC03-76SF00098. This work was performed under the auspices
of the U.S. Department of Energy by Lawrence Livermore National 
Laboratory under contract DE-AC52-07NA27344 and has been assigned an
LLNL document release number LLNL-JRNL-787982-DRAFT.
This paper has also been assigned a LANL preprint \# LA-UR-19-28512.

\bibliographystyle{mnras}
\bibliography{references}

%\onecolumn

% Don't change these lines
\bsp    % typesetting comment
\label{lastpage}

\end{document}